\documentclass[article]{aa} 

\usepackage{graphicx}

\usepackage{mwe}
\usepackage{longtable}
\usepackage{siunitx}
\usepackage[table]{xcolor}
\usepackage{textcomp}
\usepackage{lastpage}
\usepackage{chemist}
\usepackage{multirow}
\usepackage{caption}
\usepackage[flushleft]{threeparttable}
\usepackage{subcaption}
\usepackage{float}
\usepackage{txfonts}
\usepackage{tablefootnote}
\usepackage{placeins}
\titlerunning{}

\usepackage{natbib,twoopt}

\usepackage[colorlinks=true, pdfstartview=FitV, linkcolor=blue, citecolor=blue, urlcolor=blue, breaklinks=true]{hyperref} 
\bibpunct{(}{)}{;}{a}{}{,}             
\makeatletter
  \newcommandtwoopt{\citeads}[3][][]{\href{http://adsabs.harvard.edu/abs/#3}%
    {\def\hyper@linkstart##1##2{}%
     \let\hyper@linkend\@empty\citealp[#1][#2]{#3}}}
  \newcommandtwoopt{\citepads}[3][][]{\href{http://adsabs.harvard.edu/abs/#3}%
    {\def\hyper@linkstart##1##2{}%
     \let\hyper@linkend\@empty\citep[#1][#2]{#3}}}
  \newcommandtwoopt{\citetads}[3][][]{\href{http://adsabs.harvard.edu/abs/#3}%
    {\def\hyper@linkstart##1##2{}%
     \let\hyper@linkend\@empty\citet[#1][#2]{#3}}}
  \newcommandtwoopt{\citeyearads}[3][][]%
    {\href{http://adsabs.harvard.edu/abs/#3}
    {\def\hyper@linkstart##1##2{}%
     \let\hyper@linkend\@empty\citeyear[#1][#2]{#3}}}
\makeatother

\usepackage{color}

\usepackage{color}

\usepackage{cleveref}

\usepackage{chemist}

\begin{document}
\title{Metallicity estimations of MW, SMC, and LMC classical Cepheids from the shape of the $V$- and $I$-band light curves\thanks{Full Table \ref{Tab:data_set} and metallicity estimations from Sect.~\ref{sect:validity_I} are only available at the CDS via anonymous ftp to cdsarc.u-strasbg.fr (130.79.128.5) or via
http://cdsarc.u-strasbg.fr/viz-bin/cat/J/A+A/}}

\titlerunning{Metallicity estimations of classical Cepheids from the shape of the $V$- and $I$-band light curves.}
\authorrunning{Hocd\'e et al. }

\author{V. Hocd\'e , R. Smolec , P. Moskalik, O. Zi\'ołkowska, R. Singh Rathour}

\institute{Nicolaus Copernicus Astronomical Centre, Polish Academy of Sciences, Bartycka 18, 00-716 Warszawa, Poland\\
email : \texttt{vhocde@camk.edu.pl}
}

\date{Received ... ; accepted ...}

\abstract{Estimating the metallicity of classical Cepheids is of prime importance for studying metallicity effects on stellar evolution and the chemical evolution of galaxies, as well as on the period-luminosity relation used on the extragalactic distance scale.}
{Our first aim is to establish new empirical relations for estimating the iron content of classical Cepheids for short and long periods based on Fourier parameters from the $V$- and $I$-band light curves. We go on to apply these relations to Cepheids from data on the Milky Way (MW) as well as the Small and Large Magellanic Clouds (SMC and LMC) from the literature.}{We retrieved the metallicities of 586 fundamental-mode Cepheids from spectroscopic determinations in the literature and we found well-sampled light curves for 545 of them in different $V$-band catalogs. We then described the shape of these light curves by applying a Fourier decomposition and we fit the empirical relations between the Fourier parameters and the spectroscopic metallicities individually, for short-period ($2.5<P<6.3\,$days) and long-period Cepheids ($12<P<40\,$days). We verified the accuracy of these relations by applying them to $V$-band light curves of Cepheids from the Small and Large Magellanic Clouds and comparing these derived metallicities to literature values. We calibrated new interrelations of Fourier parameters to convert these empirical relations into the $I$ band. We then used these $I$-band relations to derive the metallicity of fundamental-mode Cepheids from OGLE-IV for MW, SMC, and LMC (486, 695, and 1697 stars, respectively). Finally, we mapped the metallicity distribution in these galaxies for the purpose of investigating potential applications in galactic archeology.} {For short-period Cepheids, our best fit is given for a relation based on explicit amplitude terms $A_1$ and $A_2$ of the first and second harmonic, respectively. In the $V$ and $I$ bands, these empirical relations are found with an intrinsic scatter (rms) of 0.12$\,$dex. This relation performs well for estimations of [Fe/H] between about $-0.5$ and 0.1$\,$dex, but it remains uncertain outside this range because of the lack of a spectroscopic metallicity required for the calibration. For long-period Cepheids, we found a metallicity dependence on the Fourier parameters $A_1$, $\phi_{21}$, and $R_{41}$. We found an intrinsic scatter of $0.25\,$dex when using this relation. The empirical relations in the $V$ and $I$ bands allow us to  derive the mean metallicity of a sample of MW, SMC, and LMC Cepheids that is in agreement with literature values within 1$\sigma$. We also show that these relations are precise enough to reconstruct the radial metallicity gradients within the MW from  OGLE data.}{The empirical relations in the $V$ and $I$ bands that are calibrated in this work for short- and long-period Cepheids provide a  useful new tool for estimating the metallicity of Cepheids that are not accessible via spectroscopy. The calibration can be improved with further high-resolution spectroscopic observations of metal-poor Cepheids and homogeneous photometries in the $V$ and $I$ bands.}
{}

\maketitle

\section{Introduction}\label{Intro}
Cepheids are yellow supergiant variable stars that are essential for  distance-scale determinations as a result of the correlation
between their pulsation period and their luminosity (hereafter, the PL
relation) \citep{leavitt08,Leavitt1912}. However, the PL relation is still affected by uncertainties on both the zero point and slope, which represents one of the main error contribution to the extragalactic distance scale \citep{Riess2022}. 

Chemical composition plays an important role in terms of the evolution and the brightness of Cepheids. As a result, metallicity introduces a bias on the zero point when the PL relation is calibrated in different galaxies.
The sign of the metallicity term in the period-luminosity-metallicity (PLZ) relation, which also depends on the given passband, is still under debate from theoretical and observational standpoints, as, for example, in \cite{Caputo2000,Bono2008,Fiorentino2013,Gieren2018,Ripepi2021,Breuval2021,Wielgorski2022,Breuval2022,DeSomma2022}.

Determining the metallicity for a large number of Cepheids is also crucial for constraining pulsation and evolution models.  Accurate predictions of Cepheid evolution as well as its pulsation properties are necessary to answer questions related to the mass discrepancy among Cepheids \citep[see, e.g.,][]{neilson11}. The derived stellar mass
can be up to 20\% lower in pulsation models as compared to stellar evolution models; this discrepancy is still unresolved after 50 years of research \citep[see, e.g.,][]{stobie69,bono06,keller08}.

Until now, high-resolution spectroscopic observations have led to determinations of the iron-to-hydrogen ratio [Fe/H] for several hundreds of Cepheids, which are mostly in the vicinity of the Sun (within about 5$\,$kpc). Indeed, these observations are difficult to apply to faint stars that  are distant or that are located in the line of sight of a significantly reddened environment such as the Galactic Center. However, this limitation can be overcome thanks to medium- and high-resolution near-infrared (NIR)\ spectroscopy \citep{Inno2019,Kovtyukh2019,Kovtyukh2022IR}. An interesting alternative to spectroscopy is to carrying out estimations of the metallicity of the Cepheids using the shape of the light curve, since it contains information on its physical properties such as the pulsation period, temperature, and chemical composition.  Thus, it is possible to infer these parameters using photometry. In the case of RR~Lyrae stars, \cite{Jurcsik1996} demonstrated the possibility of estimating [Fe/H] from the shape of the $V$-band light curve described by Fourier parameters. Several authors have thus extended their work to various other photometric wavelengths, such as the $I_c$-band \citep{Smolec2005,Dekany2021} or the optical and infrared \citep{Mullen2021}. Other recent approaches have made use of learning machine algorithms to infer the physical parameters of Cepheids and RR Lyrae stars from the shape of the light curve \citep{Miller2015,Hajdu2018,Bellinger2020}. Among the available photometric methods used to estimate the metallicity, we also note the possibility of inverting the PLZ relation (when the metallicity term is characterized). In addition, \cite{Bono2010} suggested that differences in distance moduli inferred by different Period-Wesenheit relation can also be used to estimate individual metallicities.

In the case of Cepheids, several studies have shown the impact of the chemical abundance on the amplitudes \citep{Klagyivik2007,Bono2000,SzabadosKla2012AMP,Majaess2013}. An attempt was made by \cite{Zsoldos1995} to calibrate a relation with Fourier parameters. The first reliable empirical relation for short-period Cepheids ($P<6.3\,$days) based on the Fourier parameters $R_{21}$ and $R_{31}$ was proposed by \cite{Klagyivik2013}. However these relations are based on a small sample of stars and do not cover the more metal-poor regime typical of Magellanic Cloud Cepheids. These relations were used by \cite{Clementini2019} to estimate the metallicity of Cepheids of \textit{Gaia} DR2. These authors found metallicity distributions for Cepheids residing in the Small and Large
Magellanic Clouds (SMC and LMC) and the Milky Way (MW) that are shifted by about $+0.2\,$dex compared to the literature values, which might hint at a calibration issue. 
On the other hand, there is still no empirical relation available for estimating the chemical abundance of intermediate ($6<P<10\,$days) Cepheids. This is due to the resonance between the fundamental and the second-overtone mode, which strongly affects the shape of the light-curve \citep{SimonSchmidt1976,Buchler1990}.  In the case of long-period ($P>10\,$days) Cepheids, \cite{scowcroft2016} established a relation based on \textit{Spitzer} infrared colors at 3.6 and 4.5$\,\mu$m. A complementary relation based on Fourier decomposition of the light curve in common photometric bands would be, however, useful and convenient for directly measuring the metallicity of extragalactic Cepheids. Thus, there is a need for a new photometric metallicity formula that offers a proper calibration and that is applicable over a wide range of pulsation periods. The estimation of [Fe/H] for a large number of Galactic and extragalactic Cepheids will be valuable in the context of carrying out thorough comparisons with evolution and pulsation theoretical models. Combining these estimations with the accurate trigonometric parallaxes of these stars provided by \textit{Gaia} can also help to recalibrate the metallicity dependence of the PL relationship and to refine the cosmic distance scale.

In this paper, we first present a set of new empirical relations for estimating the metallicity of Cepheids, based on the Fourier parameters of the $V$-band light curve. We improve (qualitatively and quantitatively) the calibration for the short-period Cepheids ($2.5<P<6.3\,$days) and we provide for the first time an empirical relation for estimating [Fe/H] of the long-period Cepheids ($12<P<40\,$days). Then, we convert these relations to the $I$-band and use them to estimate the metallicities of Cepheids in the MW, SMC, and LMC.
In Section \ref{sect:data_set}, we present the data set used for spectroscopic metallicities and for the $V$-band light curves. In Section \ref{sect:fourier}, we apply a Fourier decomposition of the light curves and we then establish empirical relations for short- and long-period Cepheids in Sections \ref{sect:Z_short} and \ref{sect:Z_long}, respectively. In Section \ref{sect:I_band}, we calibrate the interrelations of Fourier parameters and use it to convert $V$-band relations into the $I$ band. Finally, we apply these results to estimate the metallicity of Cepheids in the MW, SMC, and LMC. We then discuss our conclusions in Section \ref{sect:discussion} and \ref{sect:conclusion}.

\section{Metallicity and V-band light-curve data sets}\label{sect:data_set}
In this section, we first gather the spectroscopic metallicities of Cepheids from the literature, then we cross-match these stars with existing photometric catalogs in the $V$ band.

\subsection{Metallicity}\label{sect:data_set_Z}
\subsubsection{Milky Way Cepheids}
The two largest data sets of metallicities for Galactic Cepheids are given by \cite{Luck2018} and 
\cite{Groenewegen2018}, with 435 and 452 Cepheids, respectively (and with all pulsation modes included). While \cite{Luck2018} homogeneously determined  the metallicities, \cite{Groenewegen2018} compiled different data from the literature (mostly from \cite{Genovali2014,Genovali2015}) and translated them into the same scale.

In this work, we first re-scaled the [Fe/H] abundance of \cite{Luck2018} to solar abundance value of A(Fe)$_\odot$=7.50, given by \cite{Asplund2009}. The iron-to-hydrogen ratio is given as:
\begin{equation}
    \mathrm{[Fe/H]}=\mathrm{A(Fe)_\star-A(Fe)_\odot}
,\end{equation}
where A(Fe) = log(NFe/NH) + 12 is the logarithmic iron abundance with respect to hydrogen.
In the following, all the metallicity data sets discussed in this paper will be consistently re-scaled to this solar reference if needed. Then we cross-matched these two data sets to obtain a sample of 478 Milky Way Cepheids. There are 409 stars in common between these two catalogs and we derived a mean offset of $-0.07\,$dex$\pm \,$0.06, which is used to correct the iron abundance of \cite{Groenewegen2018}. We rejected 17 stars out of our final sample because they are discrepant at more than 2$\sigma$ between these two data sets and, thus, the measurements may not be reliable. When possible, metallicity values from \cite{Luck2018} are preferred, since they have been determined from a single homogeneous approach. When abundances are not available (from this same source), we then drew abundance values from \cite{Groenewegen2018}. We also followed the preference order among the different references as defined by this author. This merge data set constitutes our main metallicity sample.  Additional spectroscopic determination were found in \cite{Trentin2023} (44 fundamental-mode Cepheids), \cite{Kovtyukh2022} (54 fundamental-mode Cepheids), and \cite{Ripepi2021} (19 fundamental-mode Cepheids). From \cite{Kovtyukh2022}, we found 13 stars in common with our main metallicity sample. We derived a mean offset of $-0.06\pm0.11$ that is used to correct the iron abundance of \cite{Kovtyukh2022}.
Finally, the fundamental-mode classification is verified directly from a Fourier decomposition of the $V$-band light curves presented in the next section. To this end, stars that are discrepant with respect to the Hertzsprung progression of fundamental-mode Cepheids are discarded since they can belong to other pulsation modes \citep{AntonelloPoretti1986,AntonelloPoretti1990}. 

The final metallicity sample consists of 472 fundamental-mode MW Cepheids with spectroscopic metallicities. An overview of the calibration sample construction is given in Table \ref{Tab:bilan}. Although our sample might be inhomogeneous because of the different sources for the metallicity, we emphasize that about 90\% of metallicities for the MW Cepheids of this sample are taken from a single source (specifically, from \cite{Luck2018}, Table~ \ref{Tab:data_set}).

\subsubsection{LMC and SMC Cepheids}
We retrieved the iron abundances of 89 stars in the recent work of \cite{Romaniello2022}. However, we note that \cite{Romaniello2022} re-analyzed 21 LMC Cepheids from \cite{Romaniello2008} and found systematic errors in this latter study, yielding an additional offset of $-0.11\,$dex to obtain the new values. Therefore, we assumed that the 14 SMC Cepheids from \cite{Romaniello2008} are affected by the same error, so we retrieved these stars and we corrected by applying this offset.
We also found 4 additional Cepheids from the SMC and 5 stars from the LMC cluster NGC~1866 given by \cite{Lemasle2017}. From NGC~1866, 2 more Cepheids were retrieved from \cite{Molinaro2012}.

The final metallicity sample combining MW, LMC, and SMC Cepheids consists of 586 Cepheids which span a [Fe/H] range from about $-1.0$ to $+0.4\,$dex. It is, however, difficult to attribute consistent uncertainties for each measurement. Indeed, a range of observations and methods have been used to measure the iron abundance and the uncertainties may not be homogeneous between studies. For example, \cite{daSilva2022} studied the impact of different systematics on metallicity measurements. In addition, we cannot derive any systematic difference of [Fe/H] between the two largest data sets, namely, the MW sample of \cite{Luck2018} and the LMC sample of \cite{Romaniello2022}, since they have no stars in common. Therefore, we associated a conservative uncertainty of $\pm0.15\,$dex with every abundance in our sample. This value represents an upper limit for most of the uncertainties and is also consistent with the mean uncertainty from the sample of \cite{Luck2018}.
\begin{figure}
\begin{center}
\includegraphics[width=0.50\textwidth]{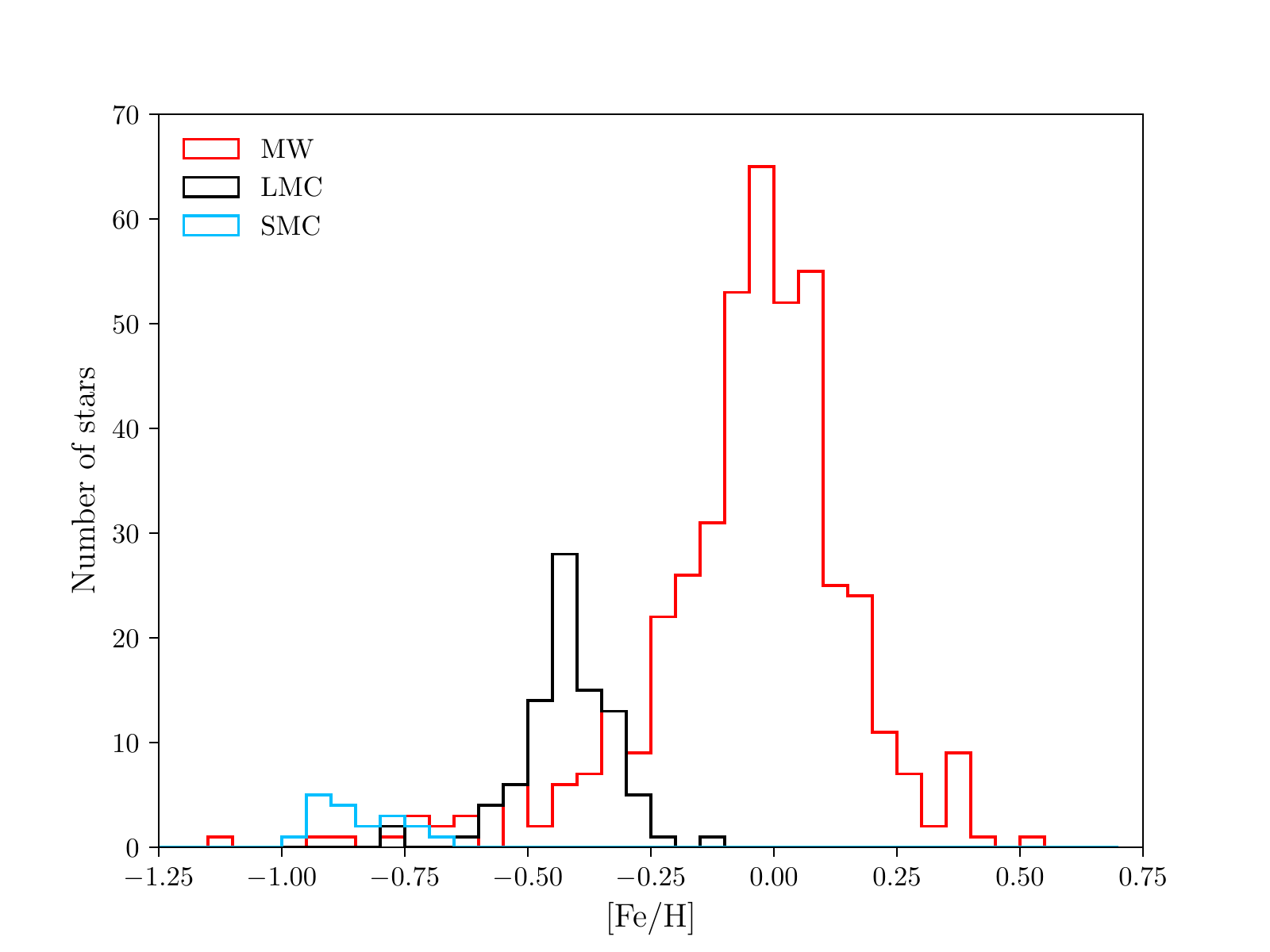}
\caption{\small Histogram of the iron abundance of the calibration sample (545 fundamental-mode Cepheids with $V$-band light-curve) gathered and re-scaled from the literature. \label{fig:histo_Z}} 
\end{center}
\end{figure}

\begin{table*}[]
\caption{\label{Tab:bilan} Summary of the data sets used to compose the calibration sample for empirical metallicity relation from the $V$ band (see Sects.~\ref{sect:data_set} and \ref{sect:fourier}).}
\begin{center}
\begin{tabular}{l|l|l|l|c}
\hline
\hline
        &MW     &       LMC     &       SMC     &               \\
\hline
                &\cite{Luck2018} (435)          &       \cite{Romaniello2022}(89)       &       \cite{Romaniello2008}(14)       &               \\
Metallicity (Sect.~\ref{sect:data_set_Z})     & \cite{Groenewegen2018} (452)    &       \cite{Lemasle2017} (5)     &       \cite{Lemasle2017} (4)  &               \\
                &  \cite{Ripepi2021} (19)       &       \cite{Molinaro2012} (2)     &                                   &           \\
                &  \cite{Trentin2023} (44)       &              &                                   &            \\
                &  \cite{Kovtyukh2022} (54)       &             &                                   &            \\
                \hline
Total  cross-match [Fe/H]  &  fundamental only : 472 &                          96      &       18                         &       586     \\
                \hline
                & ASAS (169)        &   OGLE IV (72)            &       OGLE III (4)         &               \\
$V$ band   (Sect.~\ref{sect:data_set_V})     & ASAS-SN (438)    &       OGLE III (3)         &       ASAS-SN (14)            &               \\
                & \cite{Berdnikov2008} (349)    &       ASAS-SN (15)             &                               &               \\
                \hline
Total ([Fe/H]+$V$ band)  & 438  &       89                          &       18                   &       545     \\
\hline
\end{tabular}
\normalsize
\end{center}
    \begin{tablenotes}
    \item \textbf{Notes :} The final data set consists of 545 fundamental-mode Cepheids and is presented in detail in \ref{Tab:data_set}. The first row describes the spectroscopic metallicities retrieved from the literature for a total of 586 stars (see Sect.~\ref{sect:data_set_Z}). The second row describes the $V$-band light curves we found in the literature after cross-matching with the precedent metallicity sample (see Sect.~\ref{sect:data_set_V}).
    \end{tablenotes}
\end{table*}

\subsection{V-band light curves}\label{sect:data_set_V}
The first objective of this work is to increase the sample of Cepheids with both spectroscopic metallicity and well-sampled $V$-band light curves. Thus, we did not limit our sample to a single catalog, as done by \cite{Klagyivik2013} in using the catalog from \cite{Berdnikov2008}; however, we chose to cross-match the metallicity sample with different catalogs in the $V$-band. For MW Cepheids, we retrieved 169 stars from the All-Sky Automated Survey (ASAS) catalog \citep{Pojmanski2002} and 438 stars from the All-Sky Automated Survey for Supernovae (ASAS-SN) catalog  \citep{ASAS2018}. We also extracted 349 stars from \cite{Berdnikov2008}. Among the 472 fundamental-mode Cepheids with spectroscopic metallicities, we found 438 stars that also have a light curve in the $V$ band.

For SMC and LMC stars, we first collected the light curves from Optical Gravitational Lensing Experiment (OGLE) publicly available\footnote{\url{https://ogledb.astrouw.edu.pl/~ogle/OCVS/ceph_query.php}} data \citep{Soszynski2008,Soszynski2010}. The SMC and LMC light curves were retrieved from OGLE III and OGLE IV. For 35 stars, the pulsation cycle is poorly covered or light curves are not available. For 29 of them, we retrieved the light curves from ASAS-SN. For the remaining 6 stars, we could not find any $V$-band light curves in the literature. 
Our LMC and SMC sample thus consists of 107 fundamental Cepheids from LMC and SMC with both spectroscopic metallicities and $V$-band light curves and an excellent phase coverage. In the next section, we apply Fourier decomposition to every $V$-band light curves from the different catalog and we select the best fit for MW Cepheids among the different $V$-band catalogs.

\section{Fourier decomposition of the V-band light-curve}\label{sect:fourier}
 Fourier decomposition is a very useful technique for studying the structure of light curves. In particular, the Fourier parameters efficiently describe  the bump progression \citep{Hertzsprung1926}. The favored hypothesis to explain the bump progression is a resonance  between the fundamental and the second-overtone pulsation modes \citep{SimonSchmidt1976} that are characterized by pulsation periods of $P_0$ and $P_2$, respectively. This resonance model is qualitatively reproduced by hydrodynamical models \citep{Buchler1990} and quantitatively using the amplitude equation \citep{kovacs1989}. Alternatively, the echo model proposes that a wave pressure travels inward from He ionization zone and echoes on the stellar core, then travels back to reach the surface, and produces the observed bump on the light curves \citep{Whitney1956,christy68,Christy1975}. For arguments against the echo hypothesis, we refer to \cite{Klapp1985}.

\begin{figure}
\begin{center}
\includegraphics[width=0.52\textwidth]{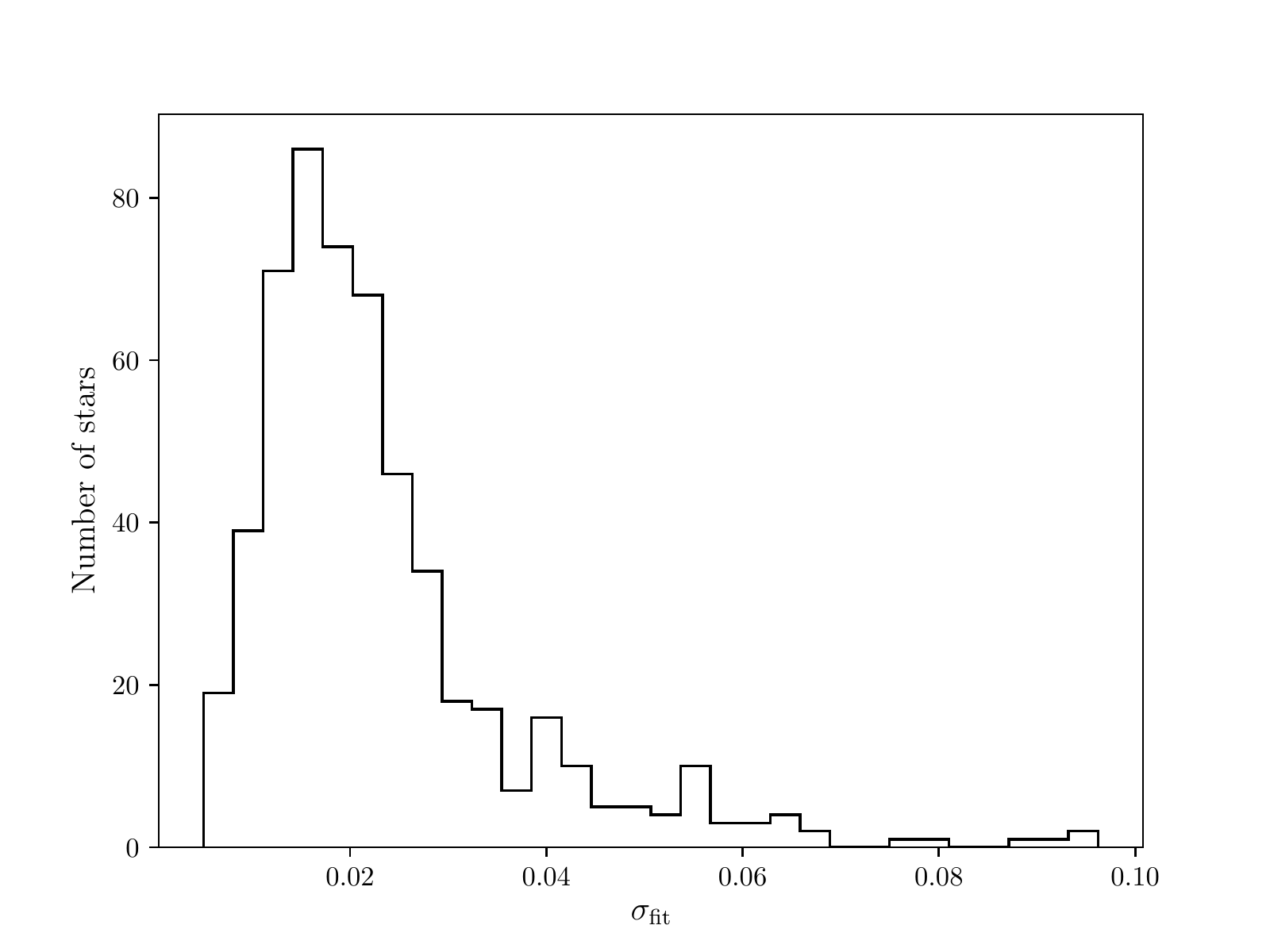}
\caption{\small Histogram of the uncertainty from the Fourier fitting of the 545 Cepheids. (see Eq.~\ref{eq:sigma}).} \label{fig:quality_fit}
\end{center}
\end{figure}

\begin{figure*} 
\begin{subfigure}{0.50\textwidth}
\includegraphics[width=\linewidth]{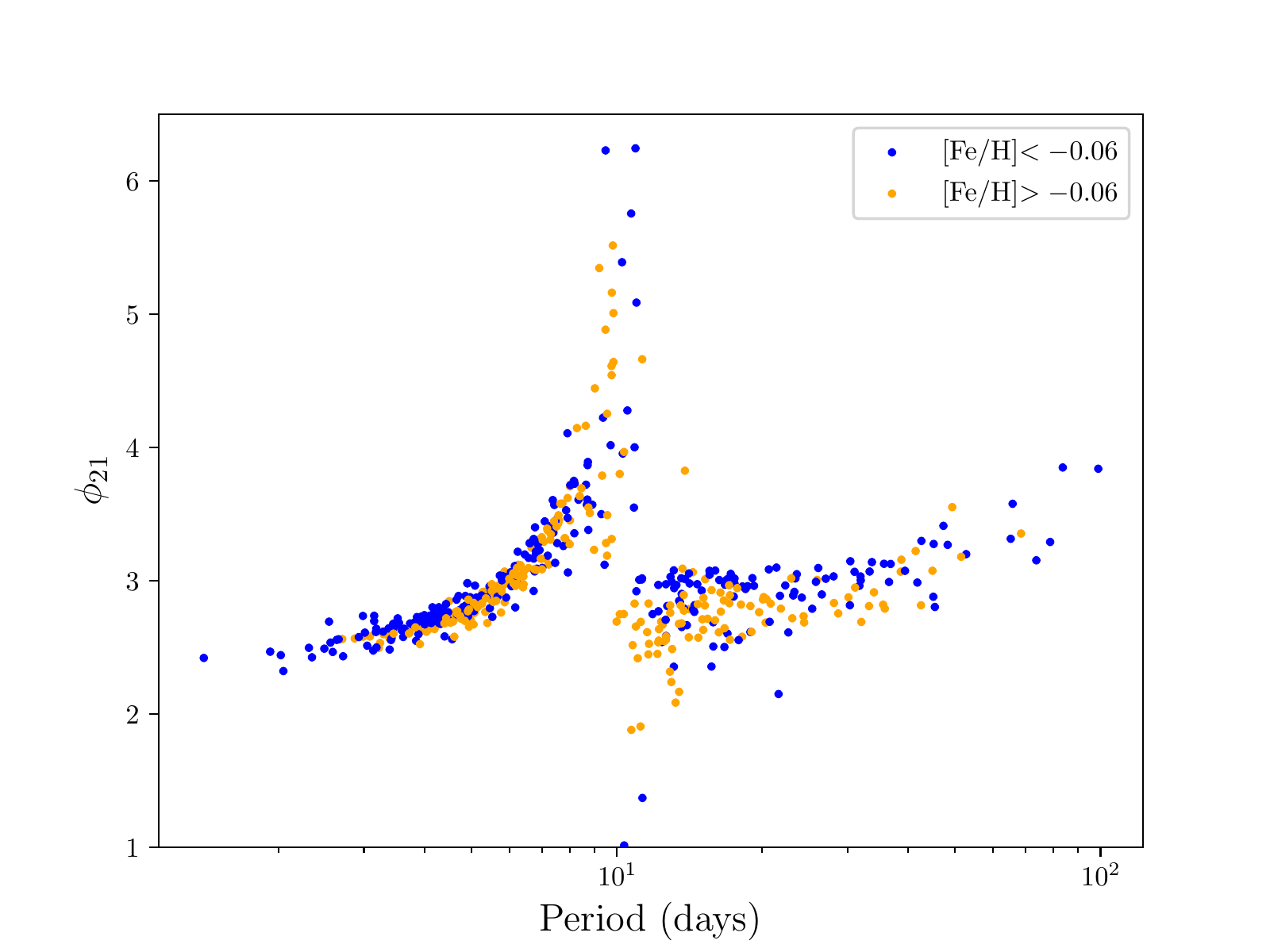}
\caption{$\phi_{21}$} \label{fig:phi21}
\end{subfigure}\hspace*{\fill}
\begin{subfigure}{0.50\textwidth}
\includegraphics[width=\linewidth]{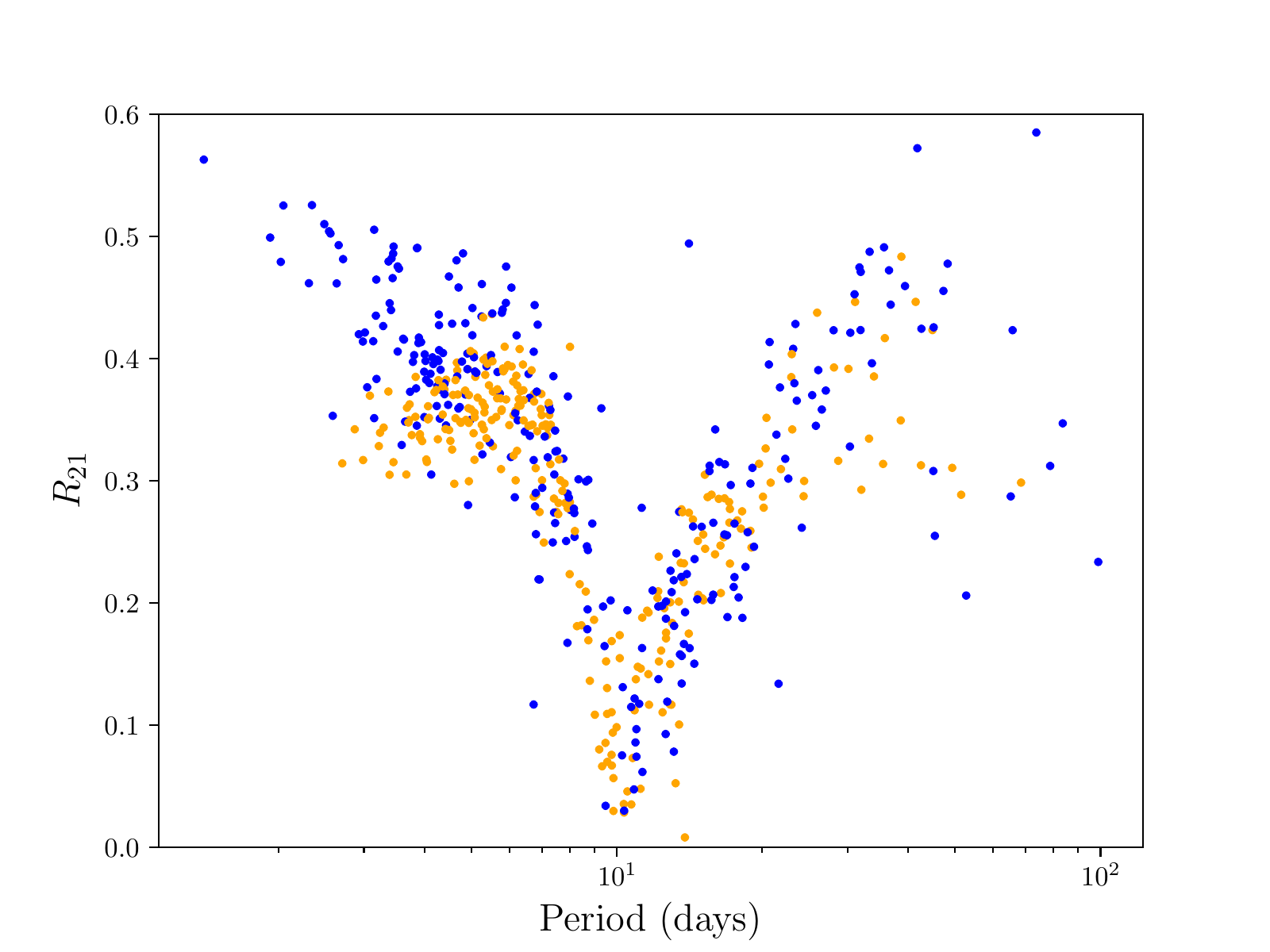}
\caption{$R_{21}$} \label{fig:R21}
\end{subfigure}

\medskip
\begin{subfigure}{0.50\textwidth}
\includegraphics[width=\linewidth]{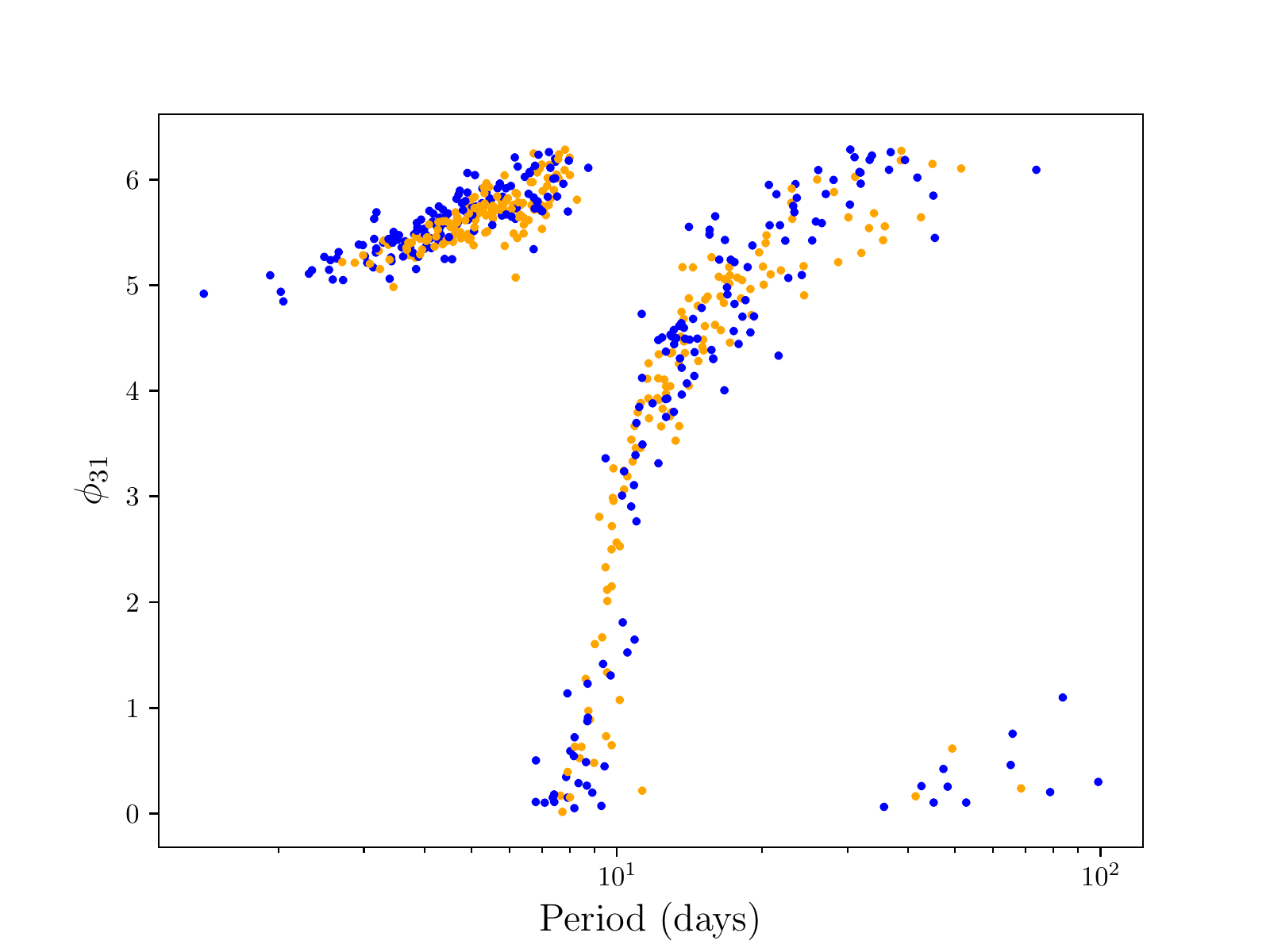}
\caption{$\phi_{31}$} \label{fig:phi31}
\end{subfigure}\hspace*{\fill}
\begin{subfigure}{0.50\textwidth}
\includegraphics[width=\linewidth]{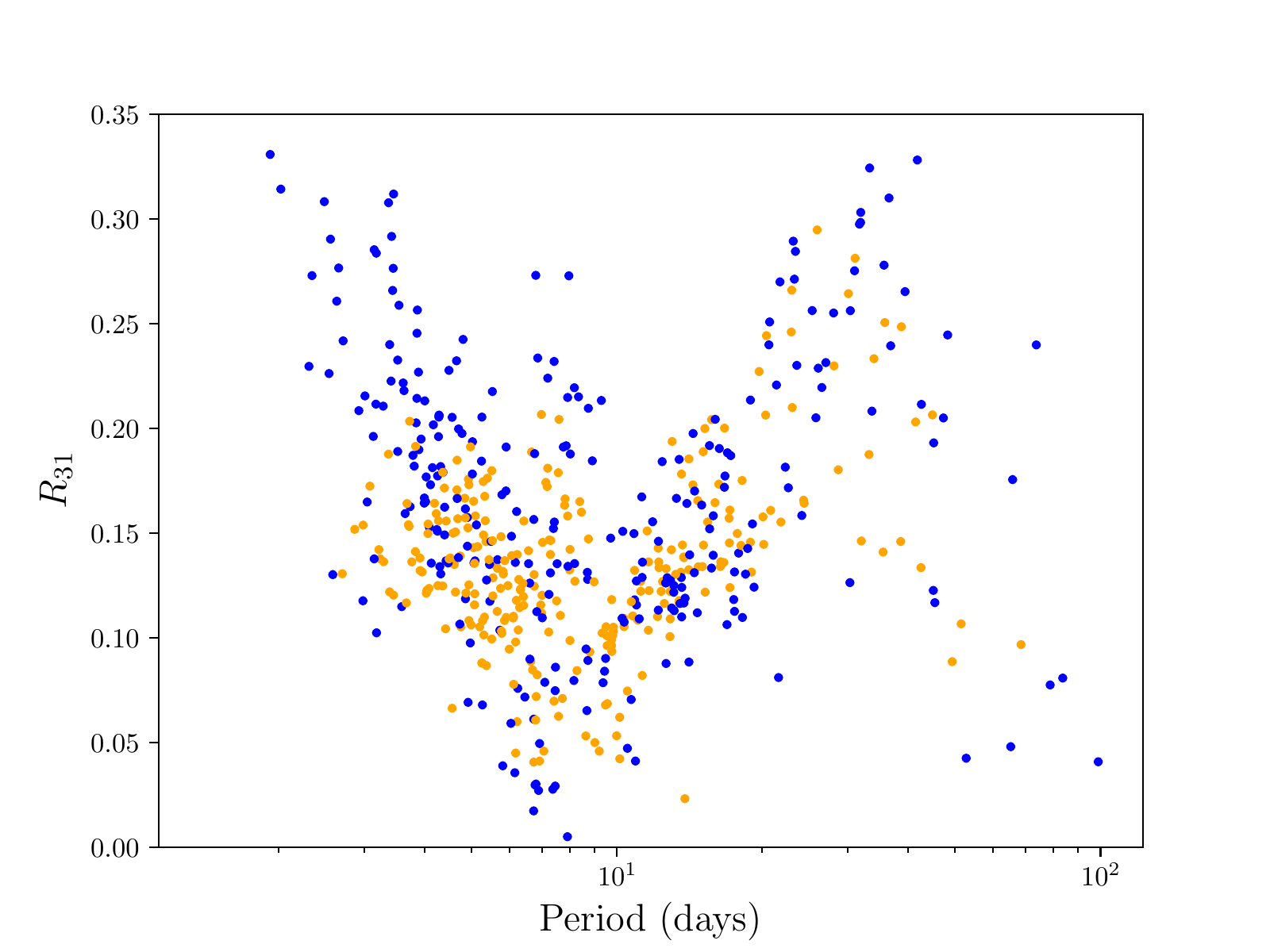}
\caption{$R_{31}$} \label{fig:R31}
\end{subfigure}

\medskip
\begin{subfigure}{0.50\textwidth}
\includegraphics[width=\linewidth]{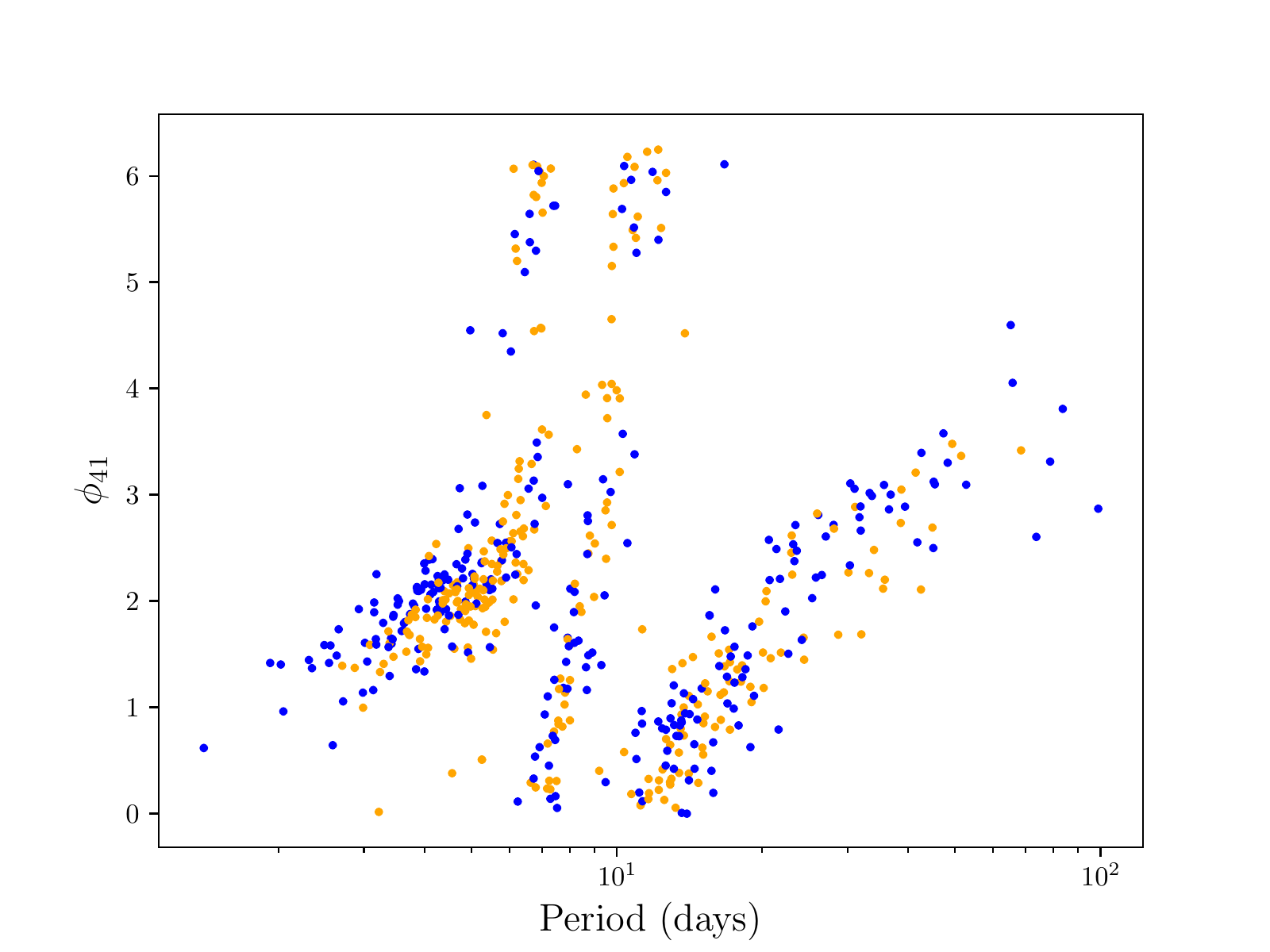}
\caption{$\phi_{41}$} \label{fig:phi41}
\end{subfigure}\hspace*{\fill}
\begin{subfigure}{0.50\textwidth}
\includegraphics[width=\linewidth]{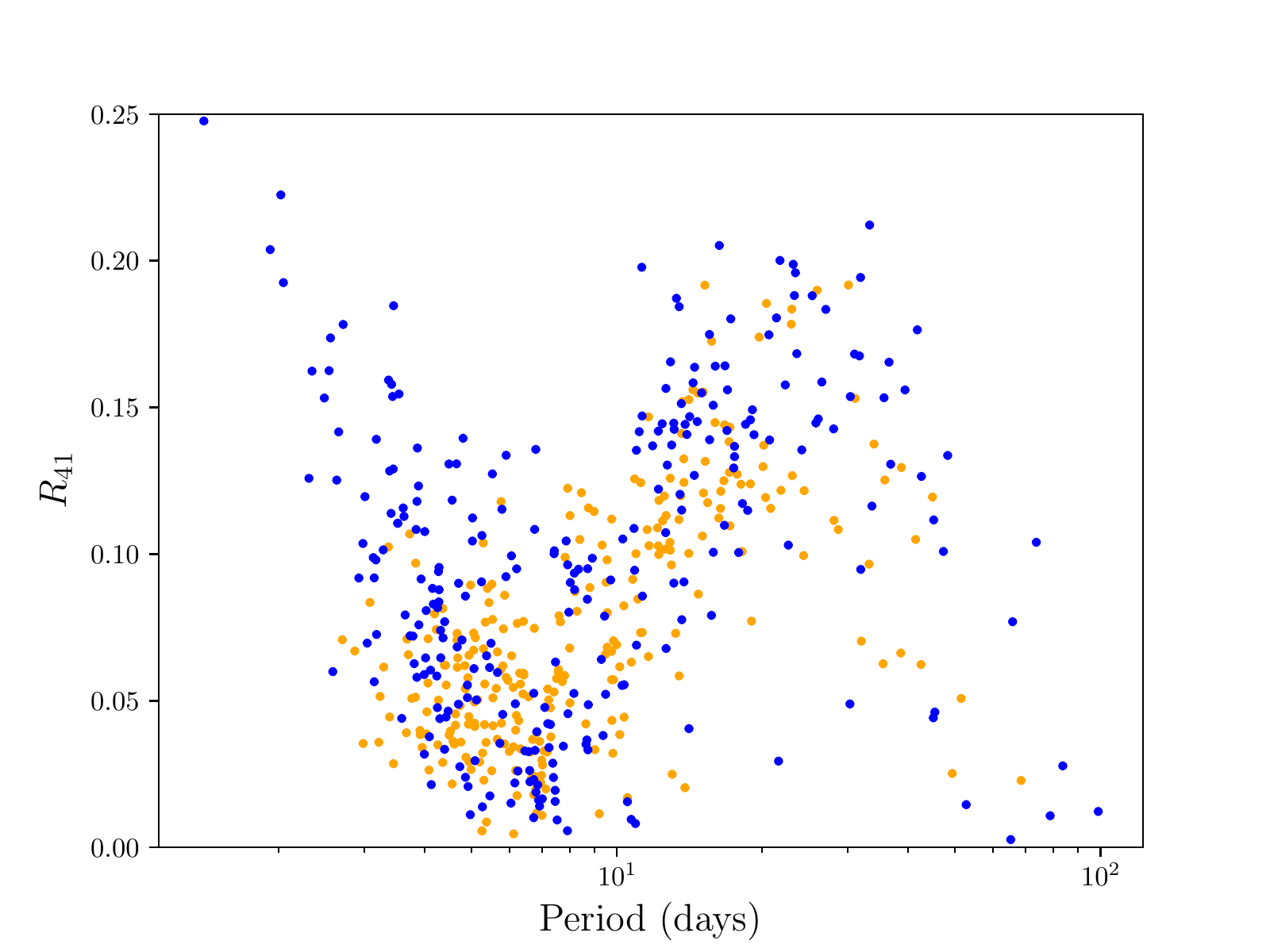}
\caption{$R_{41}$} \label{fig:R41}
\end{subfigure}
\caption{Fourier parameters of the final sample of the 545 fundamental-mode Cepheids in  the $V$ band with spectroscopic metallicities. The different colors refer to metallicity above and below the median of the sample, as indicated in Figure (a). The peculiar appearance of $\phi_{41}$ with a break around seven$\,$days was noted by \cite{SimonMoffett1985,kovacs90}. The latter proposed an unknown atmospherical effect since these authors did not observe this break in the radial velocity curves. }\label{fig:fourier}
\end{figure*}

 For each light curve, we applied a Fourier series of the form:

\begin{equation}
m(t) = A_0 + \sum_{k=1}^n A_k \mathrm{cos}[k \omega t +\phi_k],
\end{equation}
where $m(t)$ is the magnitude observed at the time $t$ and $\omega=2\pi/P$, $A_k$, and $\phi_k$ are the amplitude and the phase of the $k$ harmonic. We used the dimensionless Fourier parameters introduced by \cite{SimonLee1981}:
\begin{gather}
    R_{i1}=\frac{A_i}{A_1},\\
    \phi_{i1}=\phi_i - i \phi_1,
\end{gather}
where $\phi_{i1}$ are then adjusted to lie between 0 and 2$\pi$.
For each curve, the order of the fit, $n,$ is iterated until $A_n/\sigma_{A_n}$>4.
The uncertainty of the fit is given by:
\begin{equation}\label{eq:sigma}
\sigma^2=\frac{\chi^2}{N-2n-2},
\end{equation}
where $N$ is the number of data points, $n$ is the order of the fit, and $\chi^2$ is the sum of the squared residuals. This uncertainty is used to apply an iterative 3-$\sigma$ clipping during the fitting process. For each light curve, the period is estimated using Lomb-Scargle method \citep{Lomb1976,Scargle1982}. The uncertainty on the Fourier parameters were derived following \cite{Petersen1986}.  For every light curve of the different catalogs in the $V$-band, we selected the fit having the smallest uncertainty for Fourier parameter $\phi_{21}$. This criterion also ensures Fourier parameters are accurate enough for establishing empirical relations. A selection based on the fit uncertainty as defined by Eq.~\ref{eq:sigma} is also possible, but this criterion is less accurate when selecting the best fits. We also inspected visually the light curves to ensure the overall quality of the fits in particular for the period ranges studied in this paper in Sect.~\ref{sect:Z_short} and \ref{sect:Z_long}. The final sample of Cepheids with well sampled $V$-band light curves and spectroscopic metallicities consists of 545 stars, presented in Table~\ref{Tab:data_set}. The metallicity distribution of this sample is presented in Fig.~\ref{fig:histo_Z}, the distribution of the fit uncertainty in our sample is shown in Fig.~\ref{fig:quality_fit}, and Fourier parameters are presented in Fig. \ref{fig:fourier}.

We note that a significant fraction of Cepheids are binary or multiple systems \citep{Szabados2003,Kervella2019}. Following \cite{Klagyivik2013}, we also included binary stars in our final sample since in most cases, the companions do not affect the amplitude ratios and phases. However, the presence of a companion typically leads to smaller peak-to-peak amplitudes and Fourier amplitudes on a magnitude scale. Although most of the companions are main-sequence stars with a negligible contribution to the total luminosity \citep{Kaczmarek2022}, some Cepheids can have red giant companion that can affect significantly the luminosity and, thus, the amplitudes \citep{Pilecki2021,Kaczmarek2022}. Similarly, several studies have shown the occurrence of circumstellar envelopes (CSE) around Cepheids \citep{kervella06a,merand06,gallenne13b,nardetto16,Gallenne2021}. Thus, these envelopes could decrease the amplitude in case of constant emission in the infrared \citep{Hocde2020a,Hocde2020b,Hocde2021,Kovtyukh2022He}. In the following, we also work with Fourier amplitudes, therefore, we must keep in mind that binarity and CSE can be a source of scattering in our results.

\section{Empirical metallicity relation for short-period Cepheids (2.5 < P< 6.3 days) \label{sect:Z_short}}

\subsection{Choice of period range and Fourier parameters}\label{fig:qualitative}
\begin{figure*} 
\begin{subfigure}{0.52\textwidth}
\includegraphics[width=\linewidth]{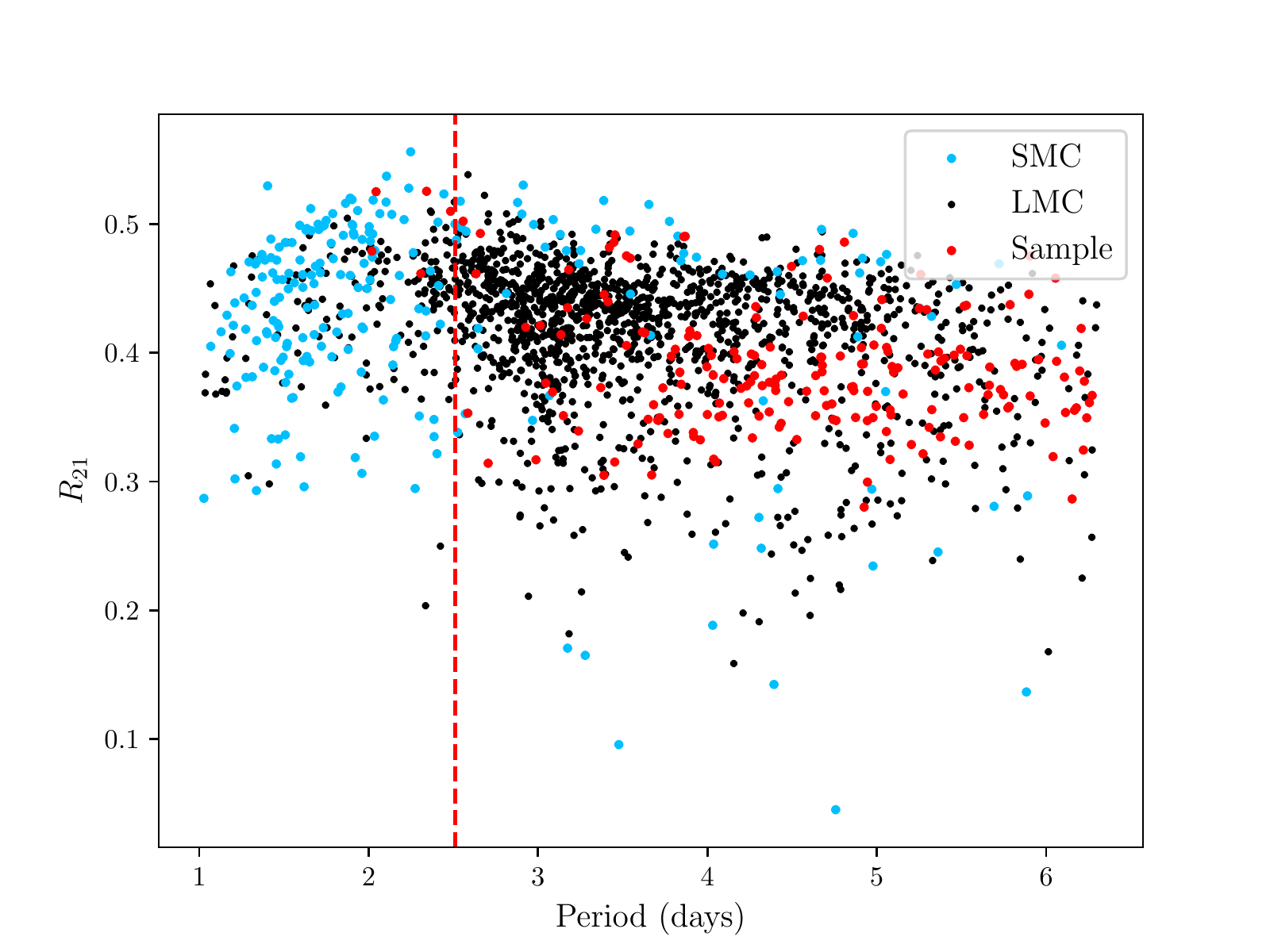}
\caption{} \label{fig:P_R21}
\end{subfigure}\hspace*{\fill}
\begin{subfigure}{0.52\textwidth}
\includegraphics[width=\linewidth]{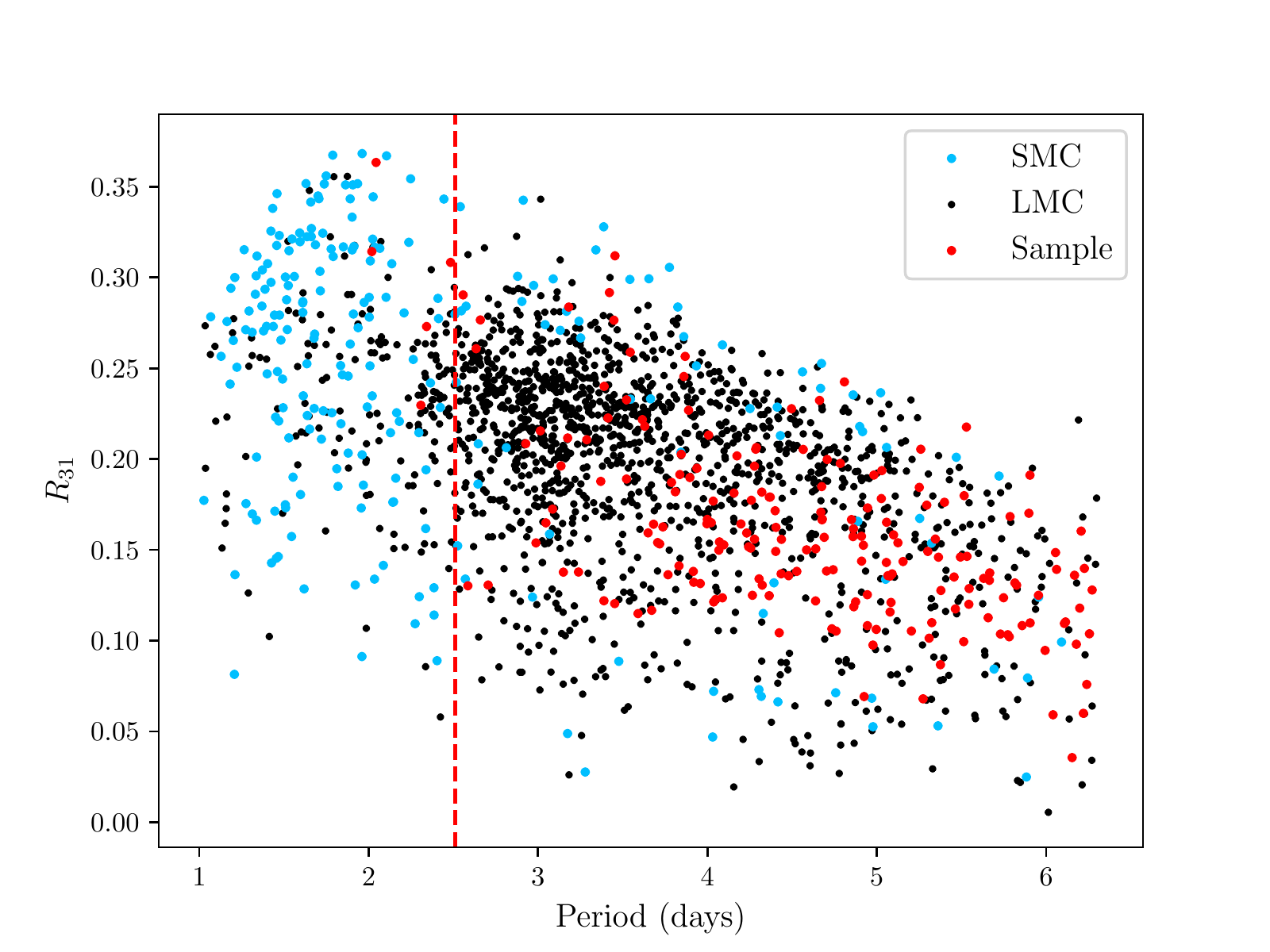}
\caption{} \label{fig:P_R31}
\end{subfigure}
 \caption{Amplitude ratios $R_{21}$ and $R_{31}$ against the pulsation periods for the OGLE SMC and LMC as (a) and (b), and with our star sample shown for consistency. The dashed vertical line shows the cut at a period of $P=$ 2.5 days.} 
\end{figure*}

In this section, we extend (quantitatively and qualitatively) the metallicity relation for short-periods Cepheids in the $V$-band, following the work of \cite{Klagyivik2013}.  In their study, the authors calibrated linear metallicity relations depending on the amplitude ratios $R_{21}$ and $R_{31}$ between 2.5 and 6.3 days. The lower boundary of these relations, defined as log $P=$ 0.4 ($\approx2.5 \,$days), corresponds to a local maxima on the amplitude ratio $R_{21}$. This can be observed for SMC and LMC Cepheids in the $I$ band in \cite{Soszynski2008,Soszynski2010}. On the other hand, above $P$ = 6.3$\,$days, the amplitude ratios are increasingly affected by the pulsation period when they approach the $P_2/P_0=0.5$ resonance at 10$\,$days. Thus \cite{Klagyivik2013} focused on Cepheids with pulsation period between 2.5 and 6.3 days where the effect of the period is mitigated in particular for $R_{21}$. This influence of the period can be better observed by displaying Cepheids of the SMC, LMC and our star sample on Figs.~\ref{fig:P_R21} and \ref{fig:P_R31}.  The Fourier parameters for the SMC and LMC were derived by applying our Fourier decomposition in Sect.~\ref{sect:fourier} to the $V$-band light curves of classical Cepheids from SMC and LMC from OGLE-IV  \citep{Udalski2015,Udalski2018,Soszynski2017}.  From these figures, we observe that $R_{21}$ and $R_{31}$ have a smooth linear dependence with the pulsation period. 

In the following, we choose to fit linear relations based on an unique period interval between 2.5 and 6.3 days. We re-analyze relations based on $R_{21}$ and $R_{31}$ but we also explore others possible linear combinations between the different Fourier coefficients and the pulsation period.

\subsection{Fit of the empirical relation with an orthogonal distance regression}\label{sect:short_result}
In this part, we fit the linear relations based on $R_{21}$ and $R_{31}$ and we also test all possible combinations of Fourier parameters on a two-by-two basis. We considered the following set of Fourier parameters: $R_{21}$,$\,R_{31}$,$\,R_{41}$,$\,\phi_{21}$,$\,\phi_{31}$,$\,\phi_{41}$. In addition we also take into account the amplitudes $A_1$ and $A_2$ and the pulsation period, $P$. We selected the stars where $\sigma_{\phi_{21}}<0.05$ to ensure the overall accuracy of the Fourier parameters. We employed an orthogonal distance regression (ODR) to take into account both the errors on the Fourier parameters and those on the metallicity. We adopted the ODR routine of the SciPy
package\footnote{\url{https://docs.scipy.org/doc/scipy/reference/odr.html}}. To summarize, we fit linear relations of the following form:
\begin{equation}
    \mathrm{[Fe/H]}=aX+b,
\end{equation}
for $R_{21}$ and $R_{31}$, and then the following:
\begin{equation}
    \mathrm{[Fe/H]}=aX+bY+c,
\end{equation}
for all possible combinations. In
order to compare the performance of the different fits to predict
accurately and precisely the metallicity, we used a Monte Carlo cross-validation analysis \citep{XU2001}. This technique consists of  randomly selecting 80\% of the data set to fit a relation, then testing its ability to estimate the metallicity on the remaining 20\% of the sample. This is done by deriving the average difference with the expected [Fe/H] values to evaluate the bias and the precision is estimated by deriving the rms. This procedure is repeated 1000 times and the final fitted coefficients, rms, and bias are given by the average of all the results (see Table \ref{Tab:odr_short}).

\begin{figure*} 

\begin{subfigure}{0.52\textwidth}
\includegraphics[width=\linewidth]{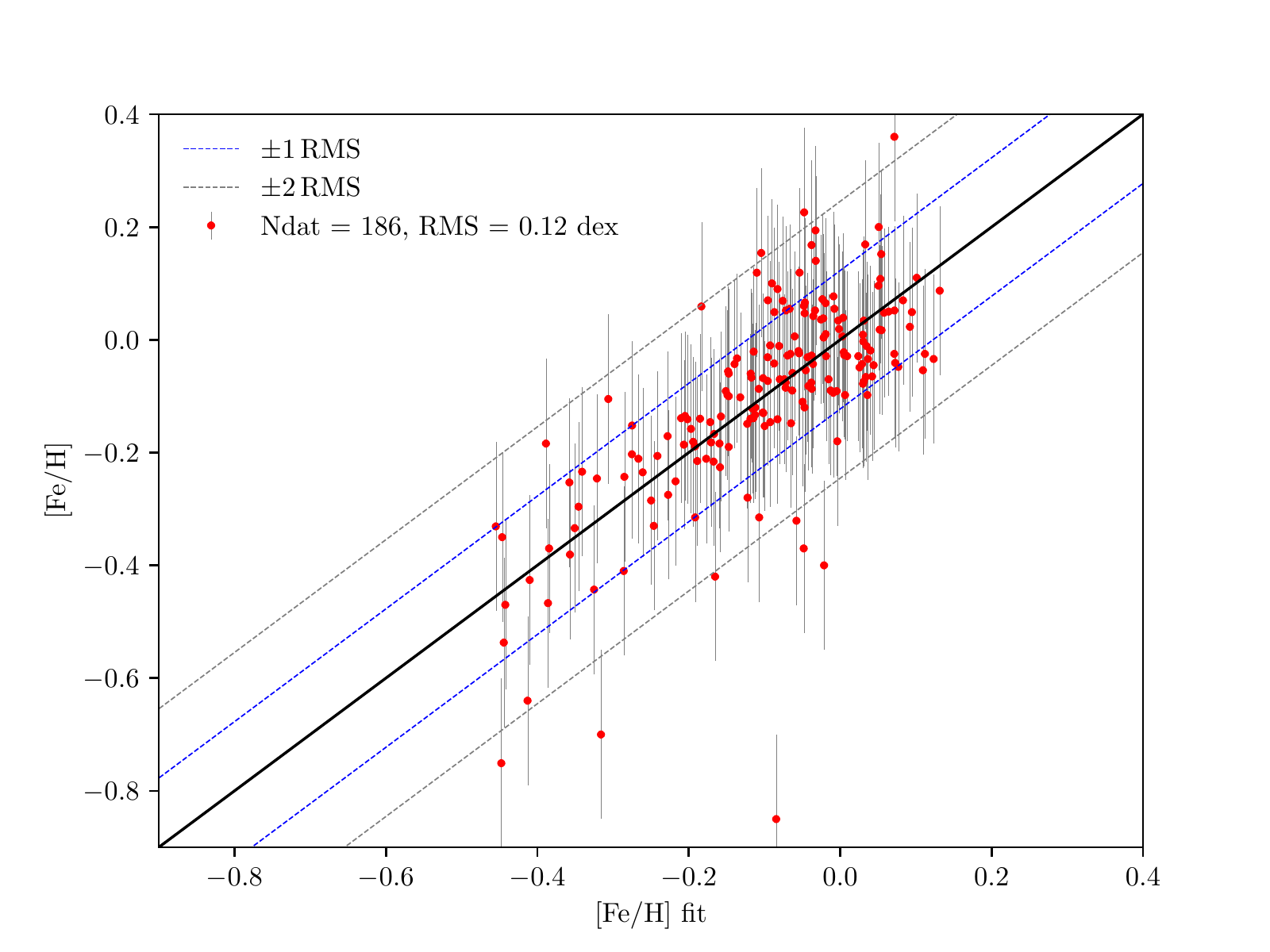}
\caption{[Fe/H]$=6.27\,A_{1} -11.73\,A_{2} -0.59$} \label{fig:fit_A1A2}
\end{subfigure}\hspace*{\fill}
\begin{subfigure}{0.52\textwidth}
\includegraphics[width=\linewidth]{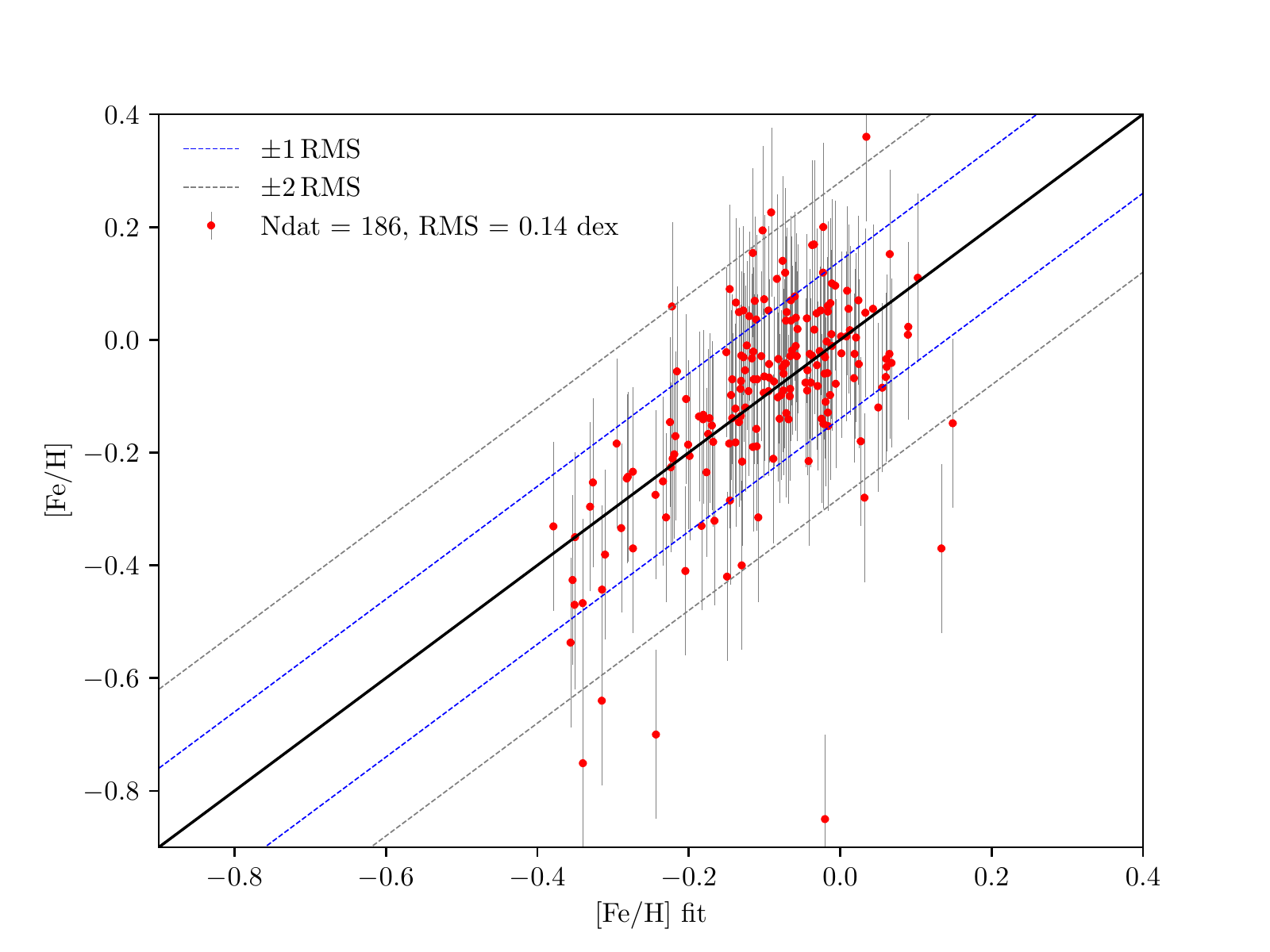}
\caption{[Fe/H]$=-2.42\,R_{21} + 0.84$} \label{fig:fit_R21}
\end{subfigure}

\caption{Comparison of [Fe/H] from the literature and from the fitted empirical relations based on (a) $A_1A_2$ and (b) $R_{21}$ for short-period Cepheids. The number of points used in the fit is indicated (Ndat) as well as the rms of the fit. Dashed lines represent the rms deviation to guide the eye. \label{fig:results_fit_short}}
\end{figure*}

From these results, the best fit is achieved by the combination
of parameters $A_1$-$A_2$: (see Fig. \ref{fig:fit_A1A2}):
\begin{equation}
    \mathrm{[Fe/H]}=(6.27\pm0.53)\,A_1 + (-11.73\pm0.93)\,A_2 + (-0.59\pm0.07),
    \label{eq:SHORT_eq}
\end{equation}

 with $\mathrm{rms}=0.12\,$dex. In Fig.~\ref{fig:results_fit_short}, we compare this result with a linear relation based on $R_{21}$ only, as in \cite{Klagyivik2013}. The relation based on $A_1A_2$ presents a lower rms than the one based on $R_{21}$ and is in better agreement for lower and higher metallicity range. Thus, the $A_1A_2$ relation performs better to recognize the light curves features sensitive to metallicity.
 In order to further assess the capability of the $A_1A_2$ relation to estimate the metallicity, we test it in the next section on the SMC and LMC sample and compare the results to the literature.

\begin{table}[]
\caption{\label{Tab:odr_short} \small Results of the ODR fitting and Monte Carlo analysis. Only the best results with a final rms below 0.145$\,$dex are presented.}
\begin{center}
\begin{tabular}{l|c|c}
\hline
\hline
Param.& rms (dex) & bias (dex) \\ 

\hline
$A_1$-$A_2$& 0.123 & -0.001 \\
$A_1$-$R_{21}$& 0.126 & -0.000 \\
$A_2$-$R_{21}$& 0.128 & -0.000 \\
$A_1$-$R_{41}$& 0.129 & -0.001 \\
$A_1$-$A_3$& 0.129 & -0.000 \\
$A_1$-$R_{31}$& 0.134 & -0.002 \\
$R_{21}$-P& 0.134 & -0.000 \\
$R_{21}$&0.139&-0.001\\
$R_{21}$-$\phi_{21}$& 0.139 & -0.001 \\
$R_{31}$-$R_{21}$& 0.139 & -0.000 \\
$A_2$-$R_{41}$& 0.140 & -0.001 \\
$A_3$-$R_{31}$& 0.141 & -0.002 \\
$R_{31}$-$\phi_{31}$& 0.141 & -0.002 \\
$R_{31}$-$\phi_{41}$& 0.142 & -0.003 \\
$R_{41}$-$R_{21}$& 0.142 & -0.002 \\
$R_{31}$&0.143&-0.001\\
$A_2$-$R_{31}$& 0.143 & -0.000 \\
$A_3$-$R_{21}$& 0.143 & -0.001 \\
$R_{41}$-P& 0.143 & -0.000 \\
$R_{21}$-$\phi_{31}$& 0.143 & -0.002 \\
$R_{21}$-$\phi_{41}$& 0.144 & -0.002\\
$R_{31}$-P& 0.144 & -0.000 \\
$R_{31}$-$\phi_{21}$& 0.145 & -0.001 \\
$R_{41}$-$\phi_{41}$& 0.145 & -0.002 \\
\hline
\end{tabular}
\normalsize
\end{center}
\end{table}

\begin{figure*} 

\begin{subfigure}{0.52\textwidth}
\includegraphics[width=\linewidth]{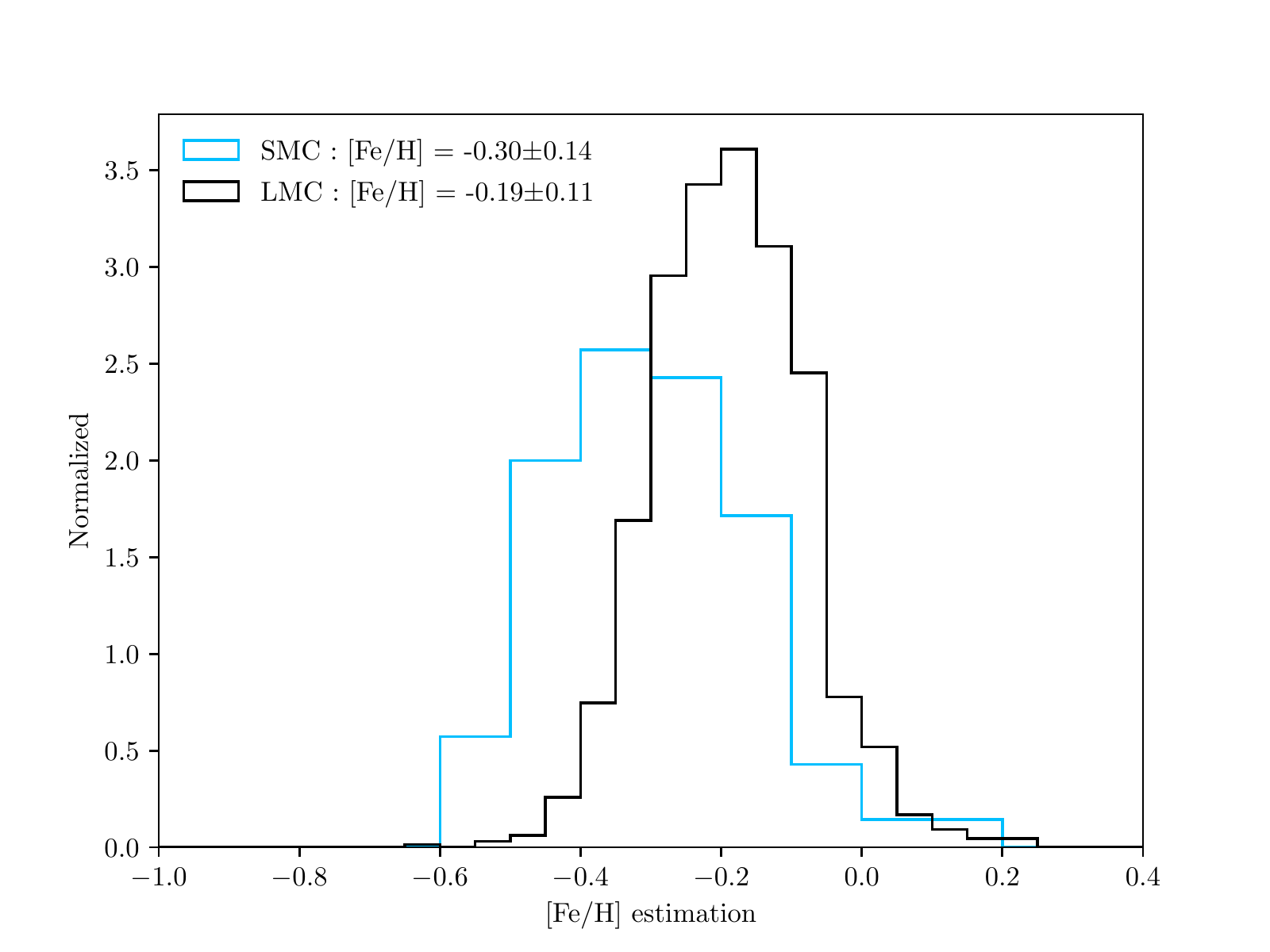}
\caption{Without cuts} \label{fig:histo_pred_A1A2}
\end{subfigure}\hspace*{\fill}
\begin{subfigure}{0.52\textwidth}
\includegraphics[width=\linewidth]{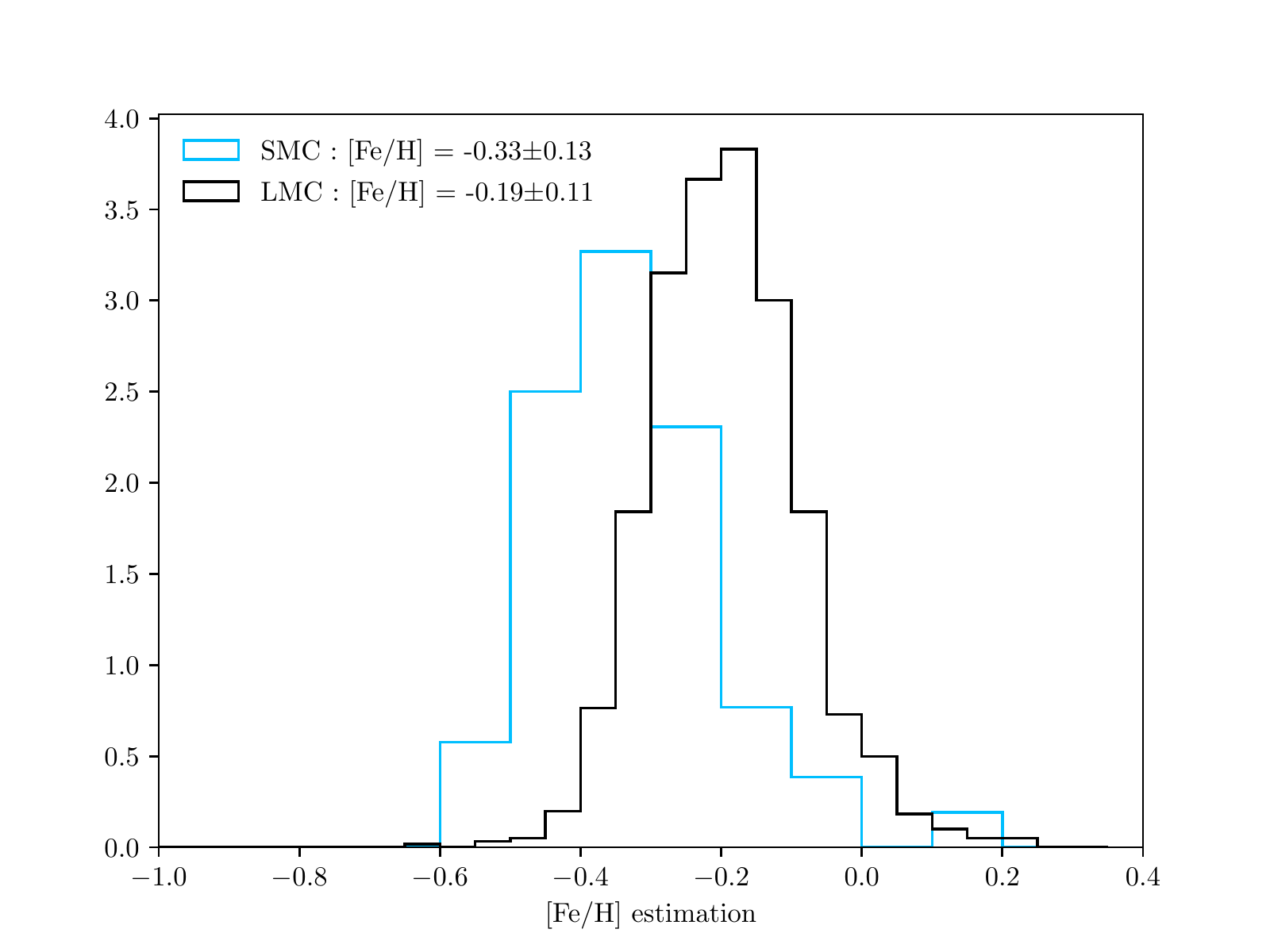}
\caption{Cut applied: $A_1>0.20\,$mag} \label{fig:histo_pred_A1A2_cuts}
\end{subfigure}
\caption{Normalized histograms (unit area)  of the metallicity estimation of the SMC (blue, 52 stars) and LMC (black, 1206 stars) short-period Cepheids from the relations established in Sect. \ref{sect:short_result}. The mean metallicity of these distributions and the standard deviation around the mean are indicated in the legend. \label{fig:result_pred_short}}
\end{figure*}

\begin{figure*}

\begin{subfigure}{0.52\textwidth}
\includegraphics[width=\linewidth]{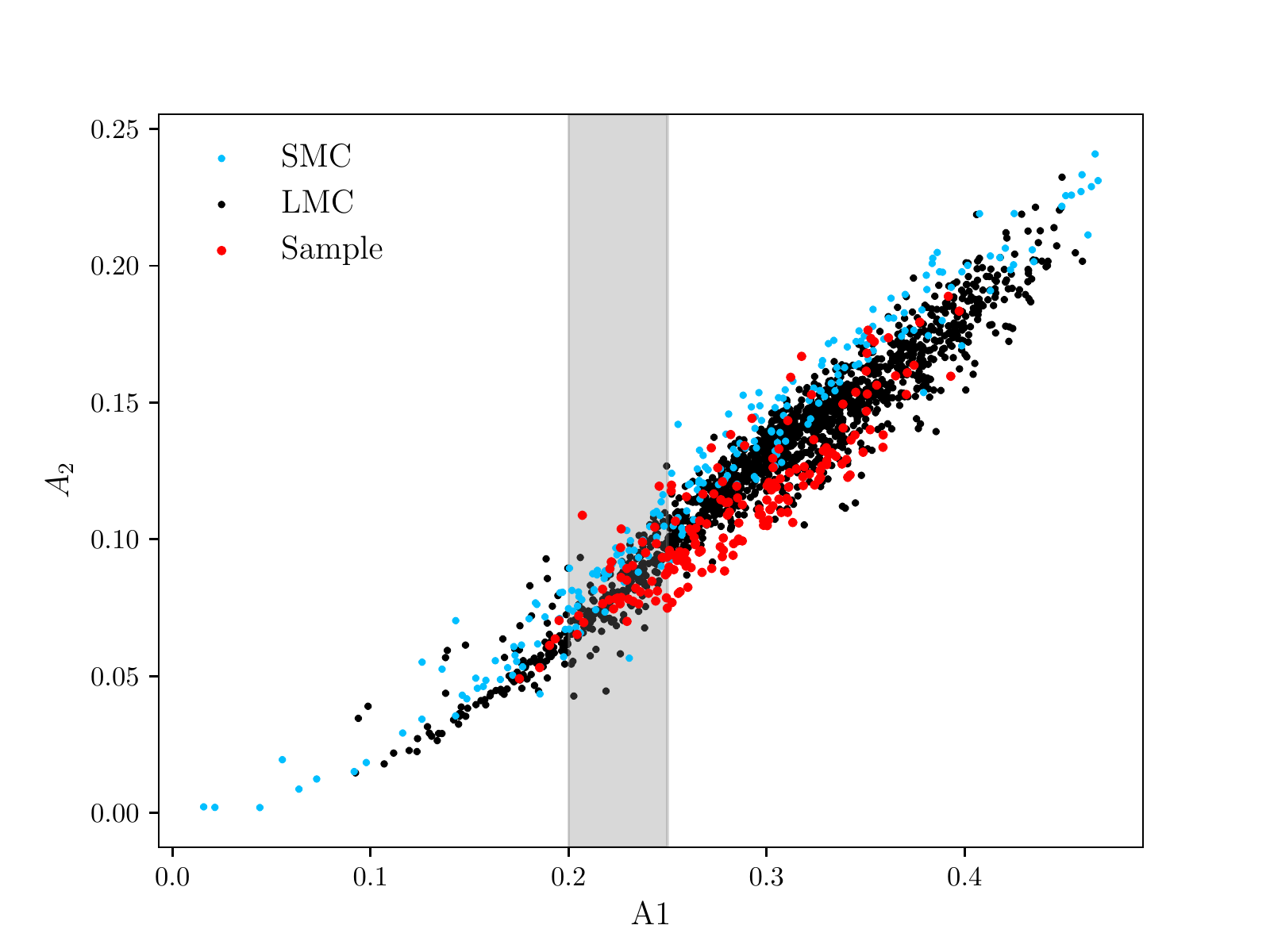}
\caption{} \label{fig:A1_A2}
\end{subfigure}\hspace*{\fill}
\begin{subfigure}{0.52\textwidth}
\includegraphics[width=\linewidth]{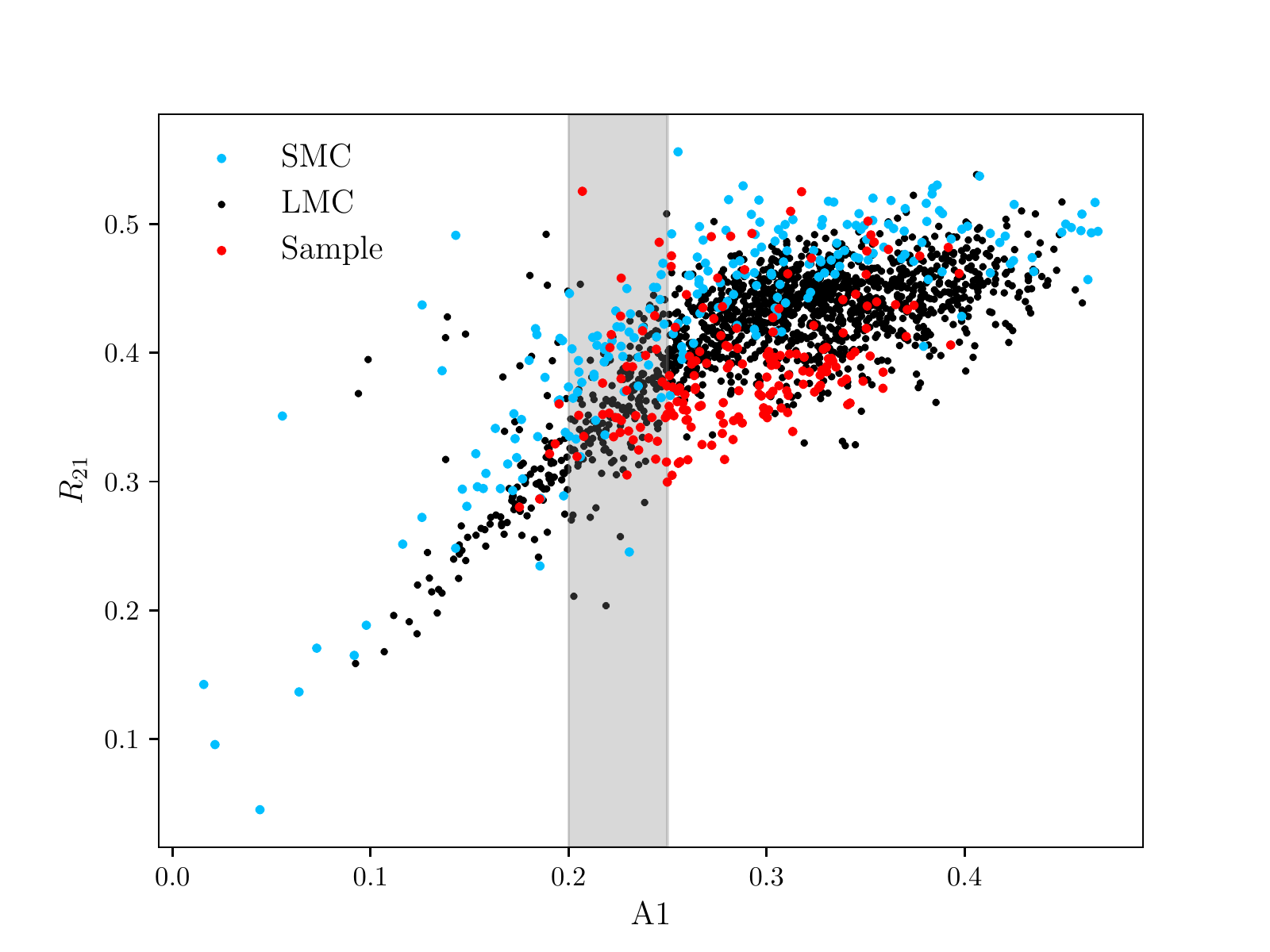}
\caption{} \label{fig:R21_A1}
\end{subfigure}
\caption{ Relations between the first and second harmonic $A_1$ and $A_2$ used in the empirical relation for short-period Cepheids between 2.5 and 6.3 days shown in (a) and (b). The vertical strip represents the values of $A_1$ where a cut can be applied (for $A_1$ between 0.20 and 0.25$\,$mag) to remove the stars of low amplitudes to mitigate the effect of location inside the instability strip.\label{fig:A1A2HISTO}}
\end{figure*}

\subsection{Testing the empirical relations to the SMC and LMC sample}\label{sect:validity}

We retrieved $V$-band light curves of classical Cepheids from SMC and LMC from OGLE-IV \citep{Udalski2015,Udalski2018,Soszynski2017}. We first removed the stars that overlap with our calibration sample. We then consistently applied the same Fourier decomposition algorithm as established in Sect.~\ref{sect:fourier}. We also considered only the stars having more than 30 photometric measurements and with an uncertainty of $\sigma_{\phi_{21}}$<0.05$\,$. We used the Fourier parameters obtained for the Cepheids with a pulsation period between 2.5 and 6.3 days to estimate the metallicity of these stars using the empirical relation previously determined. 

 The final sample consists of 52 stars from the SMC and 1206 stars from the LMC in the period range of $2.5-6.3\,$days. As noted by \cite{Soszynski2008,Soszynski2010}, there are fewer Cepheids in this period range in the SMC since the distribution of the pulsation periods is dependent on metallicity \citep{Becker1977,Bono2000}. Indeed, metal-poor Cepheids in the SMC have a lower mass cut-off to cross the instability strip than the more metal-rich Cepheids from the LMC. As a result, the majority of short-period Cepheids in the SMC have pulsation period below 2.5 days. The estimated [Fe/H] for SMC and LMC Cepheids of our sample are presented in Fig.~\ref{fig:histo_pred_A1A2}. 

\begin{figure}
\begin{subfigure}{0.50\textwidth}
\includegraphics[width=\linewidth]{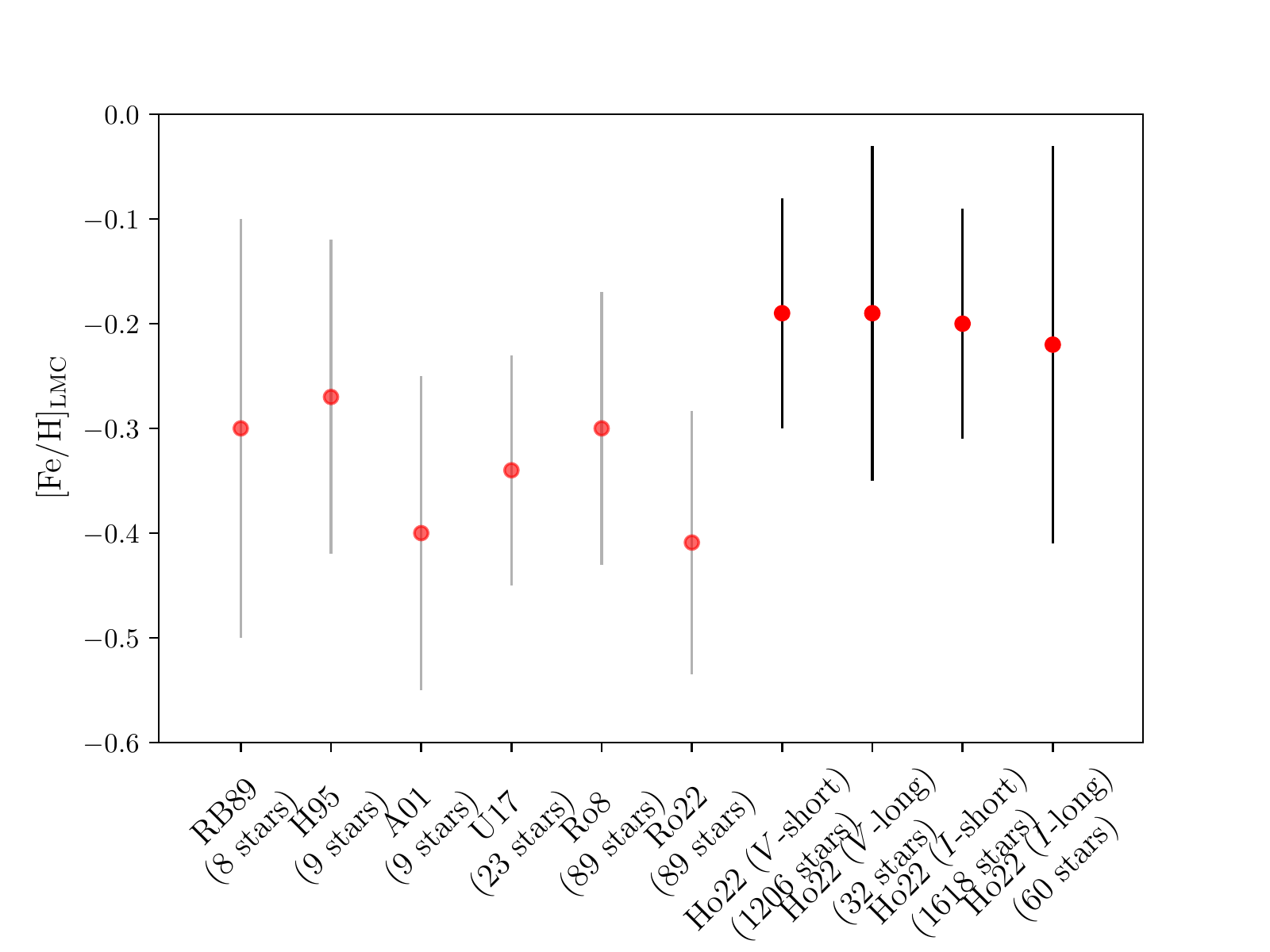}
\caption{LMC} \label{fig:comp_lmc}
\end{subfigure}

\begin{subfigure}{0.50\textwidth}
\includegraphics[width=\linewidth]{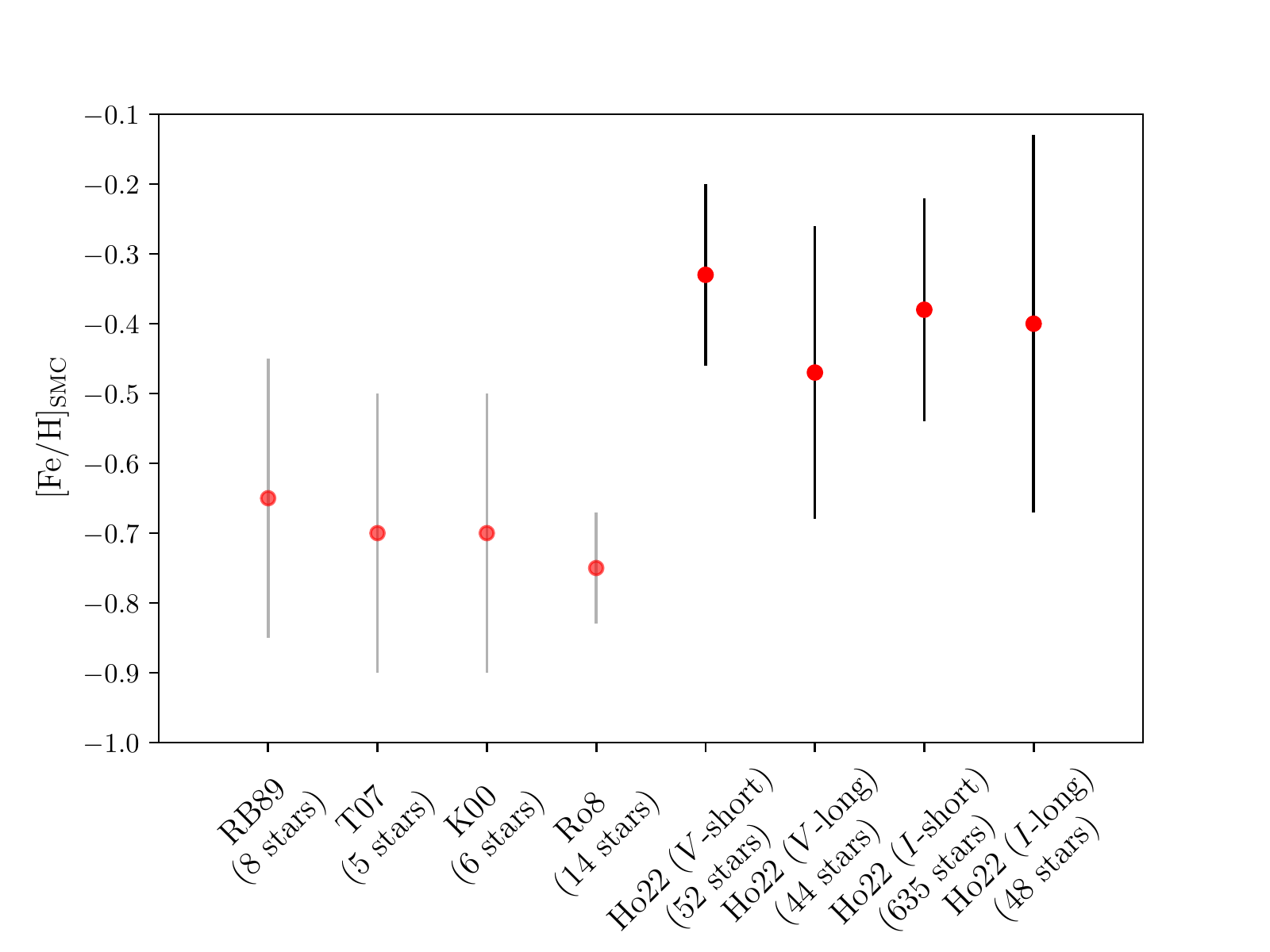}
\caption{SMC} \label{fig:comp_smc}
\end{subfigure}
\caption{\small Comparison of mean [Fe/H] derived from the literature (grey bars) and from [Fe/H] estimations presented in this paper (black bars). For each reference, the number of stars used to derive the mean is indicated. LMC metallicity from \cite{Romaniello2022} is presented taking account systematics. For details, see Sect. \ref{sect:validity} for short-period Cepheids in the $V$ band, Sect. \ref{sect:validity_long} for long-period Cepheids in the $V$ band, and Sect.~\ref{sect:validity_I} for the $I$-band relation. \label{fig:comp}}
\end{figure}

From this histogram, we can see that the relation is able to distinguish the populations of the SMC and the LMC. The mean metallicity of the LMC objects is $-0.19\pm0.11\,$dex, which is within the uncertainties and in agreement with values found in the literature based on supergiant measurements, such as $-0.30\pm0.20\,$dex \citep{Russel1989}; $-0.27\pm0.15\,$dex \citep{Hill1995};$-0.40\pm0.15\,$dex \citep{Andrievsky2001};$-0.34\pm0.11\,$dex \citep{Urbaneja2017}, and $-0.409\pm0.076\,$(stat.)$\pm 0.10\,$(syst.)$\,$dex \citep{Romaniello2022}.
Moreover, based on 1206 Cepheids from the LMC, we found a narrow metallicity distribution of the LMC of $\sigma$=0.11$\,$dex (see Fig.~\ref{fig:comp_lmc}). This is in agreement with the findings of \cite{Romaniello2022} who found a dispersion of $\sigma$=0.076$\,$dex from a sample of 89~Cepheids, and the dispersion of 0.069$\,$mag observed in the near-infrared Hubble Space Telescope LMC period-luminosity relation \citep{riess2019}.

However, this relation seems to overestimate the metallicity for the SMC with a mean [Fe/H] of $-0.30\pm0.14\,$dex. In the case of the SMC, the overestimation is significantly higher than for spectroscopic measurements of 14 long-period SMC Cepheids, giving [Fe/H]=$-0.75\pm0.08\,$dex \citep{Romaniello2008}, while it is in marginal agreement within about $1\sigma$ with other mean metallicity estimations based on giant and supergiant stars (see Fig.~\ref{fig:comp_smc}): $-0.65\pm0.20\,$dex \citep{Russel1989}; $-0.62\pm0.14\,$dex, \citep{Trundle2007}; and B stars with $-0.70\pm0.20\,$dex \citep{Korn2000}. We note however that these studies are based on few stars only, whereas we derived the metallicity from a larger sample of 52 Cepheids.

This overestimation of [Fe/H] could be attributed to the poor coverage of the low metallicity regime in our calibration sample (only four stars lower than $-0.6\,$dex). Moreover, a linear relation might not be valid at low metallicity, as has been noted by \cite{Klagyivik2013}. It is also possible that the SMC testing sample is (statistically) not large enough to properly determine the mean and the dispersion of the metallicity distribution. A more significant result for the mean metallicity of the SMC Cepheids can be obtained from $I$-band OGLE-IV light curves (explained in Sect.~\ref{sect:I_band}), which are available for a much larger sample of stars.

Finally, part of the spread observed in these distributions may be caused by the influence of other physical effects than metallicity such as the location in the instability strip. Indeed the amplitude of Cepheids, as well as their amplitude ratios, depend on the location in the instability strip \citep{Sandage1971,Sandage2004}. The amplitudes are higher close to the blue edge and decrease toward the red edge as the damping from convection increases.

\begin{figure}
\begin{center}
\includegraphics[width=0.52\textwidth]{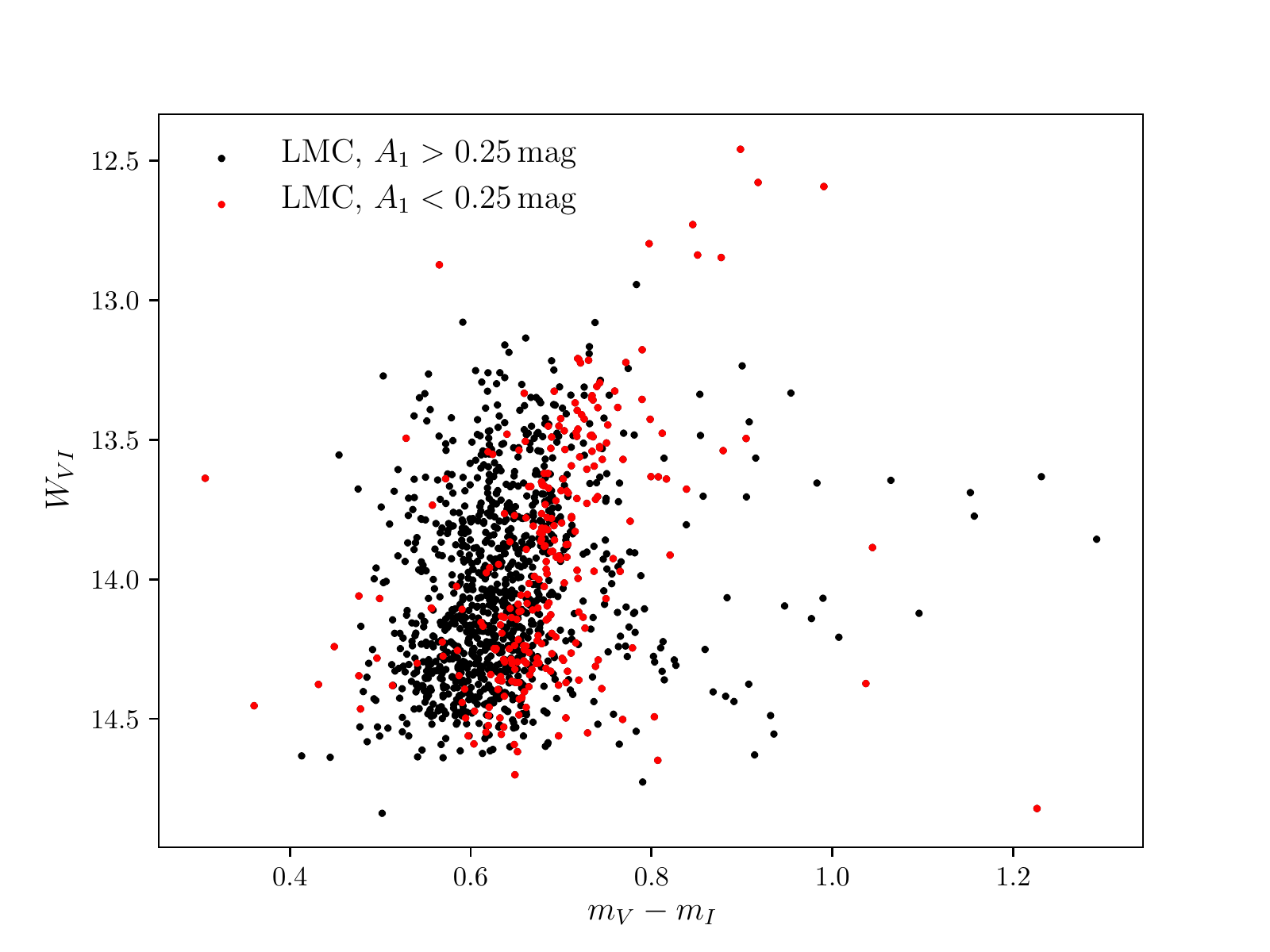}
\caption{\small Luminosity-color diagram of LMC Cepheids between 2.5 and 6.3 days, using $V$- and $I$-band magnitudes.  \label{fig:HR}} 
\end{center}
\end{figure}

In order to illustrate this effect, we display the LMC Cepheids with pulsation period between 2.5 and 6.3 days in the instability strip in Fig.~\ref{fig:HR}. We used the Wesenheit index, $W_{I,VI}$, as a function of the color $m_V-m_I$ corrected by extinction \citep{Skowron2021}. From this figure we clearly see that lower amplitude stars, with $A_1<0.25\,$mag, are concentrated at the red edge of the strip.
Hence, the stars which are closer to the blue or the red edge have different amplitudes, thus potentially affecting the empirical metallicity estimation based on $A_1A_2$. The relation between $A_1$ and $A_{2}$ can be seen in Fig.~\ref{fig:A1A2HISTO}. From this figure, we can observe that the relation between $A_1$ and $A_2$ is not linear anymore below about $A_{1}$=0.20-0.25$\,$mag. On the contrary, above $A_1$=0.20-0.25$\,$mag, the amplitude ratio, $R_{21}$, is saturated to about 0.45 and most of the dispersion is attributed to a metallicity effect.

\begin{figure}
\begin{center}
\includegraphics[width=0.50\textwidth]{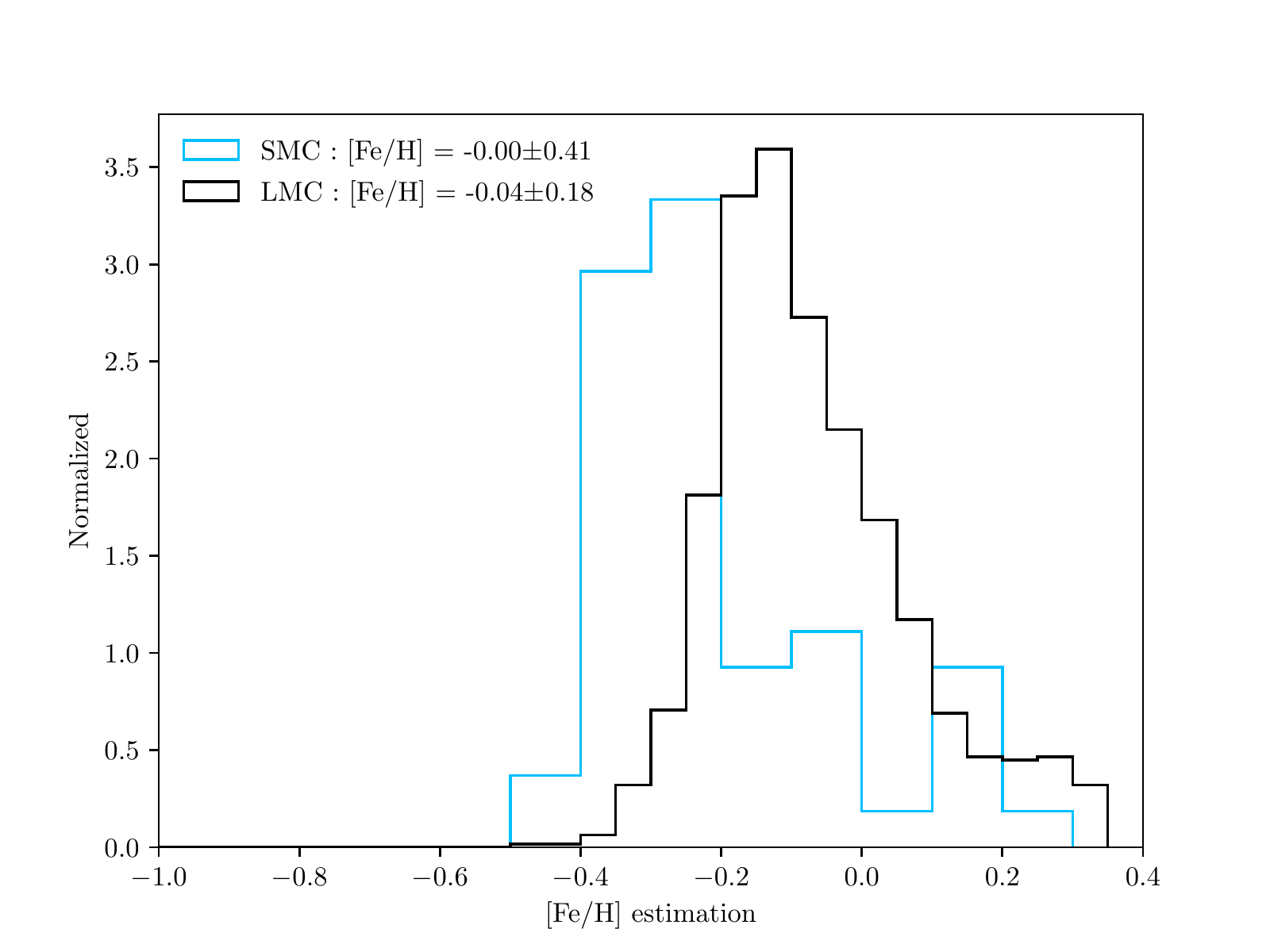}
\caption{\small $V$-band metallicity estimation using \citep{Klagyivik2013} \label{fig:histo_KLA_V}}
\end{center}
\end{figure}

Thus, we propose to cut all stars with $A_1<0.20\,$mag from the sample  in order to mitigate the effect due to the location inside the instability strip. This limit is defined only approximately in order to show the influence on the derived metallicity. A more restrictive cut at $A_{1}$=0.25$\,$mag can be applied if the objective favors the precision of individual measurement over the statistical size of the sample. By removing stars for which $A_1<0.20\,$mag, we can obtain a slightly better distinction between the SMC and LMC, as we see in Fig.~\ref{fig:histo_pred_A1A2_cuts}. However, the mean metallicity and the deviation from the mean are not significantly affected because the majority of the stars are above this threshold. Although the mean of the distribution remains unchanged, applying this cut to the empirical relation based on $A_1$ and $A_2$ prevents contaminations from stars with intrinsically low pulsation amplitudes and stars that would be significantly blended by a companion in the visible. Therefore, we recommend that this cut be  applied when using the empirical relations based on $A_1$, $A_2$.
We also note that some stars of our calibrating sample are slightly below this threshold. However, removing them does not change the result of the fit and we decided to keep them in the calibration set. We discuss the use of these cuts in Sect.~\ref{sect:discussion}. Finally, we applied the relation obtained by \cite{Klagyivik2013} in the $V$-band for the same period range (see Fig.~\ref{fig:histo_KLA_V}). As discussed in the previous section, relation based on $R_{21}$ and $R_{31}$ result in an overfitting of the metallicity around $0\,$dex; thus, we are not able to generalize for low and metal-rich stars. For that reason, the distributions observed in Fig.~\ref{fig:histo_KLA_V} are shifted by about $+0.2\,$dex compared to our results, in agreement with verifications carried out by \cite{Clementini2019}.

\section{Empirical metallicity relation for long-period Cepheids (12 < $P$< 40 days)}\label{sect:Z_long}
\subsection{Fit of the empirical relation with ODR technique}
Finding a correlation between the Fourier parameters and the metallicity for long-period Cepheids is a challenge mainly because there are fewer spectroscopic observations of metal-poor long-period Cepheids. Moreover these stars are colder, on average, and thus the convection affects the photosphere and the resulting light curves. However, the recent spectroscopic metallicities obtained by \cite{Romaniello2022} for 89 metal-poor Cepheids from the LMC allow us to compare with metal-rich Cepheids from the Milky Way for
the first time. In the case of long-period Cepheids between 10 and 40 days, a visual inspection of Fig.~\ref{fig:phi21} shows that there is a concentration of metal-poor Cepheids at the top of the $\phi_{21}$ branch, while metal-rich Cepheids seem to have lower values. Some evidence in the literature supports this observation. Observationally, this effect was noticed by \cite{Pont2001}, who studied metal-poor Cepheids in the outer part of the MW. Theoretically, this phenomenon was reproduced by \cite{Buchler1997,Buchler1998}, who used radiative pulsation models. However, it could not be compared to observations at that time. The amplitudes of long-period Cepheids may also be affected by the metallicity. The models computed by \cite{Bono2000} indicate that metal-rich Cepheids with a pulsation period of 10–30$\,$days pulsate with higher peak-to-peak amplitudes than metal-poor ones.  This estimation was observationally supported by  \cite{Majaess2013}, who found that metal-rich, long-period Cepheids display larger $V$-band amplitudes than
their counterparts in very metal-poor galaxies.

In order to quantitatively assess whether or not the shape of the $V$-band light curves depends on the metallicity, we followed the method of the previous section by fitting several possible combinations of Fourier parameters. We considered the following set of Fourier parameters: $A_1$,$\,R_{21}$,$\,R_{31}$,$\,R_{41}$,$\,\phi_{21}$,$\,\phi_{31}$, and$\,\phi_{41}$, as well as the pulsation period $P$, and [Fe/H].
 We selected the Cepheids with pulsation periods from 12 days and no longer than 40 days. Indeed, our data set contains too few stars above $P$=40 days, and longer periods could behave differently, as we can guess from Fig.~\ref{fig:fourier}, or be affected by possible resonance effects \citep{Antonello1998}. We also selected Cepheids with $\phi_{21}$ between 2.5 and 3.3 to remove the stars which could be affected by the resonance around 10~days and with $\sigma_{\phi_{21}}<0.05$.  Our final data set consists of 120 long-period Cepheids.

\begin{figure} 
\begin{subfigure}{0.50\textwidth}
\includegraphics[width=\linewidth]{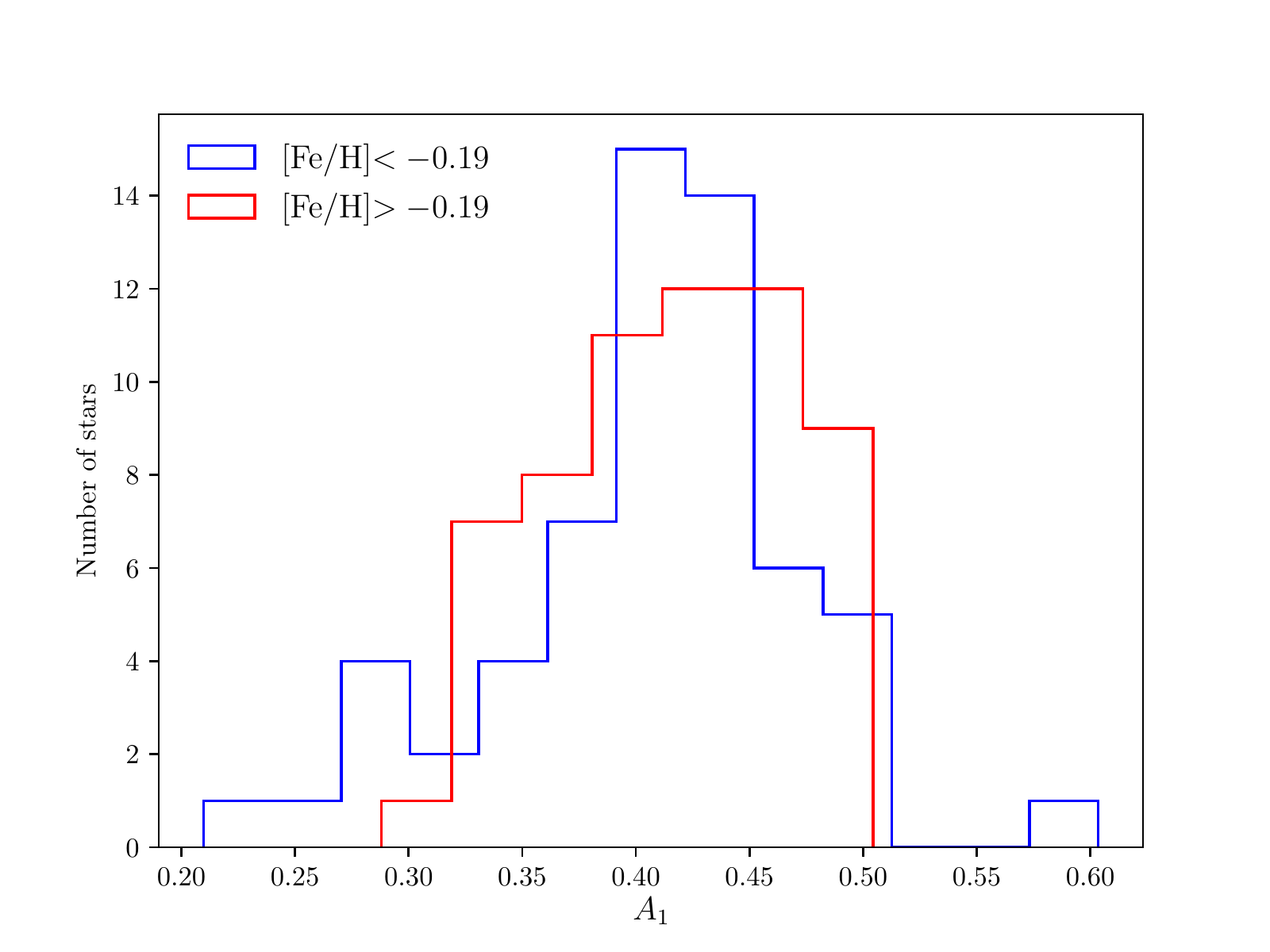}
\caption{$A_1$}
\end{subfigure}\hspace*{\fill}

\begin{subfigure}{0.50\textwidth}
\includegraphics[width=\linewidth]{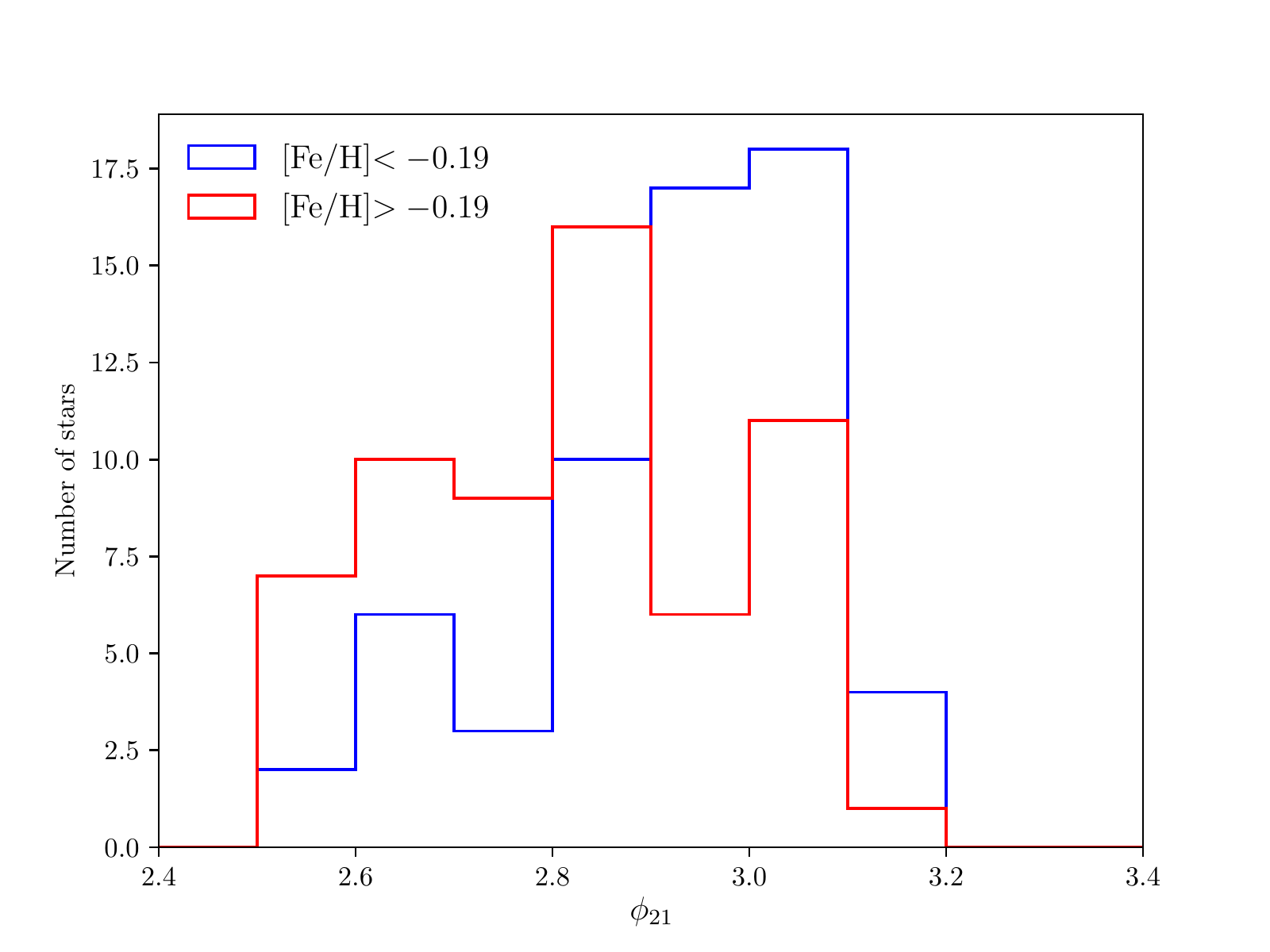}
\caption{$\phi_{21}$} 
\end{subfigure}

\begin{subfigure}{0.50\textwidth}
\includegraphics[width=\linewidth]{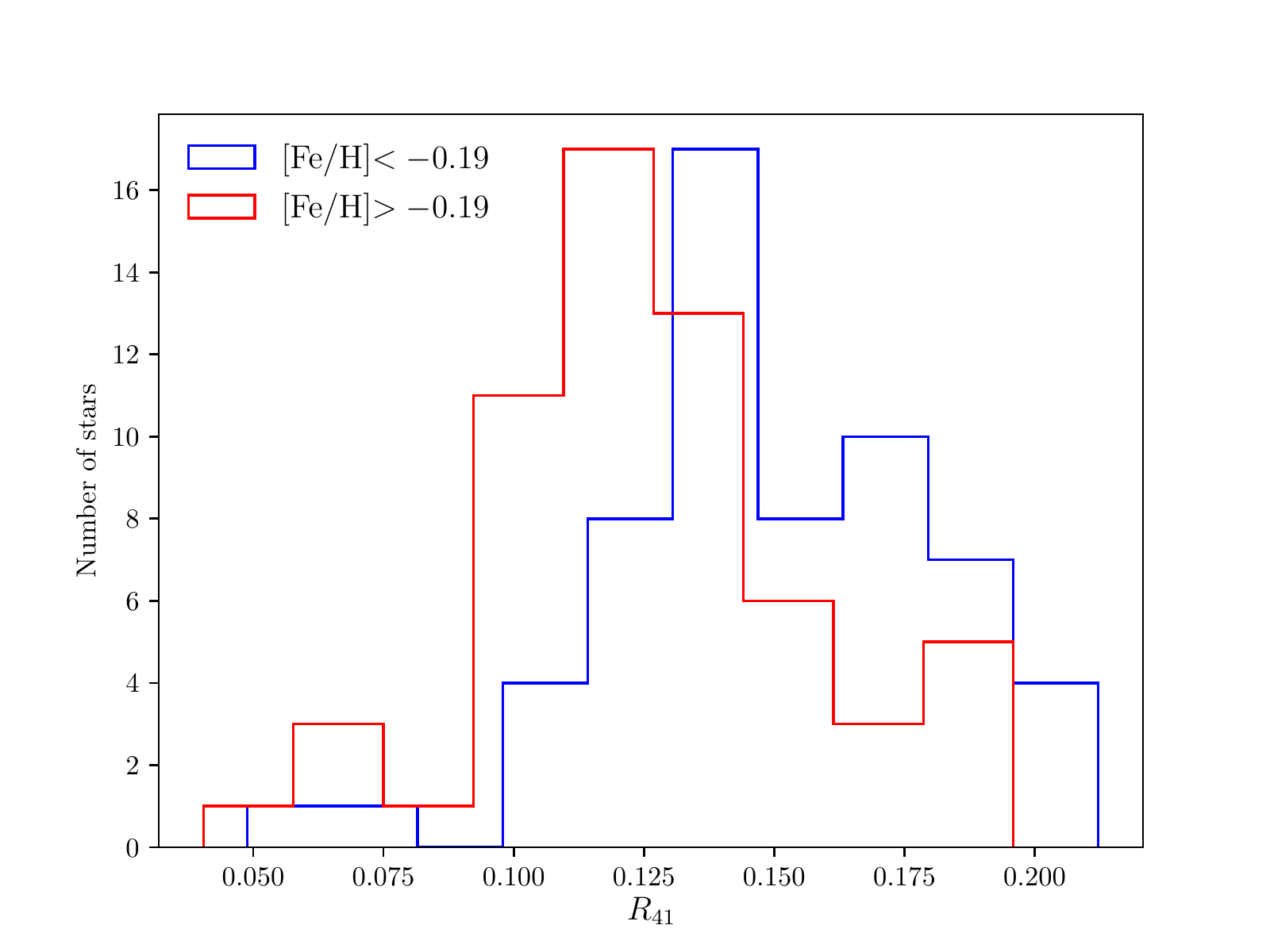}
\caption{$R_{41}$} 
\end{subfigure}
\caption{\small Histogram of $A_1$, $\phi_{21}$ and $R_{41}$ for Cepheids with pulsation period between 12 and 40 days for metal-rich ([Fe/H]>$-0.19$) and metal-poor ([Fe/H]<$-0.19$) stars. [Fe/H]=$-0.19\,$dex represents the median of the metallicity sample of long-period Cepheids.}\label{fig:histo_long}
\end{figure}

\begin{figure}
\begin{center}
\includegraphics[width=0.50\textwidth]{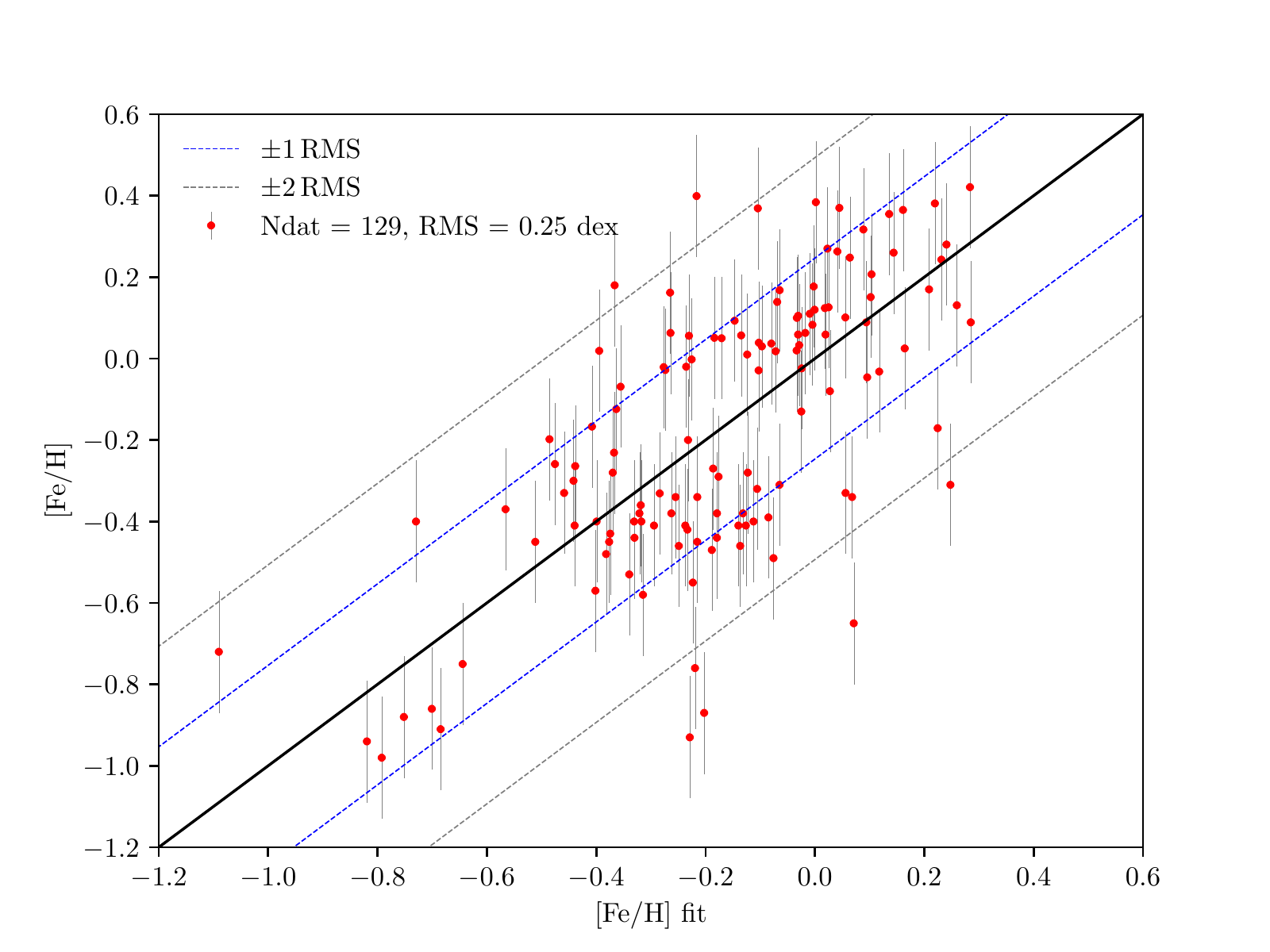}
\caption{\small Comparison of the fitted [Fe/H] with [Fe/H] from literature for long-period Cepheids between 12 and 40 days.} \label{fig:fit_long}
\end{center}
\end{figure}

As in the case of short-period Cepheids, we employed ODR technique to take into account errors on both Fourier parameters and metallicity. We tested all combinations of parameters, two by two, and we also tested all possible combinations, three by three. We estimated the rms and the bias of this relation following the method used in the previous section. We randomly selected  80\% of the sample to fit a relation and tested it on the 20\% remaining objects; then we repeated the procedure 1000 times. We found a significantly better result in the case of the triplet $A_1-R_{41}-\phi_{21}$. The histograms of their distribution are presented in Fig.~\ref{fig:histo_long}. From these figures, we can see that $R_{41}$ and $\phi_{21}$ decrease with larger metallicity, while the amplitude of the harmonic $A_1$ seems to increase with metallicity. The [Fe/H] from the literature and [Fe/H] obtained from the fit is presented in Fig.~\ref{fig:fit_long}. We obtained the following equation:

\begin{equation}
\begin{split}
    [\mathrm{Fe/H}]=(3.94\pm0.46)\,A_1+(-5.80\pm0.85)\,R_{41}\\
    +(-0.93\pm0.16)\,\phi_{21} + (1.67\pm0.39)
\end{split}\label{eq:LONG_eq}
.\end{equation}

 We find rms$=0.25\,$dex for estimating individual metallicity and the relation is accurate with a bias of 0.001$\,$dex.

\subsection{Testing the empirical relations within the SMC and LMC samples}\label{sect:validity_long}
\begin{figure}
\begin{center}
\includegraphics[width=0.50\textwidth]{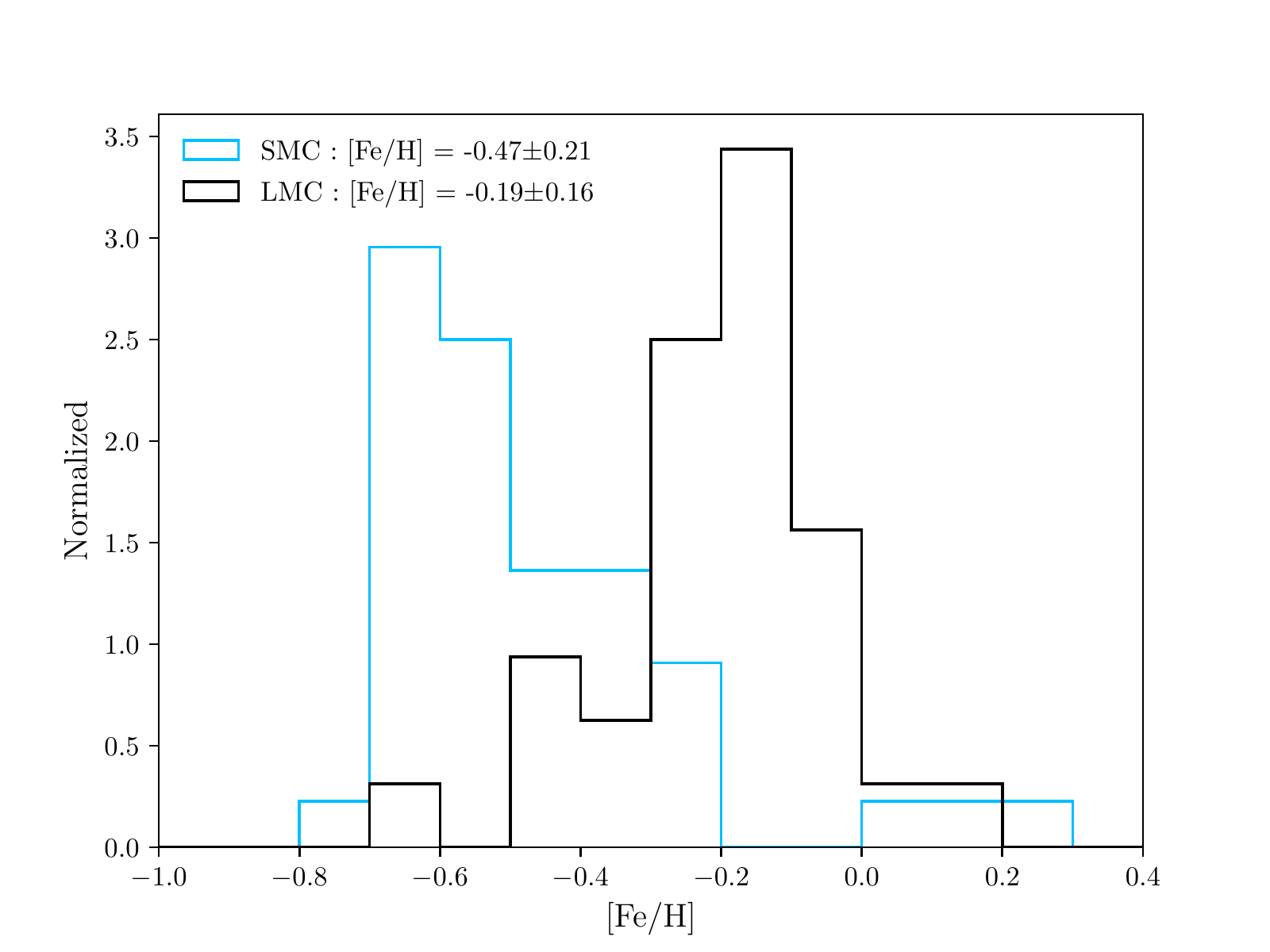}
\caption{\small  Normalized histogram (unit area) for estimated metallicity of SMC and LMC long-period Cepheids (44 and 32 stars, respectively) based on an empirical relation in the $V$ band.} \label{fig:check_phi21_Z}
\end{center}
\end{figure}

We followed the method employed in Sect.~\ref{sect:validity} to check the validity of this empirical relation on a sample of SMC and LMC stars. The Cepheids of the LMC were retrieved from OGLE-IV. However, the long-period Cepheids of the SMC are poorly sampled in OGLE-IV overall. Thus, we compiled the light curves from OGLE III for Cepheids in the SMC instead. We applied a Fourier decomposition (see Sect.~\ref{sect:fourier}) and we removed the stars with $\sigma_{\phi_{21}}$>0.05 as well as those that were  already used for the calibration. We selected Cepheids with $\phi_{21}$ between 2.5 and 3.3, consistently with the calibration sample. The final sample consists of 44 stars from the SMC and 32 stars from the LMC. The result is presented in Fig.~\ref{fig:check_phi21_Z}. As it can be seen from this figure, the empirical relations for long-period Cepheids are able to aptly distinguish between SMC and LMC populations of Cepheids. The mean of the distributions is $-0.19\pm0.16$ and $-0.47\pm0.21\,$dex for the LMC and SMC, respectively. While the tested sample is relatively small, the mean distributions are in agreement within 1$\sigma$ with the mean values from the literature (presented in Sect.~\ref{sect:validity}  and Fig.~\ref{fig:comp}.
In conclusion, the empirical relation based on Fourier parameters $A_1-R_{41}-\phi_{21}$ is found to perform well in deriving accurate results for the mean metallicity for a sample of long-period Cepheids between 12 and 40 days. The estimation of individual metallicity, however, has to be associated with an uncertainty of $\pm0.25\,$dex.

In the next section, we establish empirical relations in the $I$-band and we further test them thanks to a larger sample of stars with well-sampled light curves in the $I$ band (from OGLE-IV).

\section{Empirical relations in the I-band using Fourier interrelations}\label{sect:I_band}
\subsection{Conversion using interrelations}
The near-infrared $I$-band is particularly interesting for observing obscured galactic or extragalactic Cepheids since it is less affected by extinction than in the visible. Therefore empirical calibrations in the $I$-band could be useful for studying the effect on the metallicity on the PL relation and also for galactic archeology. However, it is more difficult to calibrate this relation in the $I$ band because there are fewer $I$-band light curves available with a good sampling in the MW for short- and long-period Cepheids. Although the empirical relations are yet to be calibrated for the $I$ band, we can use interrelations between Fourier parameters in the $V$ and $I$ bands to transform $V$-band relations into the $I$-band. Interrelations consist of linear relations that exist between the same Fourier parameters in different bands and they can be used to convert the $V$-band Fourier parameters into the $I$-band and vice versa \citep{Hendry1999,2000Kanbur,Ngeow2003}. Moreover, \cite{Ngeow2003} emphasized that the change in metallicity does not affect significantly the interrelations between the Fourier coefficient in different bands. Hence, applying these interrelations is shown to be convenient in the case of Cepheids, however, it is less so in the case of RR~Lyrae stars, where the interrelations are affected by metallicity \citep{Skowron2016}.  Thus, our empirical relations (Eqs.~\ref{eq:SHORT_eq} and \ref{eq:LONG_eq}) can be converted to the $I$ band. 
We chose to re-calibrate interrelations in the $V$ and $I$ bands using the sample of SMC and LMC OGLE data. In particular, we focus on the relevant Fourier parameters needed for the conversion, within the specific period ranges adopted by the empirical relations. For the short-period Cepheids between 2.5 and 6.3 days, we crossed-matched the stars in common between OGLE LMC for the $V$ and $I$ bands (1342 stars) and we fit the linear (ODR) relations between $A_1(V)$ and $A_1(I)$ as well as $A_2(V)$ and $A_2(I)$ amplitudes (see Fig.~\ref{fig:inter}). We obtain the following interrelations:

\begin{align}
A_1(V)&=(1.7035\pm0.0066)\,A_1(I) + (-0.0023\pm0.0011), \label{eq:A1}\\
A_2(V)&=(1.7011\pm0.0046)\,A_2(I) + (-0.0002\pm0.0003).
\end{align}

For the long-period Cepheids between 12 and 40 days, we used both SMC and LMC OGLE data to maximize the number of stars for the calibration (giving 120 stars in total; see Fig.~\ref{fig:inter}) to obtain the following interrelations:
\begin{align}
A_1(V)&=(1.6332\pm0.0303)\,A_1(I) + (0.0060\pm0.0070),\\
R_{41}(V)&=(1.0399\pm0.0239)\,R_{41}(I) + (-0.0056\pm0.0034),\\
\phi_{21}(V)&=(0.8389\pm0.0184)\,\phi_{21}(I) + (0.0693\pm0.0633).
\end{align}

Then, there are two equivalent strategies for converting the $V$-band metallicity relations into the $I$-band using interrelations. One possibility is to convert the Fourier parameters of our calibrating sample and then to re-process the fit as done in the $V$-band in the last sections.
The other idea is to directly convert the empirical $V$-band relations. We applied the latter possibility in the following, by substituting the precedent relation into $V$-band empirical relations.
For the short-period Cepheids we obtain:

\begin{equation}
    \mathrm{[Fe/H]}=(10.69\pm0.90)A_1 + (-19.96\pm1.58)A_2 + (-0.60\pm0.07).
\end{equation}
This relation is valid for $P$ between 2.5 and 6.3$\,$days. The cuts that can be applied for $A_{1}(I)$ are between 0.12 and 0.16$\,$mag (corresponding to $A_1(V)$ between 0.20 and 0.25$\,$mag) to mitigate the effect of location of the star inside the instability strip.
For the long-period Cepheids, we obtain:
\begin{equation}
\begin{split}
[\mathrm{Fe/H}]=(6.43\pm0.76)A_1+(-6.03\pm0.89)R_{41}\\
    +(-0.77\pm0.13)\phi_{21} + (1.66\pm0.40)
\end{split}
,\end{equation}
with $\phi_{21}(I)$ between 2.9 and 3.85 (corresponding to $\phi_{21}(V)$ between 2.5 and 3.3).

\begin{figure}. 
\begin{subfigure}{0.50\textwidth}
\includegraphics[width=\linewidth]{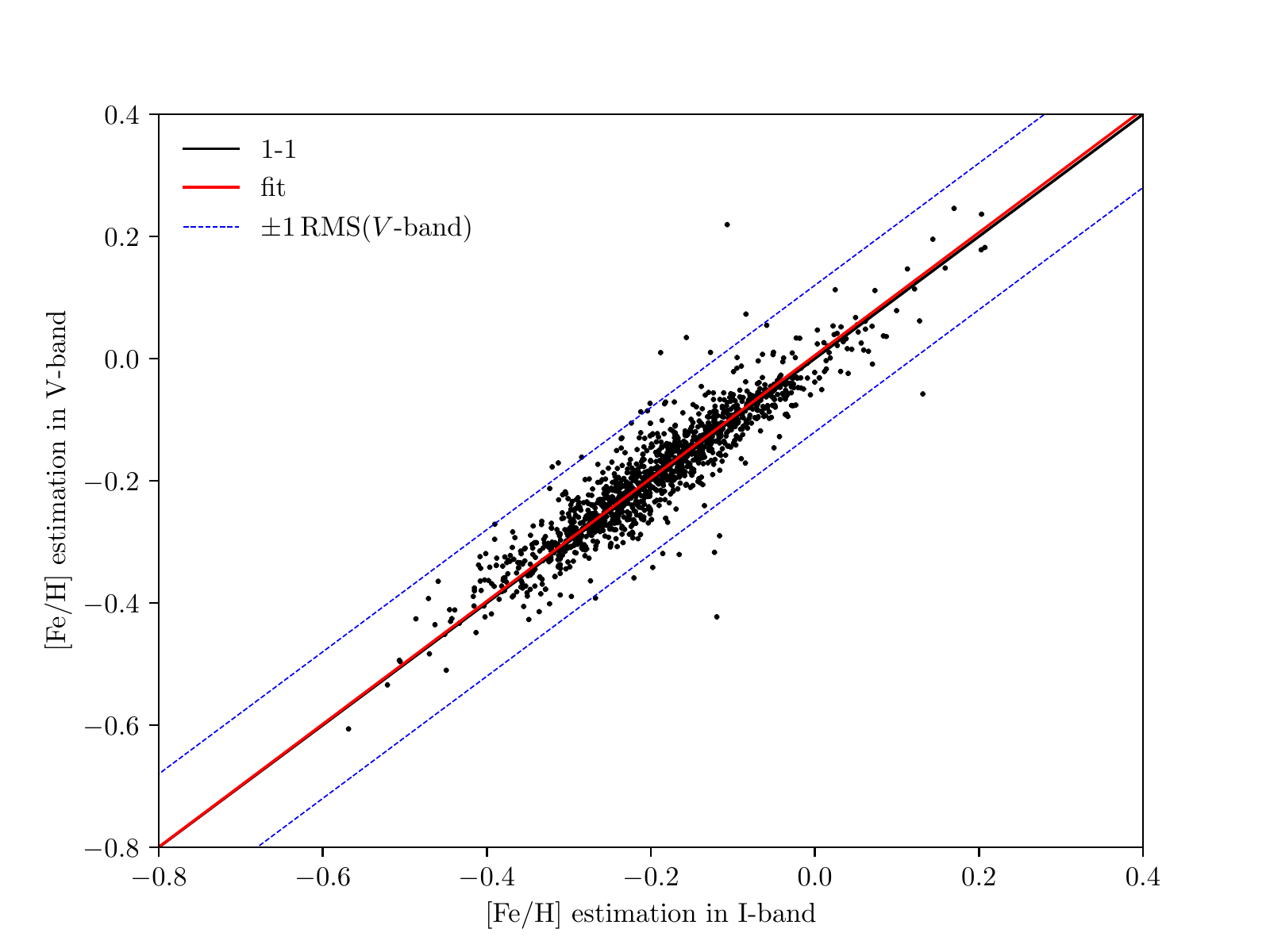}
\caption{Short-period Cepheids} \label{fig:verif_short}
\end{subfigure}

\begin{subfigure}{0.50\textwidth}
\includegraphics[width=\linewidth]{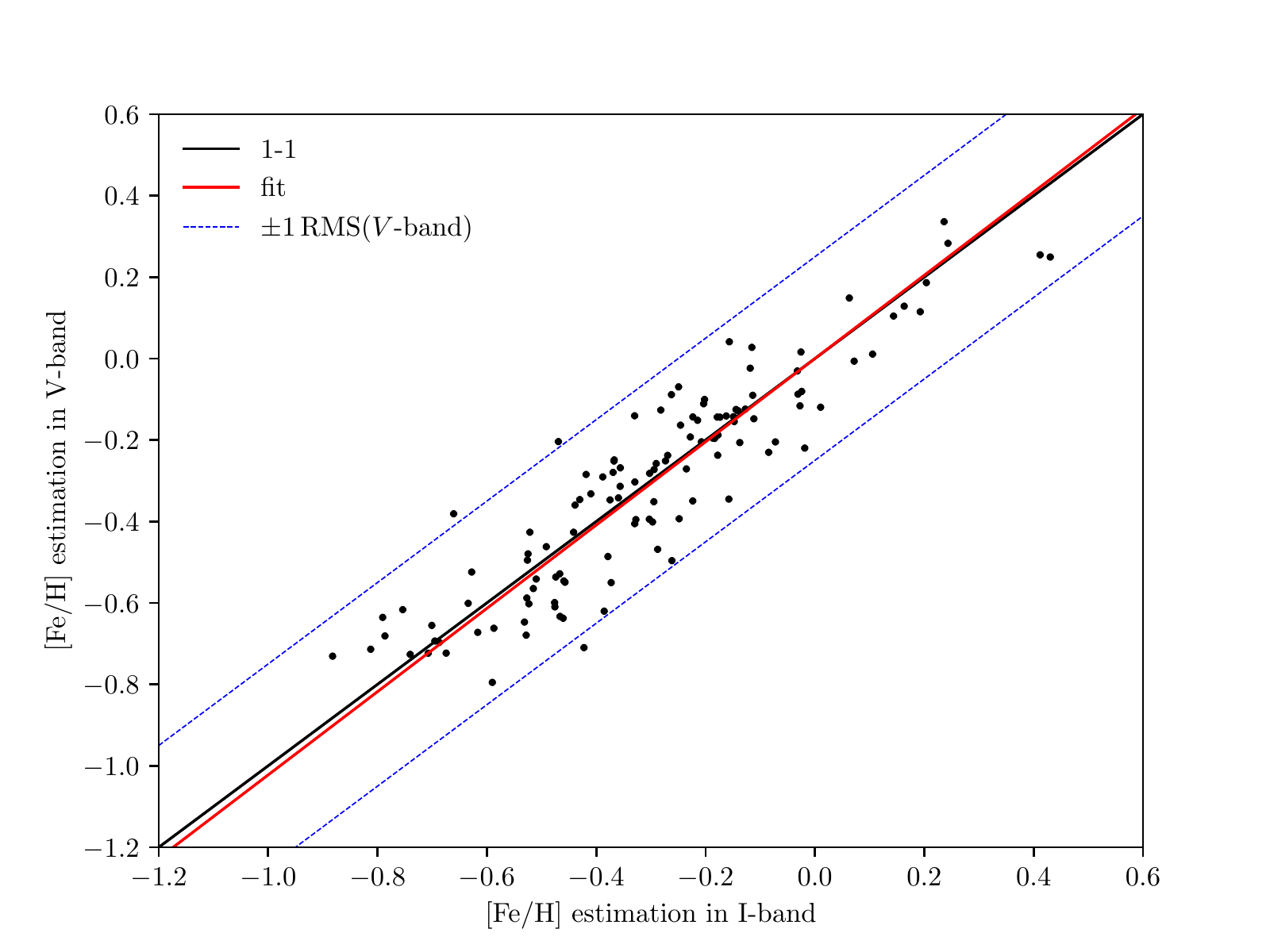}
\caption{Long-period Cepheids} \label{fig:verif_long}
\end{subfigure}
\caption{\small Comparison of metallicities derived with both the $V$-band and $I$-band empirical relations. Black line indicates a one-to-one correspondence and thick red line is the fit between the two bands. Blue dashed lines represent the rms of the empirical relation in the $V$ band.  \label{fig:verif}}
\end{figure}

As a first check, it is important to assess whether these relations in the $I$ band are consistent within the uncertainties with empirical relation in the $V$-band. In order to compare these relations, we consistently applied the empirical relations for the stars in common. The result is shown in Fig.~\ref{fig:verif_short}. Applying a linear fit to the data, we found that the [Fe/H] estimations in the $V$ and $I$ band are colinear with no offset: [Fe/H]$_V$= 1.01$(\pm0.01)$[Fe/H]$_I$ $+0.005(\pm0.002)$. The resulting [Fe/H] estimation in the $I$ band is also consistent with the $V$ band within 0$\pm$0.04$\,$dex, namely, this is below the rms of 0.12$\,$dex for $V$-band empirical relation.

We repeated the procedure for the long-period Cepheids in our sample (see Fig.~\ref{fig:verif_long}). After applying a linear fit, we obtained [Fe/H]$_V$= 1.02$(\pm0.03)$[Fe/H]$_I$ $+0.001(\pm0.014),$ which is also consistent with colinearity and no offset. The resulting estimation in the $I$ band is in agreement in the $V$ band within 0$\pm$0.10$\,$dex, namely, below the $V$-band rms of 0.25$\,$dex.

\subsection{Testing the empirical relations to MW, SMC, and LMC OGLE-IV sample}\label{sect:validity_I}
In order to further test the performance of the $I$-band relation, we first verified the metallicity distribution obtained for the MW, SMC, and LMC. Similarly to previous sections, we applied these empirical relations to OGLE-IV light curves in the $I$ band to derive the metallicity of SMC, LMC, and MW Cepheids. For the MW Cepheids, we gathered the light curves from the Galactic Disk and the Bulge.

\begin{figure}
\begin{subfigure}{0.50\textwidth}
\includegraphics[width=\linewidth]{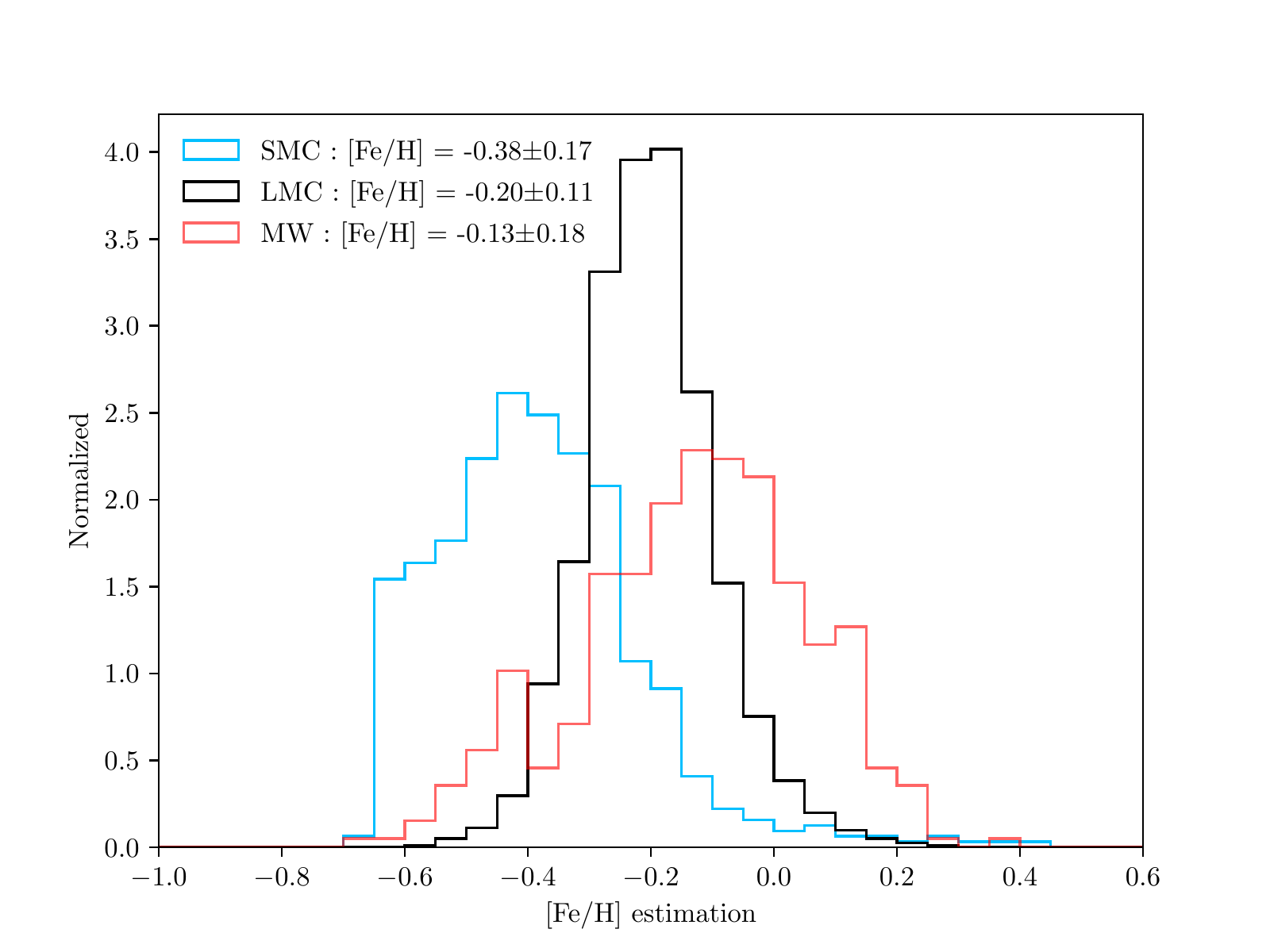}
\caption{$I$-band metallicity estimation for short-period Cepheids} \label{fig:histo_short_Iband}
\end{subfigure}

\begin{subfigure}{0.50\textwidth}
\includegraphics[width=\linewidth]{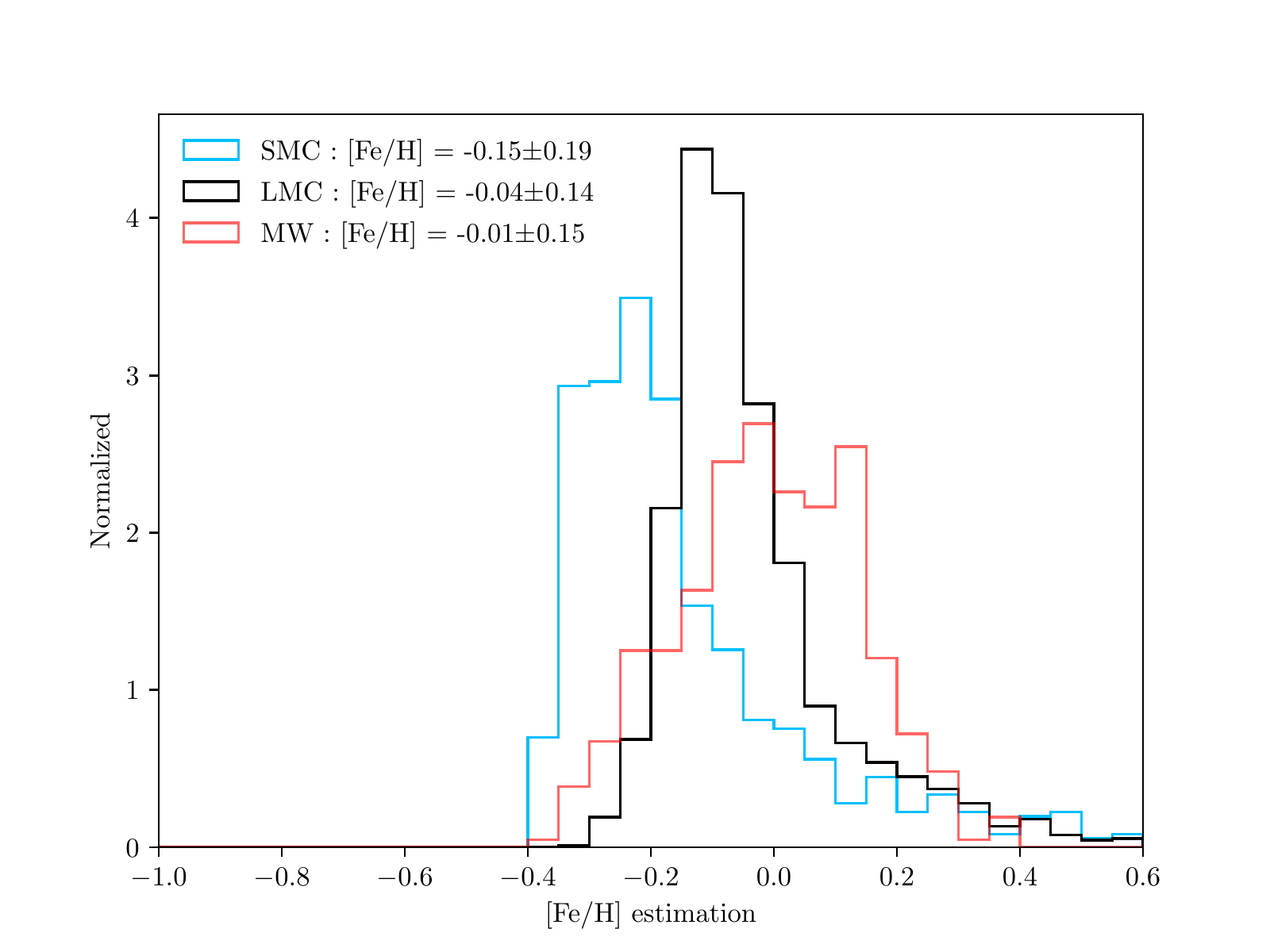}
\caption{$I$-band metallicity estimation \citep{Klagyivik2013}} \label{fig:histo_KLA}
\end{subfigure}

\begin{subfigure}{0.50\textwidth}
\includegraphics[width=\linewidth]{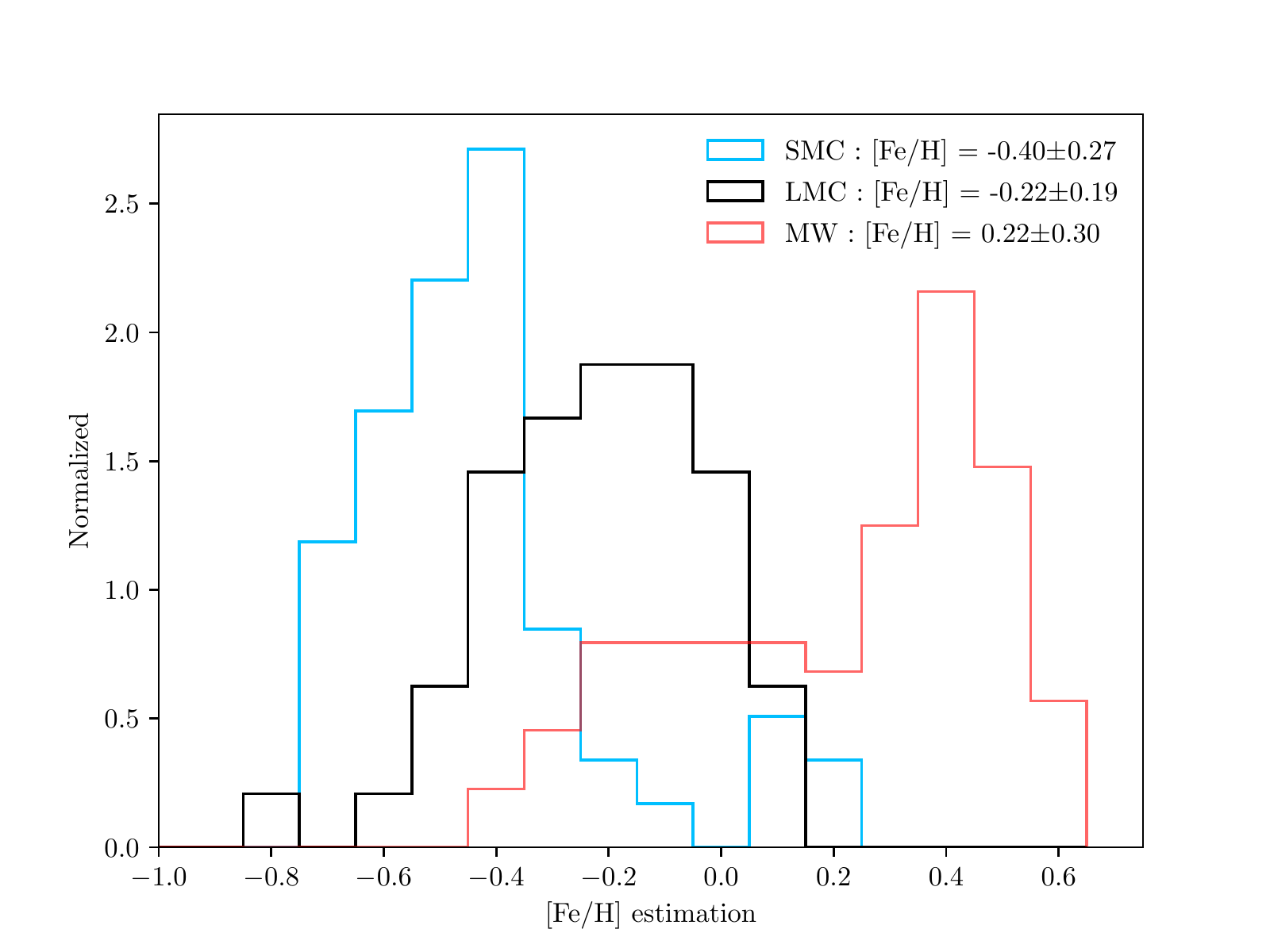}
\caption{$I$-band metallicity estimation for long-period Cepheids} \label{fig:histo_long_Iband}
\end{subfigure}
\caption{Normalized histograms (unit area)  of the metallicity estimation of the SMC (blue) and LMC (black) fundamental Cepheids and the MW (red) fundamental Cepheids (GD+BULGE)   from the shape of the $I$-band light curves. For short periods, SMC, LMC, and MW(GD+BULGE): 635, 1649, and 394 stars, respectively. For long periods, SMC, LMC and MW(GD+BULGE): 60, 48 and 92 stars, respectively.\label{fig:result_I_band}}
\end{figure}

In case of short-period Cepheids, we cut all stars with $A_1<0.12\,$mag (corresponding to $A_1<0.20\,$mag in $V$-band) from the sample. This was done to mitigate the effect of the location within the instability strip. The testing sample is thus composed of 394 stars from the MW, 635 stars in the SMC and 1649 stars in the LMC. The results are presented in Fig.~\ref{fig:histo_short_Iband}.
We can see that the $I$-band relation is able to distinguish the three populations of stars. For each distribution, we derived a mean metallicity for the SMC: $-0.38\pm0.17\,$dex, LMC: $-0.20\pm0.11\,$dex, and MW: $-0.13\pm0.18\,$dex -- these results are in agreement with the mean values from the literature. As a comparison, we also applied the empirical relation from \cite{Klagyivik2013} based on $R_{21}$ in the $I$ band in the same period range (see Fig.~\ref{fig:histo_KLA}). As noticed by \cite{Clementini2019}, this relation overestimates the metallicities by about +0.2$\,$dex. This problem is mitigated when using the relation calibrated in this paper.

For long-period Cepheids, the testing sample is composed of 92 stars from the MW, 60 stars in the SMC, and 48 stars in the LMC. The empirical relation performs well to obtain the distribution of the expected metallicity for these galaxies (see Fig.~\ref{fig:histo_long_Iband}), although the sample of star is much smaller than the short-periods. For each distribution, we derived a mean metallicity for the SMC: $-0.40\pm0.27\,$dex and LMC: $-0.22\pm0.19\,$dex; these results are in agreement within 1$\sigma$ with both the mean values from the literature and our previous results for short-period Cepheids. For the MW sample, we derived a mean [Fe/H] of $+0.22\pm0.30\,$dex, which offers a large spread, in agreement to what is observed for short-period Cepheids.
 We show in the next section that this effect is due to the metallicity gradient in the Galaxy.

\subsection{Mapping the metallicity distribution in the MW, SMC, and the LMC}
In this part we use our metallicity estimations for the fundamental-mode Cepheids in the MW, SMC, and LMC and we retrieved their coordinates from 3D maps available in the literature. Then, we tested whether our relations are precise enough to obtain information on the metallicity gradient in these galaxies for potential applications in galactic archeology. 

\begin{figure*} 
\begin{subfigure}{0.50\textwidth}
\includegraphics[width=\linewidth]{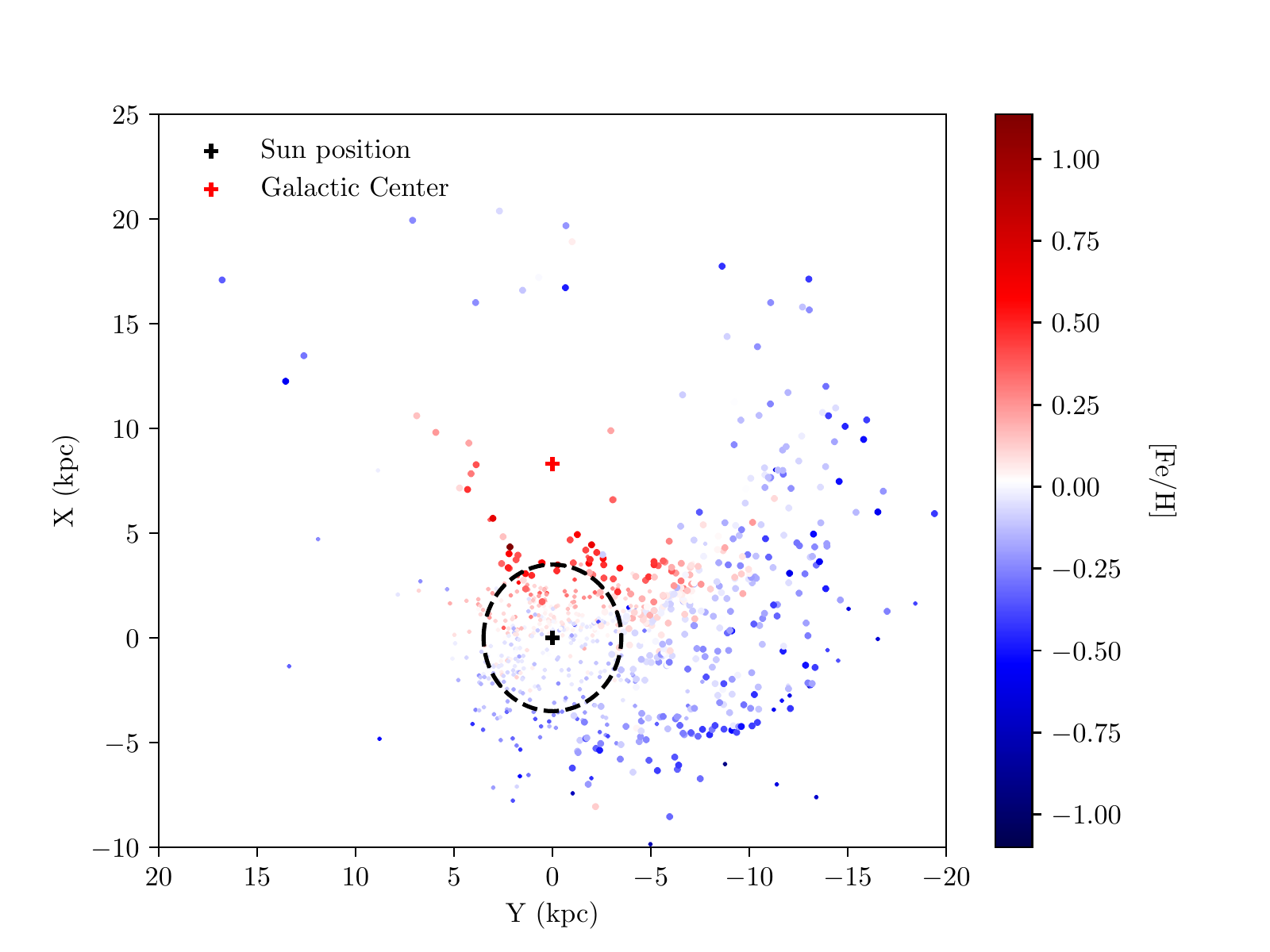}
\caption{MW on-view} \label{fig:MW_map}
\end{subfigure}\hspace*{\fill}
\begin{subfigure}{0.50\textwidth}
\includegraphics[width=\linewidth]{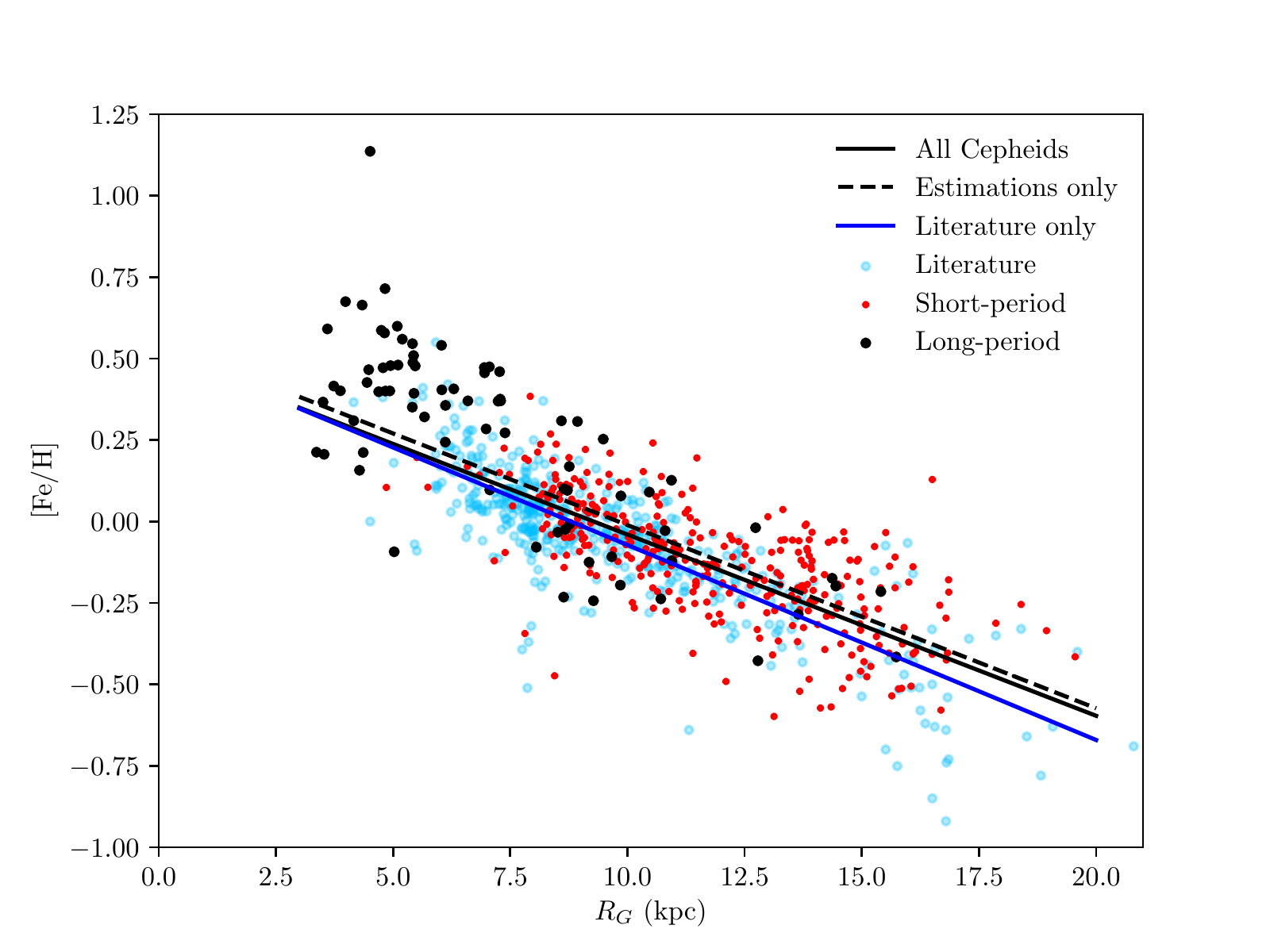}
\caption{} \label{fig:grad_MW}
\end{subfigure}

\medskip
\begin{subfigure}{0.50\textwidth}
\includegraphics[width=\linewidth]{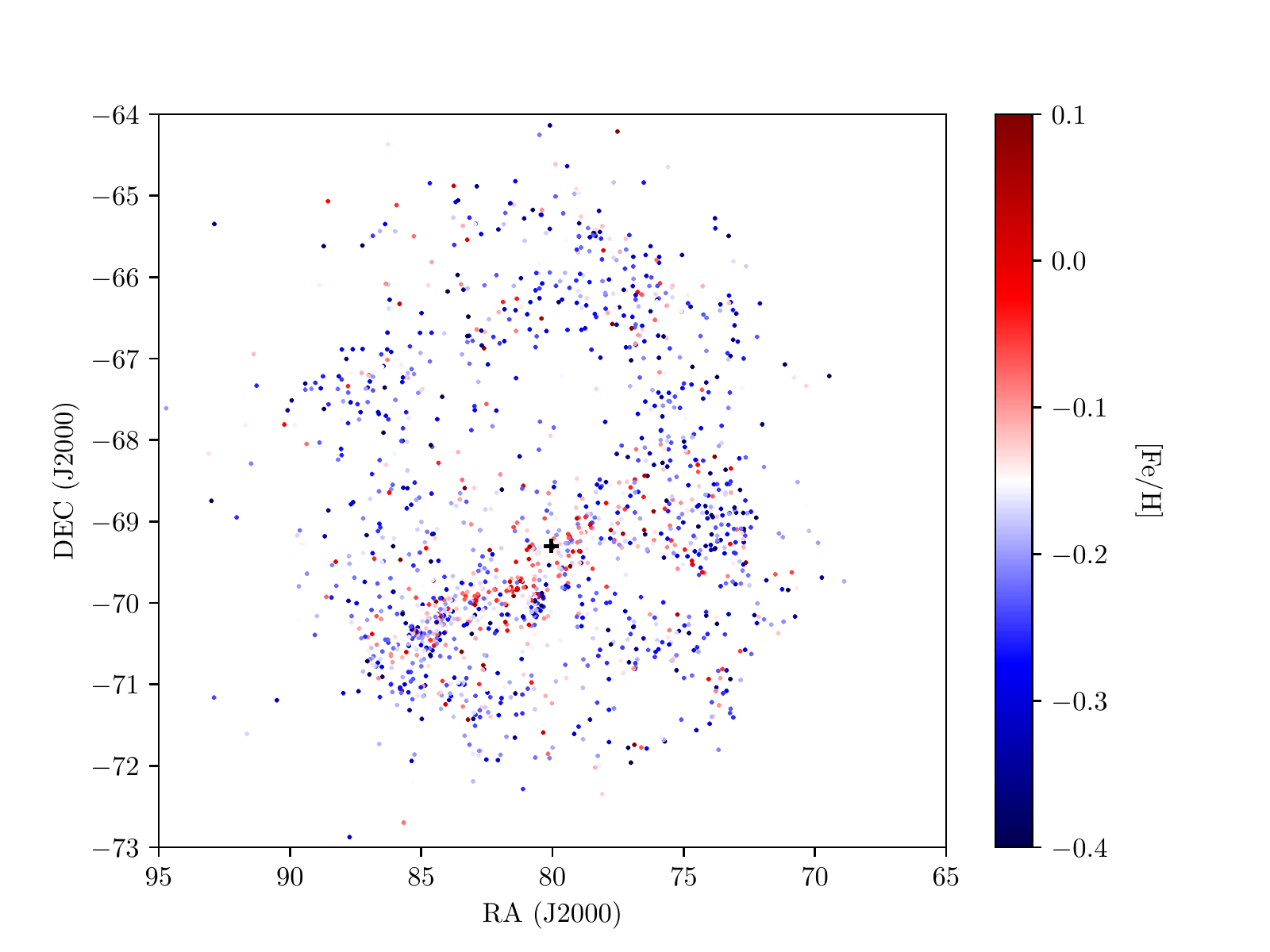}
\caption{LMC} \label{fig:LMC_map}
\end{subfigure}\hspace*{\fill}
\begin{subfigure}{0.50\textwidth}
\includegraphics[width=\linewidth]{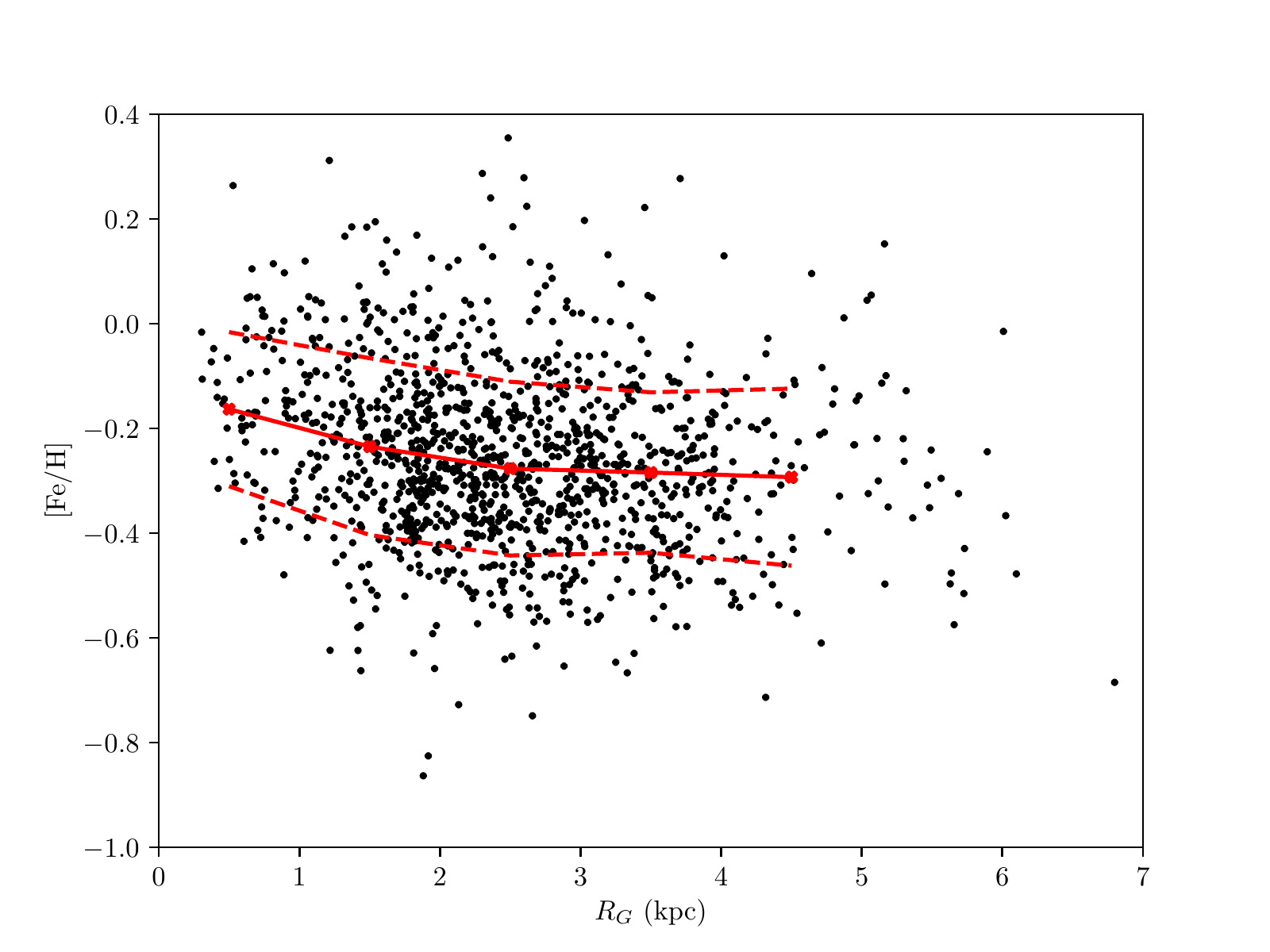}
\caption{} \label{fig:LMC_grad}
\end{subfigure}

\medskip
\begin{subfigure}{0.50\textwidth}
\includegraphics[width=\linewidth]{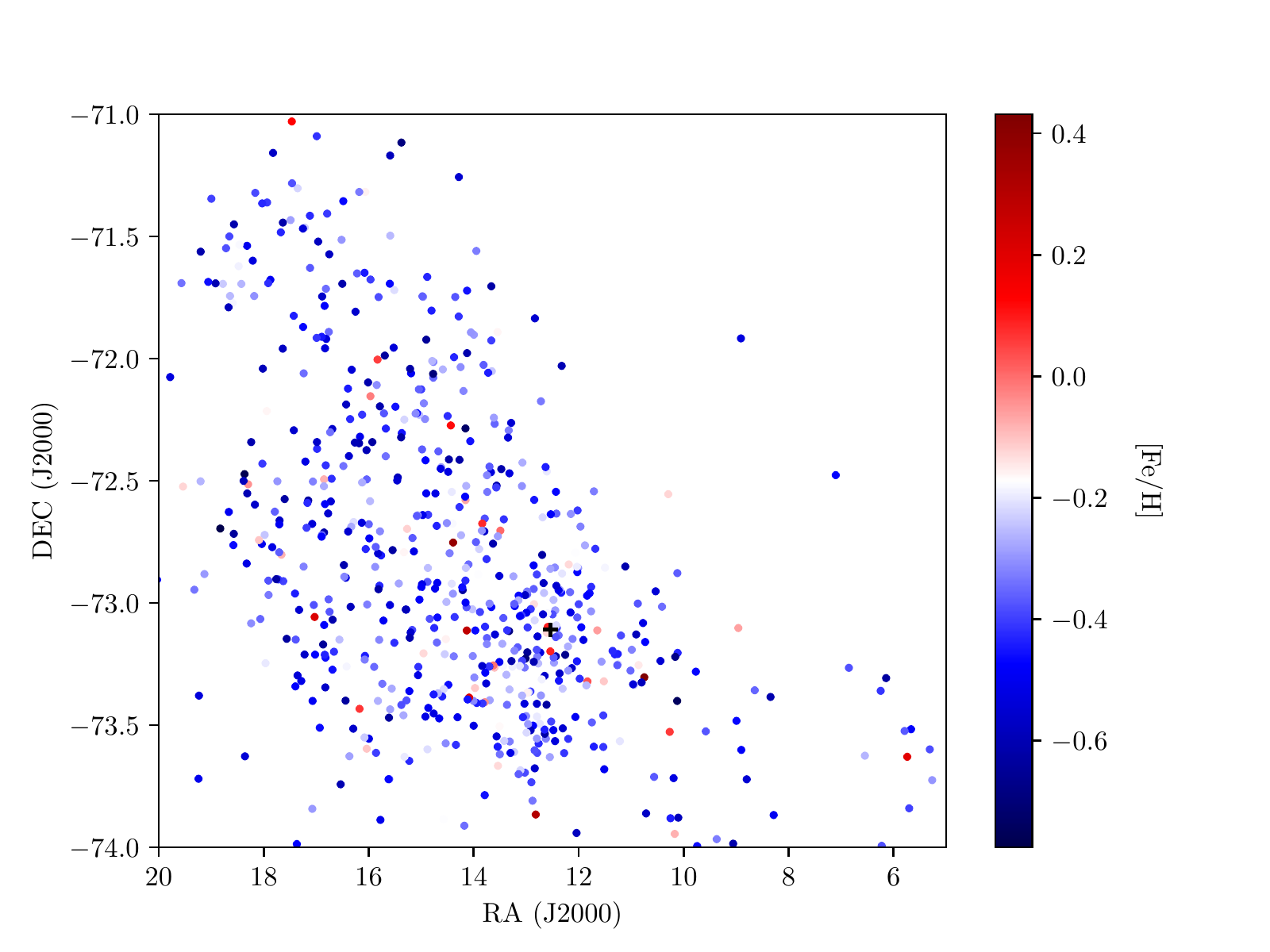}
\caption{SMC} \label{fig:SMC_map}
\end{subfigure}\hspace*{\fill}
\begin{subfigure}{0.50\textwidth}
\includegraphics[width=\linewidth]{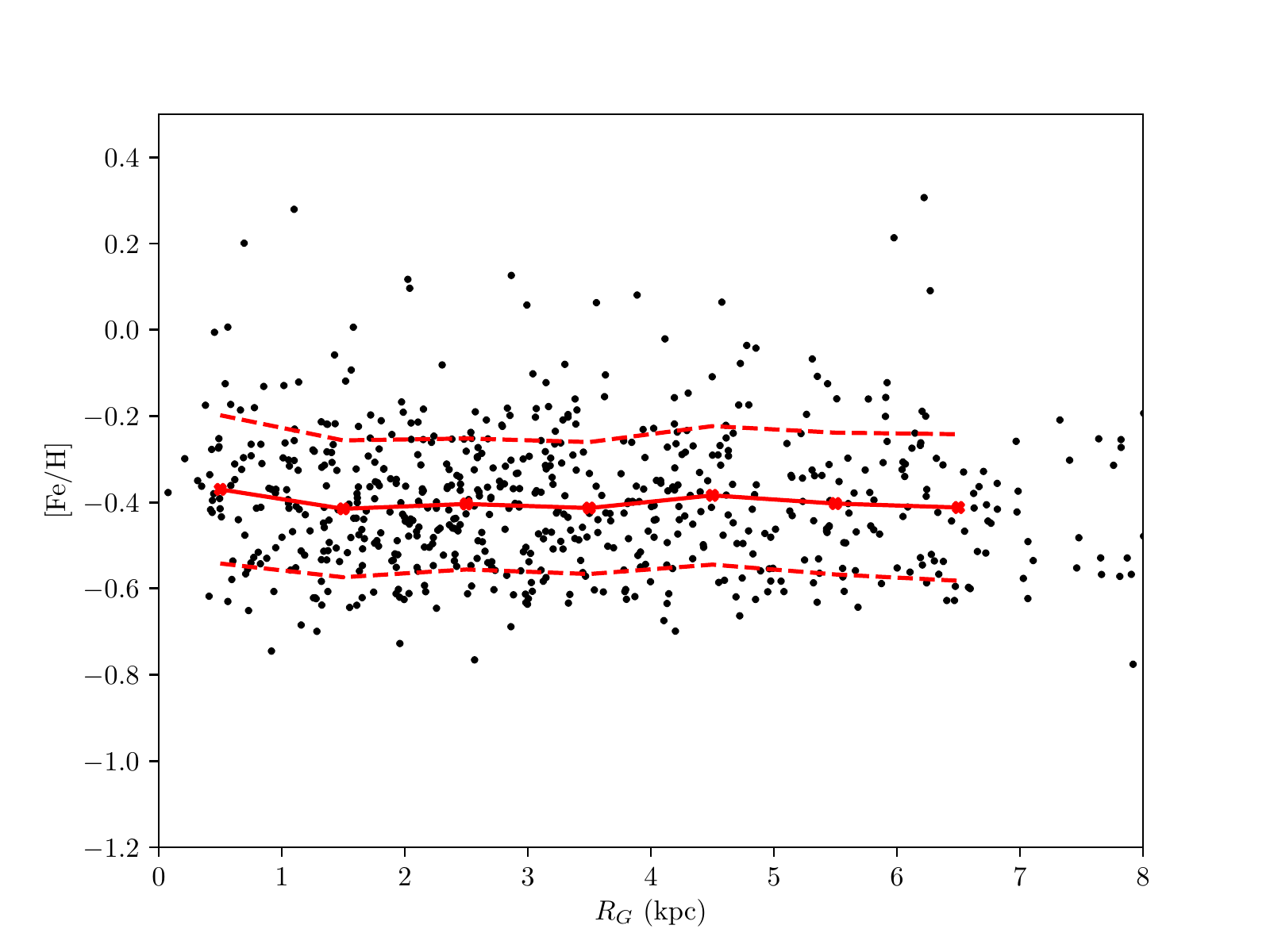}
\caption{} \label{fig:SMC_grad}
\end{subfigure}
\caption{\small Metallicity distribution of the Milky Way (868 stars), LMC (1561 stars), and SMC (674 stars) from empirical metallicity relations in the $I$ band: a) small points represent stars with metallicity from the literature while bigger points are new metallicity estimation. Dashed circle encloses the solar neighbour region which concentrates most of the spectroscopic measurements; (b), (d), and (f): Estimated Cepheid metallicities versus the galactocentric distance for MW, LMC, and SMC. Red dashed lines are the $1\sigma$ deviation in each bin of 1$\,$kpc around the median value of each bin. MW and LMC maps are plotted in a larger size in Figs.~\ref{fig:MW_map_annex} and 
\ref{fig:LMC_map_zoom}.}\label{fig:map}
\end{figure*}

For stars with a metallicity estimation of the MW Cepheids, we cross-matched the OGLE-IV sample in the $I$-band with the stars from the Galactic 3D map from \cite{Skowron2019}. For the literature sample, we cross-matched the 472 fundamental Cepheids of our metallicity sample (see Table \ref{Tab:bilan}) with the map from \cite{Skowron2019} and \cite{Kovtyukh2022}. As a result, we obtained 868 stars in the MW, with 460 having metallicity from the literature and 408 others estimated by our relations. The galactic coordinates and the distance measurements of these stars (when available) allow us to plot the metallicity distribution in the MW (see Fig.~\ref{fig:MW_map}). By adopting a Sun galactocentric distance of 8.1$\pm$0.1$\,$kpc \citep{Bobylev2021}, we can also display the metallicity as a function of the galactocentric radius $R_G$ (see Fig.~\ref{fig:grad_MW}). From these distributions we observe that spectroscopic observations are mostly given in the vicinity of the Sun while the OGLE sample covers farther distances. This is possible thanks to the sensitivity of the spectroscopy, which limits the observation to the closest (bright) stars, while saturating the OGLE detectors. We also note that for galactocentric distances below about 7$\,$kpc, it is  nearly entirely long-periods Cepheids that are observed. The higher level of interstellar extinction at these distances toward the galactic center likely limits the observation of faint short-period Cepheids. We fit a straight line for three different samples, taking into account the uncertainty on the galactocentric distance and the metallicity. For the sample of the spectroscopic metallicity that we gathered from the literature (460 stars), we obtain:
\begin{equation}
    \mathrm{[Fe/H]}=(-0.0598\pm0.0018)\,R_G + (0.5263\pm0.01848).
\end{equation}
For the metallicity estimations from $I$-band light curves only (408 stars):
\begin{equation}
    \mathrm{[Fe/H]}=(-0.0562\pm0.0023)\,R_G + (0.5514\pm0.0280);
\end{equation}
and the sample with both literature and estimations (868 stars):
\begin{equation}
    \mathrm{[Fe/H]}=(-0.0556\pm0.0015)\,R_G +(0.5156\pm0.0163).
\end{equation}
The relation obtained only from the literature sample is in agreement with previous results for the near- and far-side of the disk \citep{Genovali2014, Luck2018,Minniti2020}, which provide a gradient between $-0.05$ and $-0.06\,$dex$\,$kpc$^{-1}$. The empirical relation in the $I$ band performs well to reconstruct a metallicity gradient in the galaxy that is on the same order of magnitude as reported in the literature, with $-0.056\,$dex$\,$kpc$^{-1}$.
It is difficult to adequately compare the slope obtained from the empirical relations with the results from the literature. Indeed, the OGLE-IV MW sample encompasses regions that are different from the spectroscopic observations, in terms of both the radial and azimuthal angle (see small vs. big points in Fig.~\ref{fig:MW_map}; also, a larger figure is presented in the annex for visualization Fig.~\ref{fig:MW_map_annex}). Our result suggests that empirical relations in the $V$ and $I$ bands are useful tools for obtaining information on the metallicity distribution in the Galaxy, particularly in regions that are not accessible by spectroscopy up to 20$\,$kpc from the Sun.

We also mapped the SMC and the LMC with our metallicity estimation by using the distances and coordinates of Cepheids for the SMC and the LMC \citep{Jacyszyn2016} (see Figs.~\ref{fig:SMC_map} and \ref{fig:LMC_map}).  After cross-matching, we obtained 1561 LMC Cepheids and 674 SMC Cepheids with estimated metallicities. Interestingly, we observe higher metallicities in the LMC bar in the LMC map. We display the metallicity as a function of the galactocentric radius in Figs.~\ref{fig:SMC_grad} and \ref{fig:LMC_grad}. Since the distribution of the relatively young population of Cepheids is highly non-uniform, in contrast to older RR Lyrae, for example, displaying the metallicity as a function of the galactocentric radius has the sole purpose of showing non-correlations.
In order to observe the metallicity gradient for the SMC and the LMC, we followed the method described in \cite{Skowron2016} for the case of RR Lyrae stars. For each bin of 1$\,$kpc, we derived the median of the metallicity and we derived the standard deviation (see red crosses and dashed lines in Figs.~\ref{fig:LMC_grad} and \ref{fig:SMC_grad}). We find similar results to those obtained from \cite{Skowron2016}, with an apparent decrease of the median metallicity within 4$\,$kpc from the center in the case of the LMC, while no slope is visible in the SMC. Because of the large scatter of the metallicity gradient observed in these galaxies, as observed in a similar was for RR Lyrae metallicities \citep{Skowron2016}, our findings on the slope are only qualitative in scope.

In conclusion, our metallicity estimates are able to not only reproduce the peak of metallicity distributions of galaxies (as shown in the previous sections), but they can also reproduce the trends of the metallicity gradient in MW. Therefore, the empirical relations calibrated in this paper can be applied to further studies in galactic archeology.

\subsection{Quantitative test: Investigation of systematics using Cepheid-cluster pairs}
A relevant quantitative test can be performed by comparing Cepheid metallicity with open-cluster metallicity in order to investigate possible systematics. A recent paper investigated the open-cluster (OC) housing MW classical Cepheids \citep{Hao2022} from the Gaia DR3 release. They identified 39 probable open cluster-and-Cepheid pairs. Most of these Cepheid-OC pairs are in the Sun's vicinity and thus are already part of our calibrating sample. For that reason, this MW sample cannot be independently used to check systematics in our relations.

Alternatively, we can use OC-Cepheid pairs in the LMC or SMC to perform those tests at a lower metallicity. This is also interesting since we expect systematics at the more metal-poor regime of our empirical relations. The two massive LMC clusters NGC 1866 and NGC 2031 have the largest known Cepheid population \citep{WelchStetson1993,Testa2007,Musella2016}. In NGC 1866 we found light-curves for five Cepheids from OGLE. Among these stars three are already part of our calibration sample \citep{Lemasle2017} and one has a low amplitude (LMC-CEP-3731, $A_1$(I)$<$0.12$\,$mag). Thus, we only have an estimation for LMC-CEP-3721, with [Fe/H]$=-0.14\pm0.12\,$dex. Although one photometric estimation is not enough to make a comparison, this measurement seems to be biased by at least +0.2 dex compared to various NGC 1866 spectroscopic measurements \citep[see][]{Lemasle2017}. We note, however, that the light curves for other Cepheids in NGC 1866 might be obtained by combining photometry from different instruments \citep[see][]{Musella2016}.

In NGC 2031, we found five others Cepheids that have their light curves available from OGLE and we determined the following metallicities using our empirical relations: OGLE LMC-CEP-2377 ($-0.08\,$dex);
LMC-CEP-2371 (0.03$\,$dex);
LMC-CEP-2391 ($-0.17\,$dex);
LMC-CEP-2385 ($-0.18\,$dex);
and LMC-CEP-2375 ($-0.27\,$dex), which yields the average [Fe/H] of $-0.13\pm0.10\,$dex.
To our knowledge, the only available spectroscopic metallicity value for NGC 2031 is $-0.52\pm0.21\,$dex \citep{Dirsch2000}. On the other hand, \cite{Chilingarian2018} obtained [Fe/H]$=-0.14\pm$0.02$\,$dex from a fitting of integrated optical spectra. They also found that the metallicity of NGC~2031 is in excellent agreement with age-metallicity relation and models of the LMC's chemical enrichment history. Our empirical value is discrepant with respect to the spectroscopic measurement, but it is in agreement with the latter. Hence, it is unclear whether or not our estimations are affected by systematics with regard to this cluster.

In the case of the SMC, the cluster NGC 330 is the brightest young cluster that also is host to several Cepheids \citep{SeboWood1994}. Among the classical Cepheids in NGC~330, our empirical relation is valid (period and amplitude ranges, light-curve quality, etc.) only for OGLE-SMC-CEP-2634 and we derived
 [Fe/H]$= - 0.61\pm0.12\,$dex. Several metallicity estimates obtained from different methods are available for NGC~330, which also complicates the comparison. Using high-resolution spectroscopy \cite{Hill1999} found that NGC~330 is metal-deficient compared to the field stars ($-0.82\pm0.11\,$dex vs $-0.69\pm0.10\,$dex). More recently, \cite{Narloch2021} found $-0.98\pm0.08\,$(stat.)$\pm 0.10\,$(syst.)$\,$dex from stromgren photometry. Based on these values, our empirical estimation might by biased by about +0.2 dex or more at low metallicity. However we cannot firmly conclude since we have based our comparison on only one empirical estimation.
 
In conclusion, it is difficult at this stage to check the potential systematics in empirical relations by comparison with cluster metallicity. Clusters  generally host one Cepheid (see \cite{Dinnbier2022} and reference therein), thus, identifications of a significant number of OC-Cepheid pair in MW and beyond, together with reliable metallicities measurements, are necessary for further investigations of these systematics.

\section{Discussion}\label{sect:discussion}

In this paper, we provide new empirical relations for estimating the metallicity for short and long-period Cepheids in the $V$-band, and we used interrelation from the literature between $V$ and $I$ bands to convert these relations into the $I$-band.
 For individual metallicity determination, uncertainties of $\sigma$=0.12$\,$dex and $\sigma$=0.25$\,$dex have to be attributed for short- and long-period Cepheids respectively. As noted by \cite{Cacciari2005} in the case of RR Lyrae stars, photometric formulae are good tools to study mean metallicities in a population, but are less precise when determining the metallicity of individual stars. Hence our empirical relations can be more adapted to characterise large population of Cepheids. Finally, we also tested the relations in the $I$-band to map the metallicity distribution of MW, SMC and LMC and we were able to recover the metallicity gradient consistent with the literature. However, our calibration sample lacks spectroscopic measurements of metal-poor Cepheids below about $-0.5\,$dex especially for the short-periods. Thus our relation has to be used with caution when measuring metallicity of metal-poor galaxies or at the outer most region of the Milky Way. On the other hand, the calibration of such empirical relations can be still improved in the future. A way to improve the calibration would be to obtain an homogeneous data set for spectroscopic metallicity, determined with a single method which mitigates the systematics measurements \citep{daSilva2022}. Homogeneous sample of light curves in both $V$ and $I$-bands will be also helpful. In this paper we also assumed simple linear relation between the Fourier coefficients. \cite{Klagyivik2013} suggested this relation might not be linear at low metallicity in the case of $R_{21}$ and this is also likely the case for the relation based on $A_1A_2$. Further investigations are needed to find a unique relation between the Fourier coefficients that would be able to generalize from poor to rich metallic Cepheids.
 
 Apart from the assumption of a linear relation, several physical source of scattering can affect the estimation of the metallicity. As discussed in the introduction, the presence of a companion or constant CSE emission would blend the amplitudes of the harmonics used in our relations and thus can be a source of scattering for empirical metallicity relations.
 Another significant source of scattering in these relations comes from physical effect such as the location in the instability strip that we were not able to correct for. As an alternative, we proposed to use selection threshold on $A_1$ in the case of short-period Cepheids. Although this cut was defined approximately, we think it is however necessary to mitigate the dependence of the amplitude on the location within the instability strip. It is thus important to better understand the influence of the location within the instability strip on the Fourier parameters to correct for this effect.
 
 Theoretically, the impact of the metallicity on the Fourier parameters remains to be explained, just as it is for RR Lyrae. The metallicity dependence of the Fourier parameters can help to constraint hydrodynamical models of pulsation \citep{Paxton2019}. In this paper we also did not analyze the possibility of metallicity dependence between 6 and 10 days because of the strong sensitivity of the Fourier parameters to the $P_2/P_0=0.5$ resonance. A thorough understanding of the influence of the resonance on the Cepheid light curves is a prerequisite for the calibration of metallicity relation in this period range.

Using these relations for Cepheids in MW, SMC and LMC we have shown their capabilities for galactic archeology applications.
Another possible application is to determine the metallicity term in PL relation. In order to study the metallicity term authors attribute in general the average metallicity of the host galaxy (measured from a restricted sample of stars) to the sample of Cepheids used in PL relation \citep[see, e.g.,][]{Breuval2022}. Our empirical relations offer the possibility to either derive the mean [Fe/H] of the studied sample used in PL relation, or to correct individually the stars. Although the associated uncertainties for these photometric [Fe/H] are of the order of 0.12-0.25$\,$dex, our relations offer the advantage to be applicable to a significant number of stars.

This relation can be difficult to apply for galaxies beyond our local group as observed with HST or JWST. Indeed our empirical relations are limited to stars for which the photometry provides a good coverage of the pulsation cycle for applying a Fourier decomposition. This is not the case for Cepheids observed by space telescopes, which typically contain only a few epochs and make use of templates to reconstruct the light-curve. A second limitation comes from the blending, which affects the extragalactic Cepheids ($D>10\,$Mpc) more significantly, since the resolution of telescopes is not able to separate Cepheids from the stellar background \citep[see, e.g.,][]{Mochejska2000,Riess2020}. Although Fourier parameters ($R_{i1}$ and $\phi_{i1}$) are only slightly affected by the blending \citep{Antonello2002}, this is not the case of the amplitudes and the harmonics on a magnitude scale. Thus an estimation of the blending for correcting the amplitudes would be necessary when applying empirical relations.

\section{Conclusion}\label{sect:conclusion}
In this work, we gathered 586 [Fe/H] abundances for fundamental Cepheids from the literature and we transformed them into a unified scale. We then retrieved well-sampled light curves for 545 of these stars from different $V$-band catalogs. Using this data set, we performed a Fourier decomposition of the light curves. Then we fit the linear relations with all possible combinations of Fourier parameters (up to four at the same time) to provide the best empirical relations for estimating the metallicity of short-period ($2.5<P<6.3\,$days) and long-period Cepheids ($12<P<40\,$days).

In the case of short-period Cepheids, we opted to use relations based on explicit amplitudes $A_1$ and $A_2$. Testing these relations on a sample of SMC and LMC Cepheids show that the relations are able to distinguish the populations. In the case of long-period Cepheids, we found that the metallicity can be estimated from the Fourier parameters $A_1$, $\phi_{21}$, and $R_{41}$. We also found that this empirical relation can accurately derive the mean metallicity of a sample of SMC and LMC Cepheids. For individual metallicity determinations, precisions of $\sigma$=0.12$\,$dex and $\sigma$=0.25$\,$dex have to be attributed for short- and long-period Cepheids respectively. These relations can be more efficient for estimating the average metallicity of a group of Cepheids rather than isolated stars.
 We then established new interrelations between $V$ and $I$ bands using OGLE data to convert these relations into the $I$ band.
We have shown that our empirical relations are able to derive the mean metallicity of OGLE sample of the MW, SMC, and LMC with a proper agreement with spectroscopic mean estimates found in the literature. We also tested the relations in the $I$ band to map the metallicity distributions of the MW, SMC, and LMC and we were able to derive a metallicity gradient that is consistent with the literature. 
Finally, we find that the calibration of empirical relations can be still improved on the basis of  further spectroscopic observations and homogeneous photometries in the $V$ and $I$ band.

\begin{acknowledgements}The authors would like to thank Dorota Skowron and Anna Jacyszyn-Dobrzeniecka for their valuable comments and discussion.
VH, RS, OZ and RSR are supported by the National Science Center, Poland,
Sonata BIS project 2018/30/E/ST9/00598. This research made use of the SIMBAD and VIZIER databases at CDS, Strasbourg (France) and the electronic bibliography maintained by the NASA/ADS system. This research also made use of Astropy, a community-developed core Python package for Astronomy \citep{astropy2018,astropy2022}. 
\end{acknowledgements}

\bibliographystyle{aa}  
\bibliography{bibtex_vh} 

\begin{appendix}

\onecolumn
\section{Calibration data set}

\centering
\begin{longtable}{llSSccccccccc}
\caption{Final data set of Fourier parameters of $V$-band light-curves with spectroscopic metallicities. \label{Tab:data_set} }
\\
\hline
\multicolumn{1}{c}{Star} & \multicolumn{1}{c}{Source} &
\multicolumn{1}{c}{Period} &
\multicolumn{1}{c}{[Fe/H]} &
\multicolumn{1}{c}{Ref.} &
\multicolumn{1}{c}{$A_1$} &
\multicolumn{1}{c}{$\phi_{21}$} &
\multicolumn{1}{c}{$\phi_{31}$} &
\multicolumn{1}{c}{$\phi_{41}$} &
\multicolumn{1}{c}{$R_{21}$} &
\multicolumn{1}{c}{$R_{31}$} &
\multicolumn{1}{c}{$R_{41}$} &
\multicolumn{1}{c}{$\sigma$} \\ \hline 
\endfirsthead

\multicolumn{3}{c}%
{{\bfseries \tablename\ \thetable{} -- continued}} \\
\hline \multicolumn{1}{c}{Star} & \multicolumn{1}{c}{Source} &
\multicolumn{1}{c}{Period} &
\multicolumn{1}{c}{[Fe/H]} &
\multicolumn{1}{c}{Ref.} &
\multicolumn{1}{c}{$A_1$} &
\multicolumn{1}{c}{$\phi_{21}$} &
\multicolumn{1}{c}{$\phi_{31}$} &
\multicolumn{1}{c}{$\phi_{41}$} &
\multicolumn{1}{c}{$R_{21}$} &
\multicolumn{1}{c}{$R_{31}$} &
\multicolumn{1}{c}{$R_{41}$} &
\multicolumn{1}{c}{$\sigma$} \\ \hline 
\endhead
\hline
\endfoot

\hline \hline
\endlastfoot
U Aql & BER08 & 7.024 & 0.081&L18& 0.322&3.306&5.891 & 5.656 & 0.345 & 0.146 & 0.028&0.024\\
SZ Aql & BER08 & 17.139 & 0.151&L18& 0.493&2.891&5.092 & 1.426 & 0.277 & 0.161 & 0.143&0.021\\
TT Aql & BER08 & 13.755 & 0.100&L18& 0.443&2.893&4.681 & 1.000 & 0.217 & 0.139 & 0.132&0.021\\
EV Aql & ASAS & 38.576 & -0.021&L18& 0.309&3.025&6.197 & 3.086 & 0.373 & 0.137 & 0.080&0.052\\
FM Aql & BER08 & 6.114 & 0.052&L18& 0.311&2.980&5.759 & 2.017 & 0.354 & 0.110 & 0.034&0.022\\
FN Aql & ASAS & 9.482 & -0.113&L18& 0.259&5.796&3.474 & 0.192 & 0.021 & 0.079 & 0.043&0.014\\
KL Aql & ASAS & 6.108 & 0.226&L18& 0.300&3.056&5.709 & 2.639 & 0.381 & 0.109 & 0.054&0.024\\
$\eta$ Aql & BER08 & 7.177 & 0.054&L18& 0.333&3.391&5.938 & 0.237 & 0.337 & 0.172 & 0.033&0.028\\
V336 Aql & ASAS & 7.304 & 0.081&L18& 0.308&3.350&6.009 & 6.070 & 0.346 & 0.146 & 0.038&0.014\\
V480 Aql & ASAS-SN & 18.996 & 0.170&T22& 0.358&2.617&0.032 & 1.050 & 0.245 & 0.131 & 0.077&0.019\\
V493 Aql & ASAS-SN& 2.987 & -0.048&L18& 0.268&2.563&5.381 & 0.877 & 0.307 & 0.139 & 0.062&0.027\\
V600 Aql & ASAS & 7.240 & -0.110&L18& 0.278&3.409&6.258 & 0.452 & 0.361 & 0.121 & 0.034&0.018\\
V733 Aql & BER08 & 6.179 & -0.022&L18& 0.197&3.058&5.072 & 5.317 & 0.300 & 0.045 & 0.040&0.024\\
V912 Aql & ASAS-SN& 4.400 & 0.140&R21& 0.296&2.680&5.411 & 2.093 & 0.375 & 0.172 & 0.062&0.014\\
V916 Aql & ASAS-SN& 13.442 & 0.248&L18& 0.386&2.676&4.257 & 0.575 & 0.201 & 0.117 & 0.112&0.018\\
V1162 Aql & BER08 & 5.376 & -0.025&L18& 0.223&2.885&5.964 & 3.751 & 0.335 & 0.087 & 0.009&0.027\\
V1495 Aql & ASAS-SN& 8.797 & 0.550&R21& 0.283&3.508&0.890 & 2.617 & 0.136 & 0.093 & 0.089&0.011\\
V1496 Aql & BER08 & 55.917 & -0.070&R21& 0.242&3.222&2.932 & 1.641 & 0.163 & 0.100 & 0.070&0.026\\
V340 Ara & BER08 & 20.810 & 0.381&L18& 0.467&2.828&5.102 & 1.463 & 0.298 & 0.161 & 0.116&0.022\\
Y Aur & ASAS-SN& 3.860 & -0.074&L18& 0.283&2.696&5.589 & 2.124 & 0.345 & 0.214 & 0.058&0.040\\
RT Aur & ASAS-SN& 3.729 & 0.032&L18& 0.363&2.610&5.393 & 1.681 & 0.363 & 0.203 & 0.107&0.077\\
RX Aur & BER08 & 11.624 & -0.045&L18& 0.301&2.448&3.928 & 0.138 & 0.142 & 0.104 & 0.065&0.021\\
SY Aur & BER08 & 10.145 & -0.063&L18& 0.300&2.748&2.530 & 3.908 & 0.155 & 0.042 & 0.038&0.024\\
YZ Aur & BER08 & 18.193 & -0.359&L18& 0.366&2.957&4.702 & 1.285 & 0.188 & 0.110 & 0.117&0.022\\
AN Aur & BER08 & 10.289 & -0.211&L18& 0.307&3.954&1.809 & 3.574 & 0.131 & 0.151 & 0.105&0.015\\
AO Aur & ASAS-SN& 6.762 & -0.345&L18& 0.361&3.070&5.725 & 2.727 & 0.444 & 0.188 & 0.108&0.042\\
AS Aur & ASAS-SN& 3.175 & -0.203&L18& 0.268&2.616&5.308 & 1.643 & 0.435 & 0.212 & 0.098&0.016\\
AX Aur & ASAS-SN& 3.047 & -0.140&L18& 0.217&2.512&5.212 & 1.432 & 0.377 & 0.165 & 0.070&0.018\\
BK Aur & BER08 & 7.805 & -0.025&L18& 0.320&3.613&0.410 & 6.184 & 0.220 & 0.120 & 0.031&0.039\\
CY Aur & ASAS-SN& 13.845 & -0.259&L18& 0.377&3.012&4.493 & 0.944 & 0.192 & 0.119 & 0.144&0.011\\
ER Aur & ASAS-SN& 15.698 & -0.367&L18& 0.250&2.356&4.387 & 0.403 & 0.202 & 0.133 & 0.079&0.012\\
EW Aur & ASAS-SN& 2.660 & -0.537&L18& 0.293&2.561&5.313 & 1.735 & 0.493 & 0.277 & 0.142&0.017\\
GT Aur & ASAS-SN& 4.405 & -0.074&L18& 0.340&2.582&5.248 & 1.736 & 0.380 & 0.162 & 0.077&0.022\\
GV Aur & ASAS-SN& 5.260 & -0.243&L18& 0.350&2.876&5.775 & 2.366 & 0.461 & 0.205 & 0.106&0.013\\
IN Aur & ASAS-SN& 4.910 & -0.285&L18& 0.221&2.981&6.061 & 2.815 & 0.404 & 0.157 & 0.055&0.028\\
V335 Aur & BER08 & 3.413 & -0.315&L18& 0.371&2.636&5.315 & 1.658 & 0.475 & 0.256 & 0.134&0.021\\
T Ant & ASAS & 5.898 & -0.275&L18& 0.345&3.027&5.916 & 2.550 & 0.446 & 0.170 & 0.092&0.017\\
RW Cam & BER08 & 16.415 & 0.050&L18& 0.363&2.768&4.574 & 0.883 & 0.208 & 0.136 & 0.121&0.016\\
RX Cam & BER08 & 7.909 & -0.011&L18& 0.317&3.295&0.040 & 1.344 & 0.258 & 0.086 & 0.143&0.030\\
TV Cam & ASAS-SN& 5.295 & -0.056&L18& 0.371&2.878&5.696 & 2.103 & 0.434 & 0.175 & 0.104&0.013\\
AB Cam & ASAS-SN& 5.787 & -0.146&L18& 0.365&3.003&5.658 & 2.384 & 0.438 & 0.168 & 0.115&0.015\\
AC Cam & ASAS-SN& 4.157 & -0.182&L18& 0.266&2.801&5.599 & 2.395 & 0.401 & 0.181 & 0.088&0.019\\
AD Cam & ASAS-SN& 11.262 & -0.343&L18& 0.364&3.009&4.727 & 0.966 & 0.278 & 0.167 & 0.198&0.015\\
AM Cam & ASAS-SN& 3.997 & -0.189&L18& 0.232&2.714&5.532 & 2.354 & 0.389 & 0.164 & 0.059&0.012\\
CF Cam & ASAS-SN& 9.435 & -0.170&R21& 0.283&3.120&0.447 & 2.054 & 0.165 & 0.084 & 0.079&0.014\\
CK Cam & ASAS-SN& 3.295 & -0.040&L18& 0.262&2.599&5.422 & 1.410 & 0.344 & 0.136 & 0.061&0.068\\
LO Cam & ASAS-SN& 12.646 & -0.149&L18& 0.333&2.585&3.753 & 5.850 & 0.201 & 0.088 & 0.068&0.040\\
MN Cam & ASAS-SN& 8.179 & -0.084&L18& 0.260&3.727&0.722 & 2.088 & 0.254 & 0.135 & 0.088&0.024\\
MQ Cam & ASAS-SN& 6.604 & -0.163&L18& 0.247&3.282&6.070 & 5.644 & 0.368 & 0.126 & 0.026&0.017\\
OX Cam & ASAS-SN& 5.066 & -0.122&L18& 0.301&2.772&5.512 & 2.213 & 0.401 & 0.136 & 0.061&0.022\\

. . .  &  &  &  & &  &  &  & &  &  &  & \\

\end{longtable}

    \begin{tablenotes}
      \item \textbf{Notes:} This table is available in its entirety at the CDS. This data set is composed of 545 Cepheids with metallicity from about $-1.0$ to $0.4\,$dex which are from L11 \citep{luck11}; L18 \citep{Luck2018}; G14 \citep{Genovali2014}; G15 \citep{Genovali2015}; M15 \citep{Martin2015}; A16 \citep{Andrievsky2016}; L17 \cite{Lemasle2017}; R21 \citep{Ripepi2021}; R08 \citep{Romaniello2008}; R22 \citep{Romaniello2022}; T22 \citep{Trentin2023} and K22 \citep{Kovtyukh2022}. All [Fe/H] results are scaled to the solar abundance value A(Fe)$_\odot$=7.50 given by \cite{Asplund2009} and corrected from offsets if necessary; see Sect.~\ref{sect:data_set} for details. Sources of the $V$-band light curve are from ASAS-SN \citep{ASAS2018}; ASAS \citep{Pojmanski2002}; BER08 \citep{Berdnikov2008}; OGLE-III and OGLE-IV \citep{Soszynski2008,Soszynski2010}. The dispersion of the fit $\sigma$ is indicated, as defined in Eq.~\ref{eq:sigma}.
    \end{tablenotes}

\newpage
\section{Interrelations}
\begin{figure*}[htb!] 
\begin{subfigure}{0.50\textwidth}
\includegraphics[width=\linewidth]{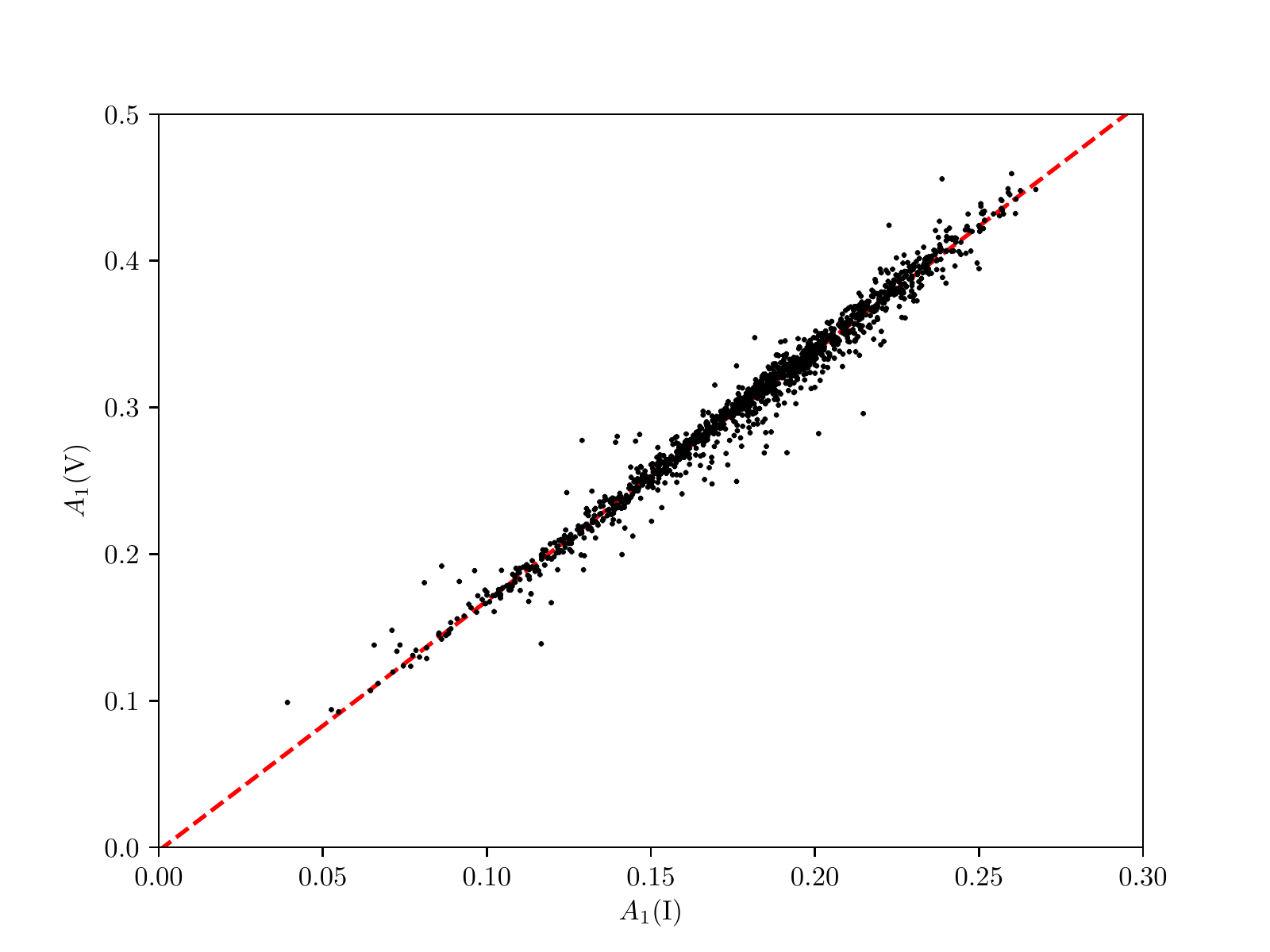}
\caption{}
\end{subfigure}\hspace*{\fill}
\begin{subfigure}{0.50\textwidth}
\includegraphics[width=\linewidth]{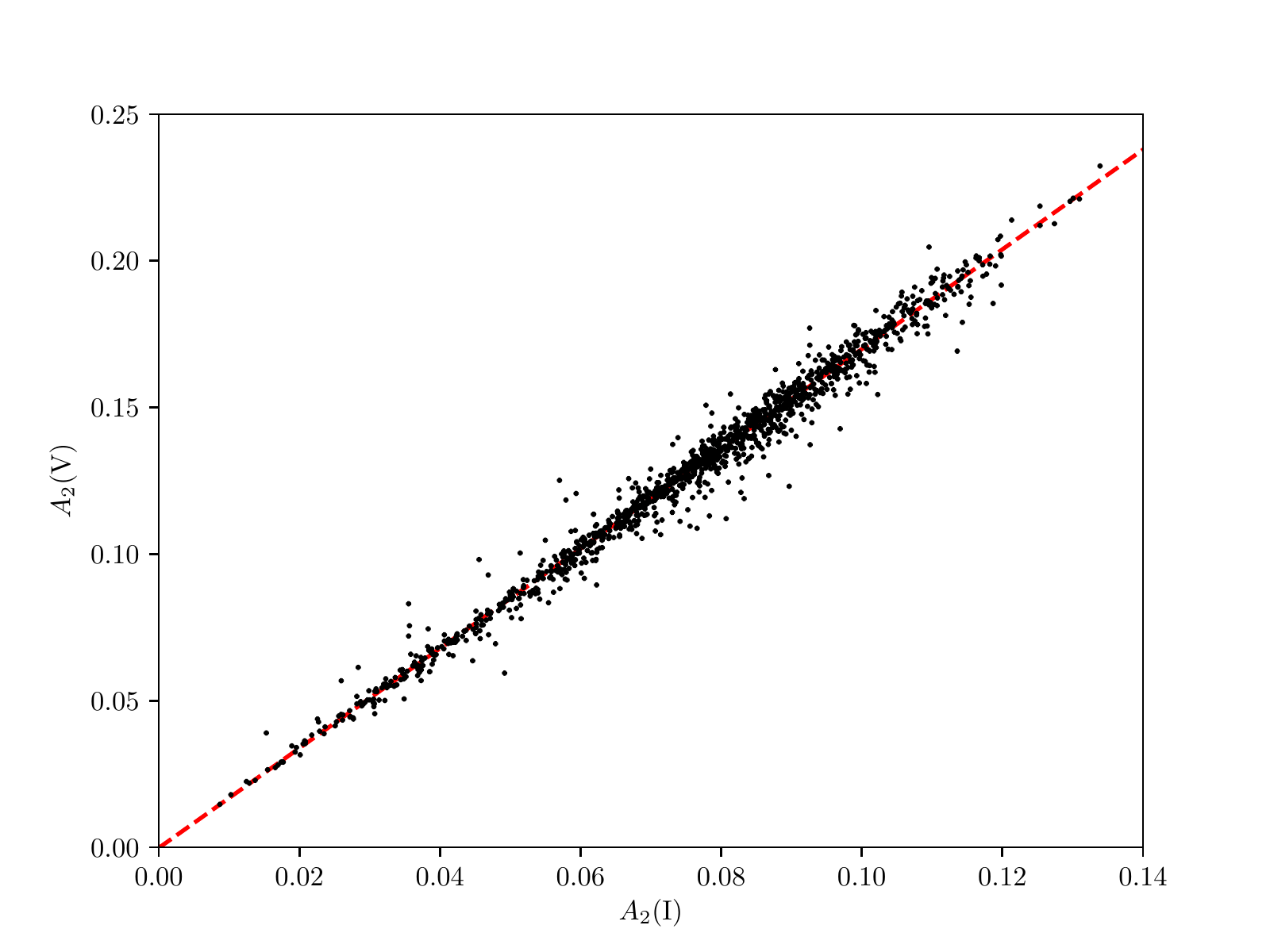}
\caption{}
\end{subfigure}
\begin{subfigure}{0.50\textwidth}
\includegraphics[width=\linewidth]{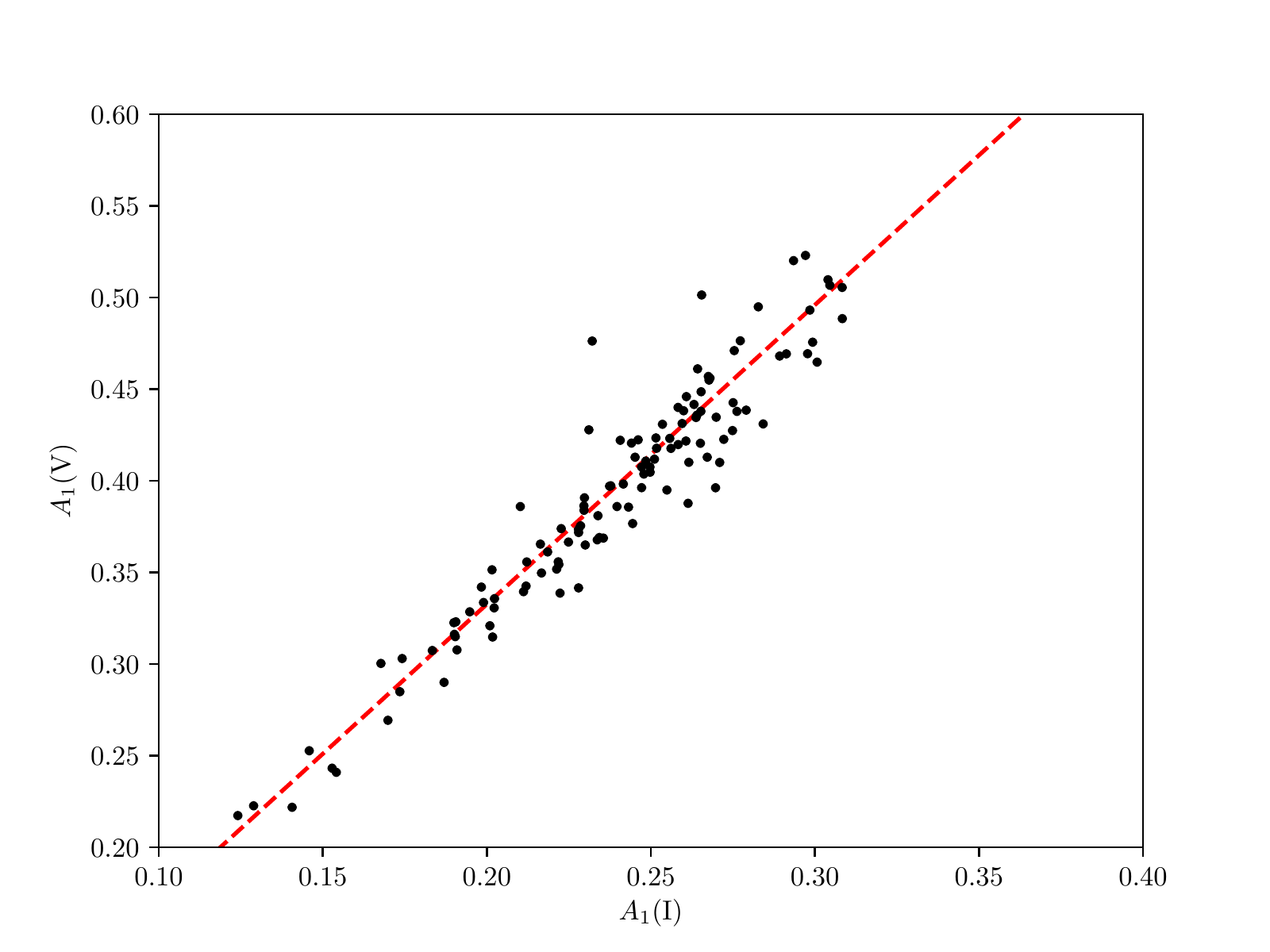}
\caption{} 
\end{subfigure}\hspace*{\fill}
\begin{subfigure}{0.50\textwidth}
\includegraphics[width=\linewidth]{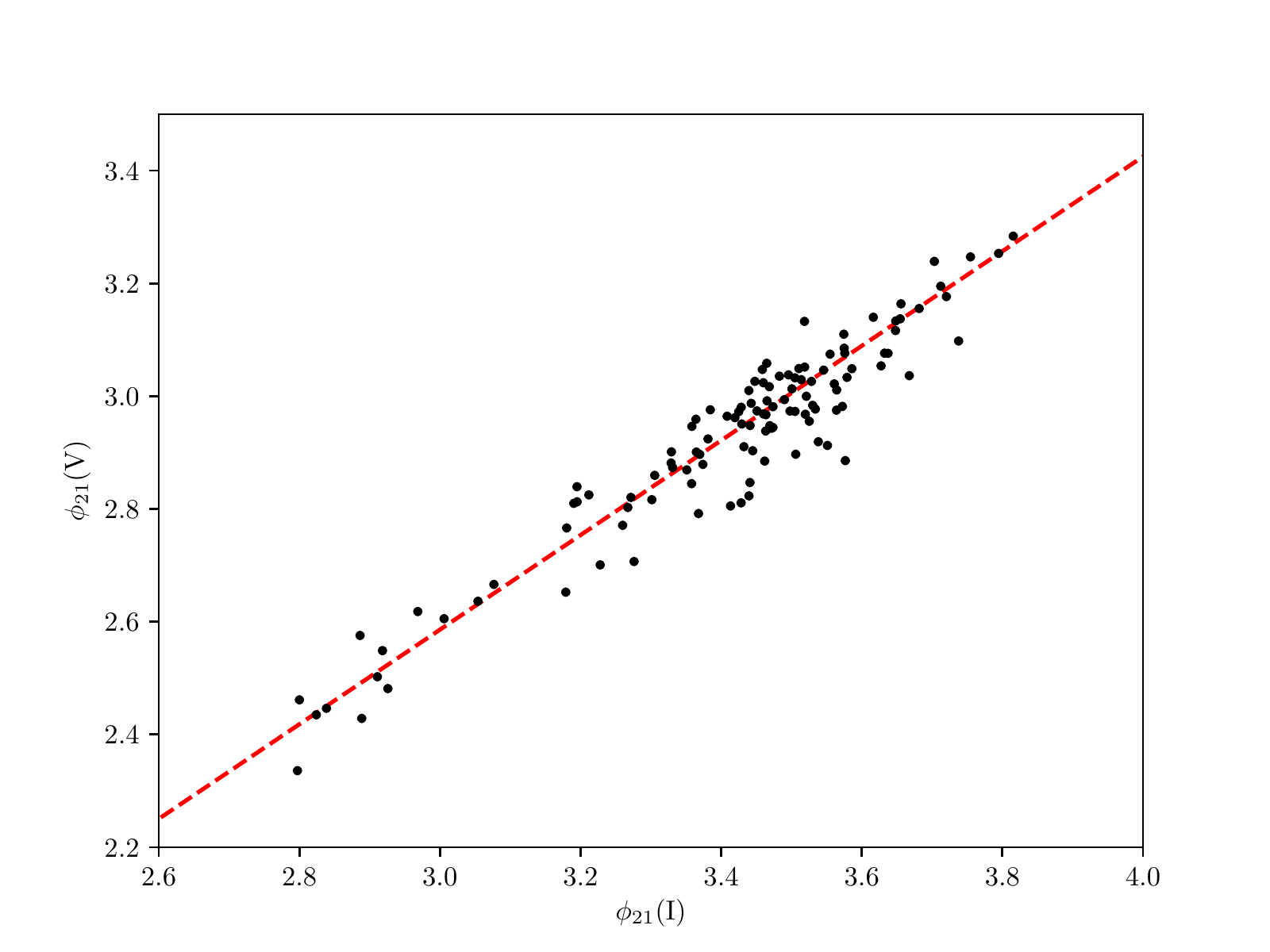}
\caption{}
\end{subfigure}
\begin{subfigure}{0.50\textwidth}
\includegraphics[width=\linewidth]{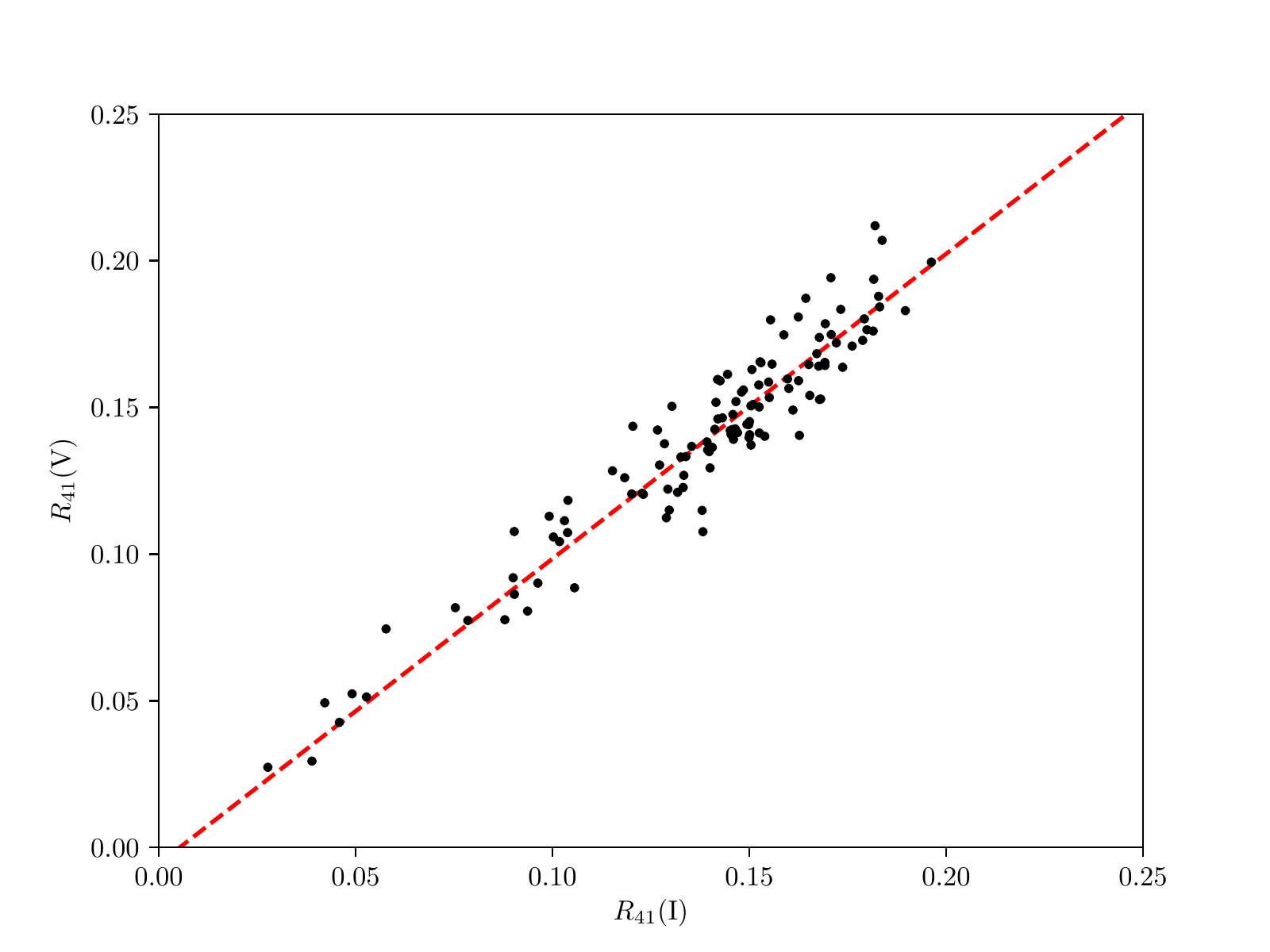}
\caption{}
\end{subfigure}\hspace*{\fill}
\caption{\small (a) and (b): Interrelations for $A_1$ and $A_2$ between $V$ and $I$-bands for short short-period Cepheids between 2.5 and 6.3 days. (c), (d) and (e): Interrelations for $A_1$, $R_{41}$, and $\phi_{21}$ between $V$ and $I$-bands for long-period Cepheids between 12 and 40 days.\label{fig:inter}}
\end{figure*}

\FloatBarrier
\section{MW and LMC maps}
\begin{figure*}
\centering
\includegraphics[width=0.75\textwidth]{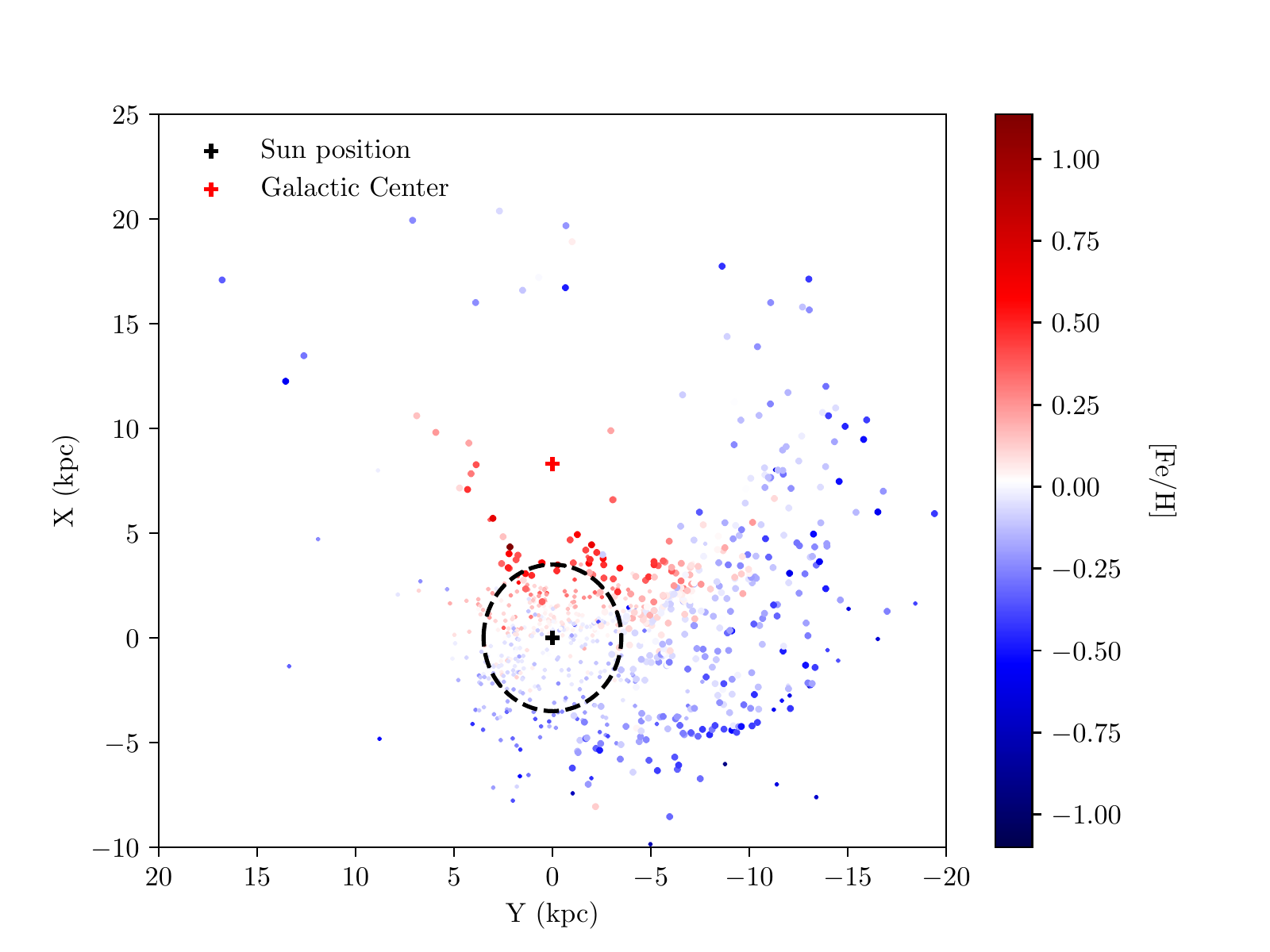}
\caption{\small MW on-view map of the metallicity distribution (868 stars) from empirical metallicity relations in the $I$ band and values from the literature. Symbols have same meanings as in Fig.~\ref{fig:map}.}\label{fig:MW_map_annex}
\FloatBarrier
\end{figure*}
\begin{figure*}[htb!]
\centering
\includegraphics[width=0.75\textwidth]{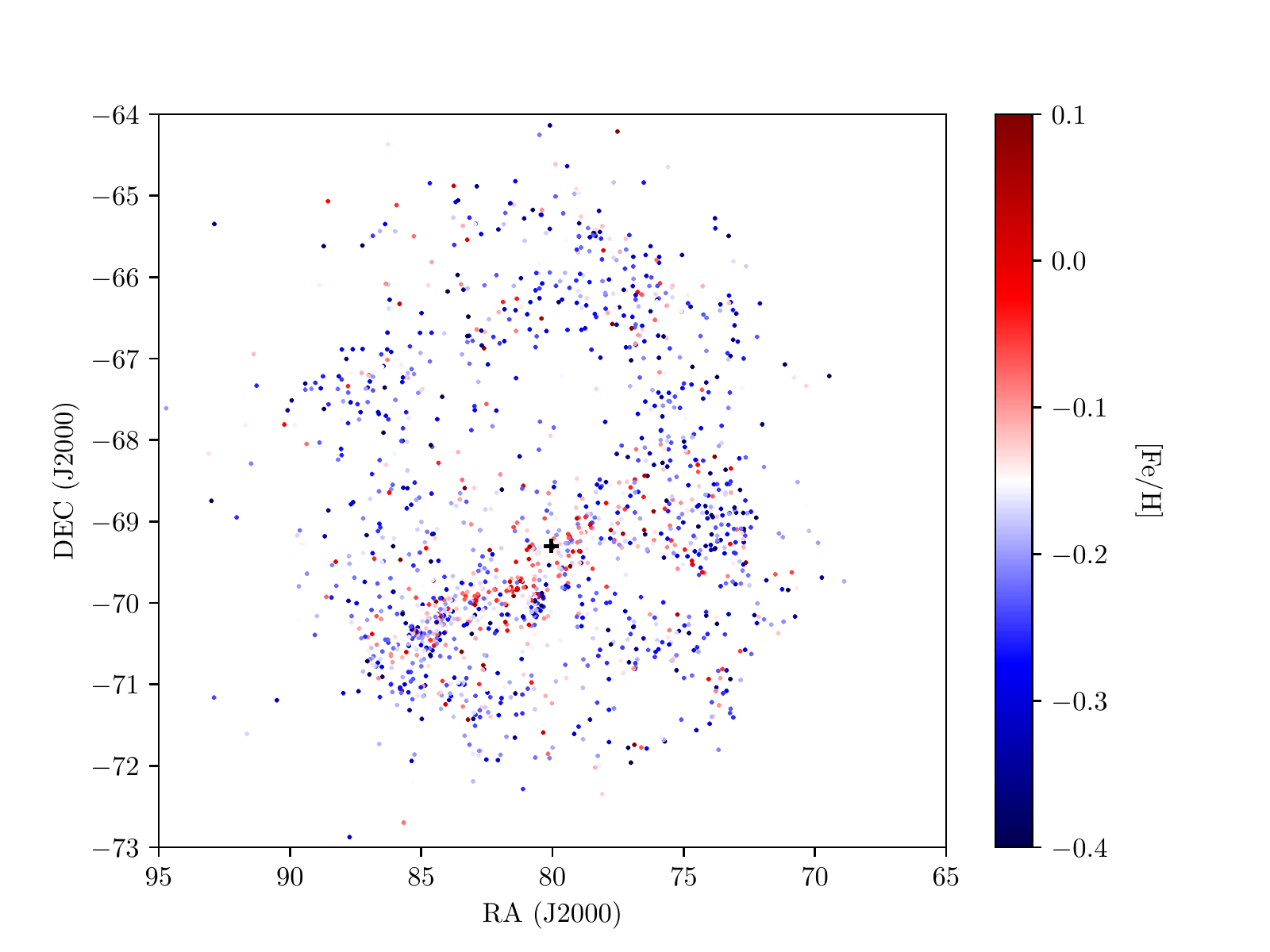}
\caption{\small LMC sky map of the metallicity distribution (1561 stars) from empirical metallicity relations in the $I$ band.} \label{fig:LMC_map_zoom}
\end{figure*}

\end{appendix}

\end{document}